\documentclass[
 twocolumn,
 preprintnumbers,
 amsmath,amssymb,
 aps,
 superscriptaddress,
]{revtex4}

\usepackage{color}
\usepackage{graphicx}
\usepackage{bm}
\usepackage{here}
\usepackage{url}
\usepackage{subfigure}
\usepackage{mhchem}

\begin{document}
\title{Information-thermodynamic characterization of stochastic Boolean networks}

\author{Shun Otsubo}
\author{Takahiro Sagawa}
\affiliation{
 Department of Applied Physics, The University of Tokyo, 7-3-1 Hongo, Bunkyo-ku, Tokyo 113-8656, Japan\\
}

\date{\today}

\begin{abstract}
Recent progress in experimental techniques has enabled us to quantitatively study stochastic and flexible behavior of biological systems.
For example, gene regulatory networks perform stochastic information processing and their functionalities have been extensively studied.
In gene regulatory networks, there are specific subgraphs called network motifs that occur at frequencies much higher than those found in randomized networks. Further understanding of the designing principle of such networks is highly desirable. In a different context, information thermodynamics has been developed as a theoretical framework that generalizes non-equilibrium thermodynamics to stochastically fluctuating systems with information. Here we systematically characterize gene regulatory networks on the basis of information thermodynamics. We model three-node gene regulatory patterns by a stochastic Boolean model, which receive one or two input signals that carry external information. For the case of a single input, we found that all the three-node patterns are classified into four types by using information-thermodynamic quantities such as dissipation and mutual information, and reveal to which type each network motif belongs. Next, we consider the case where there are two inputs, and evaluate the capacity of logical operation of the three-node patterns by using tripartite mutual information, and argue the reason why patterns with fewer edges are preferred in natural selection. This result might also explain the difference of the occurrence frequencies among different types of feedforward-loop network motifs.
\end{abstract}


\maketitle

\section{Introduction}
\begin{figure}
\includegraphics[width = \linewidth]{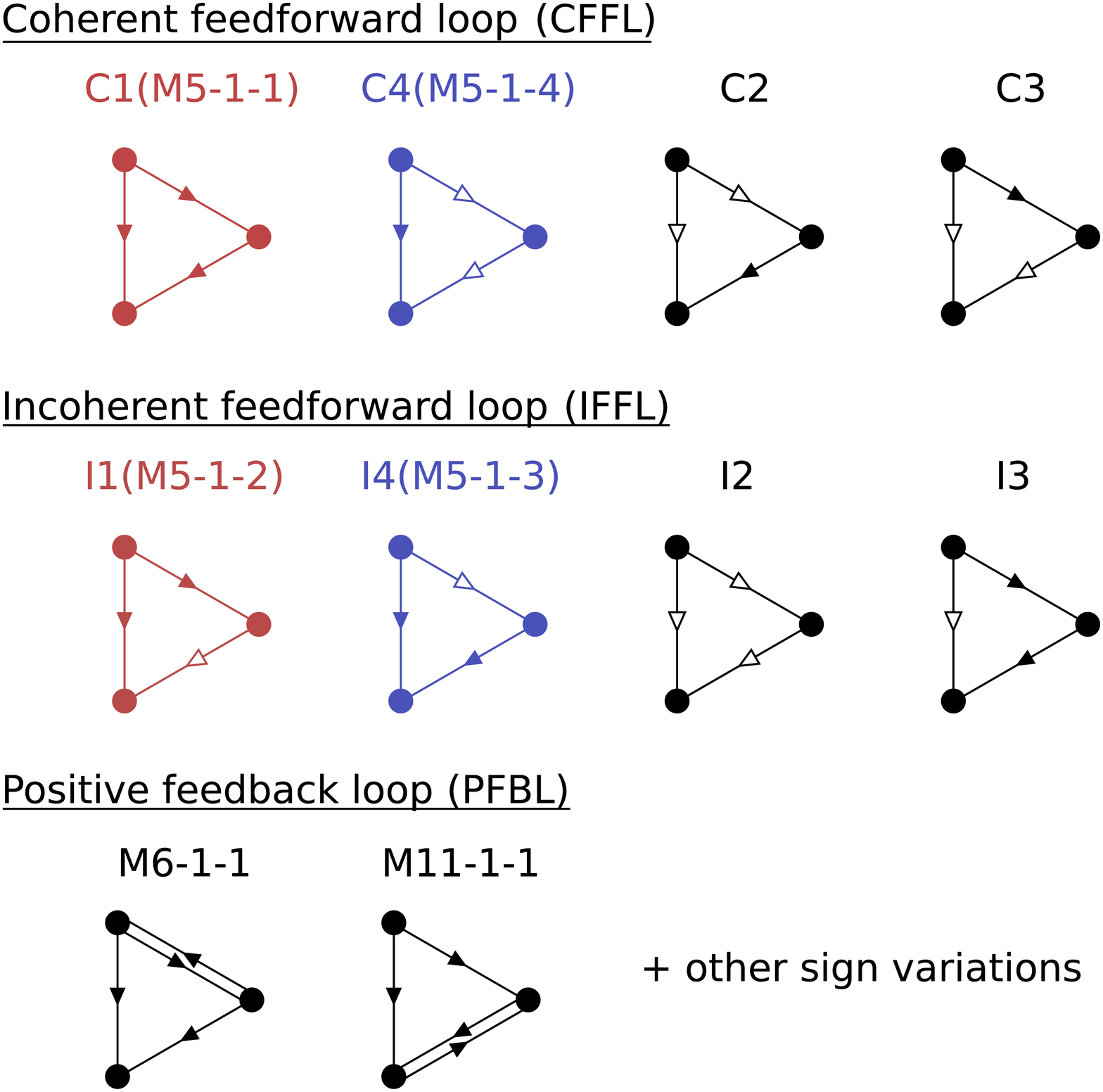}
\caption{Gene regulatory network motifs:
We list all the three-node network motifs. 
We refer to these patterns as network motifs, because according to previous studies, they occur statistically significantly in gene regulatory networks than those in random networks.
This listing and the following explanation are based on Ref.~\cite{Alon2006, Alon2007}.
A node represents a gene, and an arrow represents regulation between genes.
Black arrows represent activation, and white ones represent inhibition.
There are eight types of FFLs, which are categorized into two groups according to whether the signs of the two regulation paths from the top to the bottom node are equivalent or not. It is known that the CFFLs can act as low-pass filters and the IFFLs show adaptive behavior. There are a lot of sign variations in the PFBL network motifs (see Fig.~\ref{fig: pfblnm}), thus we list two of them above (and the others are just the sign variations of these two patterns). The PFBL network motifs are known to have two functions. One is keeping their states static, and the other is having their states bistable (see Appendix~\ref{sec: pfbl}).
In this study, we refer to a positive feedback loop that are categorized into network motif as PFBLs.
}
\label{fig: motif}
\end{figure}

\indent
In the recent development of single-cell technologies, quantitative biology has attracted much attention \cite{Taniguchi2010}, where one of the hot topics is the study of network motifs in gene regulatory networks \cite{Shen-Orr2002, Milo2002, Lee2002, Alon2007}. Complex gene regulatory networks are constituted of specific subgraphs called network motifs that occur much more frequently than those in random networks.\\ \indent
In this study, we focus on three-node patterns, which can be regarded as building blocks of gene regulatory networks. We list all the three-node network motifs in Fig.~\ref{fig: motif} (see the caption for the details).
Many studies have investigated the function of network motifs in order to reveal the reason why such specific patterns are preferred compared to the others in natural selection \cite{Shen-Orr2002,Mangan2003, Alon2006, Alon2007, Mangan2006, Kittisopikul2010, Macneil2011, DeRonde2012, Albergante2014}. Interestingly, they are commonly found in gene regulatory networks across species, including {\it E.coli}, yeast, mouse and human \cite{Lee2002, Alon2006, Alon2007, Swiers2006, Gerstein2012}. This suggests that there is a guiding principle for the network formation, while we have not yet clearly understood which properties distinguish network motifs from the others. We note that there are few researches that characterize network patterns exhaustively focusing on information quantities and thermodynamic cost, while some studies systematically investigated three-node patterns \cite{Prill2005} based on the control theory perspective (e.g., a stability analysis).\\
\indent
In statistical physics, information thermodynamics has been developed in the last decade on the basis of stochastic thermodynamics \cite{Jarzynski2000, Sekimoto2010, Seifert2012}, which clarifies the relation between thermodynamic quantities and information \cite{Sagawa2010, Toyabe2010, Sagawa2012, Ito2013, Horowitz2014, Horowitz2014a, Parrondo2015, Shiraishi2015}. Since biological systems, especially cells, can be regarded as information processing systems that operate by consuming energy sources such as ATP molecules, information thermodynamics can be applied to biological systems. In fact, several studies have revealed information-thermodynamic structure in, for example, sensory adaptation of \textit{E. coli} chemotaxis \cite{Sartori2014, Barato2014, Ito2015, Hartich2016, Ouldridge2017a, Matsumoto2018}. However, biochemical reaction networks including gene regulatory networks are yet to be further investigated.\\ \indent
Here we systematically characterize gene regulatory network patterns on the basis of information thermodynamics.
We calculate information-thermodynamic quantities such as the efficiency of information propagation and energetic dissipation to characterize all the possible three-node patterns (704 patterns), by using a stochastic Boolean model.
Since the stochastic Boolean model is a coarse-grained model, the dissipation obtained here is a lower bound of the real entropy production \cite{Esposito2012, Kawaguchi2013}.
First, we consider the case where there is a single input signal to a three-node pattern, and reveal that all the patterns can be classified into four types. These four types are characterized as dissipative, static, informative, and adaptive. The coherent feedforward loop (CFFL) network motifs belong to the informative type, and the incoherent feedforward loop (IFFL) network motifs belong to the adaptive type. The positive feedback loop (PFBL) network motifs belong to the static type or informative type, and no network motifs are categorized into the dissipative type.\\ \indent
We next consider the case where there are two input signals to a three-node pattern. We evaluate the capacity of logical operation on these two inputs by using tripartite mutual information, and reveal that the FFLs outperform the others. This result clearly accounts for network patterns that occur frequently in gene regulatory networks. 
It could also be a reasonable explanation for the fundamental difference of the occurrence frequencies among different types of FFLs \cite{Mangan2006, Kittisopikul2010}.\\ \indent
These results indicate that gene regulatory networks are efficient in terms of information propagation and thermodynamics. The key features of our work are the following.
First, information thermodynamics enables us to quantify dissipation of only a three-node pattern in a large-scale network, by taking the effect of the correlation as the learning rate. We do not make the assumption of an isolation of a three-node pattern from a network, which is the very benefit of using information thermodynamics. Second, we found that tripartite mutual information is useful to characterize the capacity of logical operation performed by three-node patterns. Our approach with these features can be applied to wide range of networks including both artificial and biochemical ones. \\ \indent
This paper is organized as follows. In Sec.~\ref{sec: setup}, we introduce the stochastic Boolean model and describe the setup of this study. We also review the fundamental concepts in information thermodynamics. In Sec.~\ref{sec: main}, we show our main results. In the first part, we consider the case of a single input and classify all the three-node patterns on the basis of information-thermodynamic quantities. 
In the second part, we consider the case of two inputs and calculate tripartite mutual information to evaluate the capacity of logical operation. In Sec.~\ref{sec: conclusion}, we make concluding remarks.

\section{Setup}
\label{sec: setup}
In this section, we formulate the setup of this study with a stochastic Boolean model. We also briefly review information thermodynamics.
\subsection{Stochastic Boolean model}
\begin{figure}
\includegraphics[width = 0.95\linewidth]{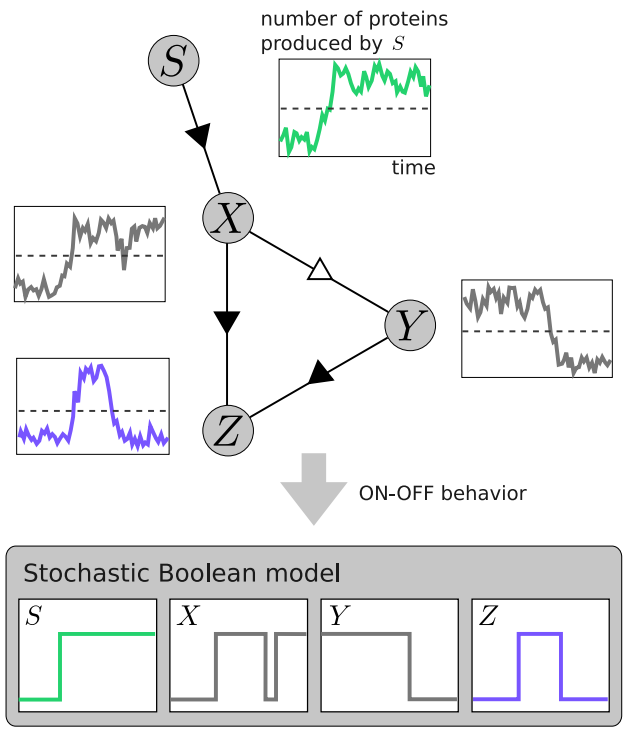}
\caption{Stochastic Boolean model: A three-node pattern has a single input from the $S$ node.
Typical dynamics of these nodes are schematically shown.
Since we can observe the on-off behavior of genes and the fluctuation around it, we model the dynamics by the stochastic Boolean model.
}
\label{fig: setup}
\end{figure}
We first discuss the general properties of gene regulation. 
In this study, we restrict ourselves to transcriptional regulation, while post-transcriptional regulation is also important \cite{Filipowicz2008, Herranz2010, Liu2015}.
A gene regulatory network is represented by a graph, whose nodes and directed edges represent genes and their regulation, respectively. In this study, we focus on three-node patterns which are elementary components of gene regulatory networks.\\ \indent
A three-node pattern receives one or two input signals as shown in Fig.~\ref{fig: setup} and Fig.~\ref{fig: setup2}, and passes information to the next nodes by processing the signals. In the single-input case a pattern just propagates information, while in the two-input case it performs logical operation. \\ \indent
There are two types of regulation: activation and inhibition. 
Therefore, there are $3^6 - 8\cdot 3 - 1 = 704$ patterns in total if we exclude auto-regulation. \\ \indent
We next formulate the stochastic Boolean model \cite{DeJong2002, Alvarez-Buylla2008, So2011}. Suppose in general that there are $N$ nodes, and each node $X^i_t~(i = 1, 2, \cdots, N)$ takes 0 or 1 at continuous time $t$. The value of each node represents whether the number of proteins produced by the gene exceeds a threshold value as shown in Fig.~\ref{fig: setup}. 
Inhibitory regulation is described by the NOT gate. 
Regulation by multiple nodes is expressed by a logic function such as AND or OR, and the state of a node $i$ is determined by the value of a regulatory function $f^i(X^j_t, X^k_t, ...)$ 
with nodes $j, k...$ regulating the node $i$. In this study, for the sake of simplicity, we assume each regulatory function as the AND gate, which is considered as one of the major regulatory functions in gene regulatory networks. (The AND and OR gates are considered to be the two major regulatory functions \cite{Alon2006}. The relation between our results and the assumption of regulatory functions is discussed in Appendix \ref{sec: app3}.) 
For example, in the case of Fig.~\ref{fig: setup}, regulatory functions take the following expressions: $f^X(S_t) = S_t, ~f^Y(X_t) = {\rm NOT}(X_t)$ and $f^Z(X_t, Y_t) = {\rm AND}(X_t, Y_t)$.\\ \indent
In real gene regulatory networks, the state of a gene $i$ (i.e., $X_t^i = 0, 1$) changes stochastically. Therefore, the time evolution of a gene expression is described by the master equation with the following transition rates:
\begin{eqnarray}
\ce{{\it X^i} = 0 <=>C[{\it\gamma_i}][{\it\gamma_i\cdot e_i}]{\it X^i} = 1} &{\rm ~~~if~~~}& f^i = 1,\\
\ce{{\it X^i} = 0 <=>C[{\it\gamma_i\cdot e_i}][{\it\gamma_i}]{\it X^i} = 1} &{\rm ~~~if~~~}& f^i = 0.
\end{eqnarray}
Here, $\gamma_i$ is a transition rate and $e_i$ is the reverse transition ratio. 
The transition matrix of the master equation is constructed by the above transition rates (see Appendix \ref{sec: app1} for an example). \\ \indent
Such Boolean models have been used to describe the behavior of gene regulatory networks composed of more than two nodes \cite{Bornholdt2008, Shmulevich2002, Alvarez-Buylla2008, Garg2009, Liang2012, Murrugarra2012}. Strictly speaking, a Boolean model might be somewhat inaccurate for describing oscillations induced by negative feedback loops \cite{Elowitz2000, Atkinson2003, Zhu2007, Hori2011, Hori2013, Tayar2017} and stationary states of positive feedback loops \cite{Gardner2000, Atkinson2003, Zhu2007, Kobayashi2003}. \\ \indent
In spite of these apparent disadvantages, we consider that the Boolean model is appropriate for our study from the following reasons.
(i) First of all, dynamics should be stochastic in order to calculate information quantities and dissipation within the framework of stochastic thermodynamics. In that sense, deterministic ODE equations \cite{Jong2002} are not suitable for our study. Since the stochastic Boolean model is the simplest stochastic and nonlinear model, we adopt it as a platform of our study. We note that the stochastic Boolean model can be derived from a stochastic ODE \cite{Meister2014} (i.e., a Langevin equation), by taking the Hill coefficients of regulatory functions large enough.
(ii) We consider that the stochastic Boolean model captures important aspects of information propagation in gene regulatory networks. It is experimentally supported that transcriptional networks perform computation by using binary states of genes \cite{Alon2006, Bornholdt2008}.
(iii) We consider that whether a positive feedback loop shows the bistable or static property depends on the regulatory functions, which is often neglected in previous studies. The Boolean model can capture the difference between these two properties.\\ \indent
We now discuss the detailed setup of this study. As shown in Fig.~\ref{fig: setup}, we consider the case where the signal source $S$ activates the input node $X$. 
The signal is then propagated to the output node $Z$ through the middle node $Y$.
Here, we assume that $X$, $Y$ and $Z$ represent genes, but $S$ is assumed to be either another gene or a signal molecule of $X$. In either case, the following discussion holds.\\ \indent
With this setup, we focus on how information flows from $S$ to $Z$. 
We assume that $S$ randomly flips between 0 and 1 with equal probabilities (i.e., $e_S=1$). 
We calculate information-thermodynamic quantities for the stationary state.
We set the parameters with the following conditions:
\begin{eqnarray}
\gamma &:=& \gamma_X = \gamma_Y = \gamma_Z \gg \gamma_S,\\
e &:=& e_X = e_Y = e_Z \ll 1.
\end{eqnarray}
Here, the condition $\gamma\gg\gamma_S$ comes from the assumption that the time scale of an external signal is slower than that of each node.
In addition, for the sake of simplicity, we assumed that the parameters of the three nodes $X$, $Y$ and $Z$ are the same. \\ \indent
We next consider the reduction of 704 patterns to 283 patterns by excluding irrelevant and equivalent patterns. 
In this study, we define the irrelevant patterns by the following criteria: (i) patterns without causal relationship from $X$ to $Z$ or $Y$ to $Z$, or (ii) patterns that have unregulated nodes.
There also exist patterns that have different edge signs but equal with respect to all information quantities, which we regard as equivalent patterns (see Appendix \ref{sec: app3} for the details). We pick up only a single pattern from equivalent ones for calculation. As a result, we actually perform calculation for 283 patterns (see Appendix \ref{sec: app4} for the details).

\subsection{Information thermodynamics}
In this section, we briefly review information thermodynamics.
The entropy production of a small subsystem can be reduced by measurement and feedback control by another subsystem. 
Information thermodynamics enables us to take into account the effect of feedback control by incorporating information quantities like mutual information, which quantifies the correlation between the two subsystems.\\ \indent
To connect information thermodynamics to our main setup smoothly,  we consider two stochastic variables $S_t$ and $Z_t$ that represent the states of node $S$ and node $Z$ at time $t$. Since they are stochastic variables, we can define the probability distribution $p(s_t, z_t)$. Here, capital letters $S_t$ and $Z_t$ describe stochastic variables and the small letters $s_t$ and $z_t$ describe their particular realizations. 
If $S_t$ and $Z_t$ form correlation, we can estimate the value of $S_t$ from the value of $Z_t$.
Such correlation between $S_t$ and $Z_t$ is quantified by the mutual information:
\begin{eqnarray}
I(S_t: Z_t) := \sum_{s_t, z_t\in{\{0, 1\}}}p(s_t, z_t)\ln\frac{p(s_t, z_t)}{p(s_t)p(z_t)}.
\end{eqnarray}
\indent
Mutual information is symmetric in terms of $S_t$ and $Z_t$, and  therefore mutual information cannot capture the directional information flow in stochastic dynamics. 
To characterize such information flow, we consider the learning rate $l_Z$ \cite{Horowitz2014, Brittain2017} and the transfer entropy $\overline{T}_{S\rightarrow Z}$ \cite{Schreiber2000}, which are respectively defined as
\begin{eqnarray}
&&l_Z(t) := \frac{I(S_t:Z_{t+dt}) - I(S_t:Z_t)}{dt},\\
&&\overline{T}_{S\rightarrow Z}(t) := \frac{I(S_t:\{Z_t, Z_{t+dt}\}) - I(S_t:Z_t)}{dt}.\label{eq: transfer}
\end{eqnarray}
These quantities are defined with the time series of the stochastic variables, and both of them characterize the increment of the mutual information during $Z_t$ evolves to $Z_{t+dt}$. Specifically, the learning rate quantifies the amount of information that the instantaneous value of $Z_t$ obtains,  while the transfer entropy quantifies the amount of information that $Z_t$ newly obtains. While the original transfer entropy \cite{Schreiber2000} is defined as the increment of the mutual information given the whole trajectory of $Z_t$ and $S_t$, the simplified version (\ref{eq: transfer}) that only considers the single-step condition is adopted in the following discussion \cite{Hartich2016, Matsumoto2018}.\\ \indent
This slight difference between the learning rate and the transfer entropy leads to an inequality \cite{Hartich2016, Matsumoto2018}
\begin{equation}
l_Z(t) \leq \overline{T}_{S\rightarrow Z}(t)\label{eq: inequality}.
\end{equation}
We note that the learning rate can take both positive and negative values, while the transfer entropy is always nonnegative.
If the learning rate becomes positive, $Z$ indeed obtains information from $S$. If it becomes negative, $Z$ consumes the correlation as a consequence of feedback control or just dissipation.\\ \indent
On the basis of inequality (\ref{eq: inequality}), it is reasonable to define the following quantity as a measure of the effectiveness of information gain by $Z$ \cite{Hartich2016, Matsumoto2018}:
\begin{eqnarray} 
\overline{C}_Z := \frac{l_Z}{\overline{T}_{S\rightarrow Z}}.
\end{eqnarray}
If the system is in the stationary state, the maximum sensory capacity, $\overline{C}_Z = 1$, is achieved if and only if $p(s_t|\{z_{t'}\}_{t'\leq t}) = p(s_t|z_t)$ \cite{Hartich2016, Matsumoto2018}, which means that $Z$ is a sufficient statistic of $S$.
This means that the latest value of $Z$ is enough for the estimation of $S$.
For example, the estimator of the Kalman filter is known to be a sufficient statistic. 
\\ \indent
So far we have discussed  information quantities, and we next consider a quantity that is more relevant to thermodynamics: information-thermodynamic dissipation \cite{Sagawa2010, Toyabe2010, Sagawa2012, Ito2013, Horowitz2014, Horowitz2014a, Parrondo2015, Horowitz2015}. 
Unlike conventional thermodynamic dissipation, information-thermodynamic dissipation explicitly includes the learning rate, which is defined for $Z$ as \cite{Horowitz2015}
\begin{eqnarray}
D_Z:= \frac{dS(Z)}{dt} - \frac{1}{T_Z}\frac{dQ_Z}{dt} - l_{Z:XY}\label{eq: DZ},
\end{eqnarray}
where the learning rate from $XY$ to $Z$ is defined as
\begin{eqnarray}
l_{Z:XY} := \frac{I(Z_{t+dt}:\{X_t, Y_t\}) - I(Z_t:\{X_t, Y_t\})}{dt}.
\end{eqnarray}
There are two types of terms in the information-thermodynamic dissipation (\ref{eq: DZ}): the entropic and the energetic terms.
The entropic terms describe the entropy change and the information flow (i.e., the learning rate). The energetic term (the second term) quantifies the consumption of chemical fuels such as ATP.\\ \indent
The first term is the derivative of the Shannon entropy $S(Z):= -\sum_{z_t}p(z_t)\ln p(z_t)$, which represents the entropy change in $Z$. 
The third term quantifies the change in the correlation between $Z$ and $XY$. This term is necessary to quantify dissipation of only a subsystem in a large network.\\ \indent
The second term is related to energetics, which in our setting quantifies a lower bound of actual consumption of chemical fuels \cite{Esposito2012, Kawaguchi2013}. In fact, the stochastic Boolean model is a coarse-grained model which only considers discretized protein numbers, where a lot of elementary processes are neglected. Specifically, dissipation from stationary protein synthesis and destruction is not taken into account in our model.  The coarse-grained entropy production (\ref{eq: DZ}) quantifies the dissipation accompanied by coarse-grained state changes.\\ \indent
$Q_Z$ is the heat absorbed by $Z$ and this term can be expressed by the transition probability of $Z$, $p(z_{t+dt}|s_t, x_t, y_t, z_t)$, and that for the backward process $p_B(z_{t}|s_t, x_t, y_t, z_{t+dt})$ through the detailed fluctuation theorem \cite{Jarzynski2000, Sekimoto2010, Seifert2012}:
\begin{eqnarray}
- \frac{1}{T_Z}\frac{dQ_Z}{dt} = &&\frac{1}{dt}\sum_{s_t, x_t, y_t, z_t, z_{t+dt}}p(s_t, x_t, y_t, z_t, z_{t+dt})\nonumber\\
&&\times\ln\frac{p(z_{t+dt}|s_t, x_t, y_t, z_t)}{p_B(z_t|s_t, x_t, y_t, z_{t+dt})}.
\end{eqnarray}
\indent
The generalized second law of thermodynamics states that the information-thermodynamic dissipation is always nonnegative: $D_Z \geq 0$. This is a tighter inequality than the conventional second law for the entire system that includes $SXYZ$, and $D_Z$ characterizes dissipation only in $Z$ by incorporating the learning rate.
\\ \indent
We define $D_X$ and $D_Y$ in the same manner, where $l_{X:SYZ}$ and $l_{Y:XZ}$ appear in their expressions, respectively.
The total dissipation of the three nodes is given by
\begin{eqnarray}
D_{\rm all} := D_X + D_Y + D_Z.\label{eq: split}
\end{eqnarray}
It immediately follows that $D_{\rm all}\geq0$. We note that $D_{\rm all}$ is rewritten in a similar form to (\ref{eq: DZ}):
\begin{eqnarray}
D_{\rm all} =\frac{dS(XYZ)}{dt} - \sum_{i = X, Y, Z}\frac{1}{T_i}\frac{dQ_i}{dt} - l_{XYZ:S}.
\end{eqnarray}
Here, $S(XYZ)$ is the joint Shannon entropy, and $l_{XYZ:S}$ is the learning rate from $S$ to $XYZ$.
As is the case for $D_Z$, if a three-node pattern learns from a signal node $S$ (i.e., $l_{XYZ:S}>0$), $dS(XYZ)/dt-\sum_i dQ_i/T_idt$ becomes positive. As in Eq.~(\ref{eq: split}), information-thermodynamic dissipation can be understood as a proper splitting of conventional thermodynamic dissipation.
A detailed derivation of these relations can be found in \cite{Horowitz2015}.
\begin{figure*}
\begin{center}
\begin{tabular}{cc}		
	\subfigure[]{
		\includegraphics[width = 0.48\linewidth]{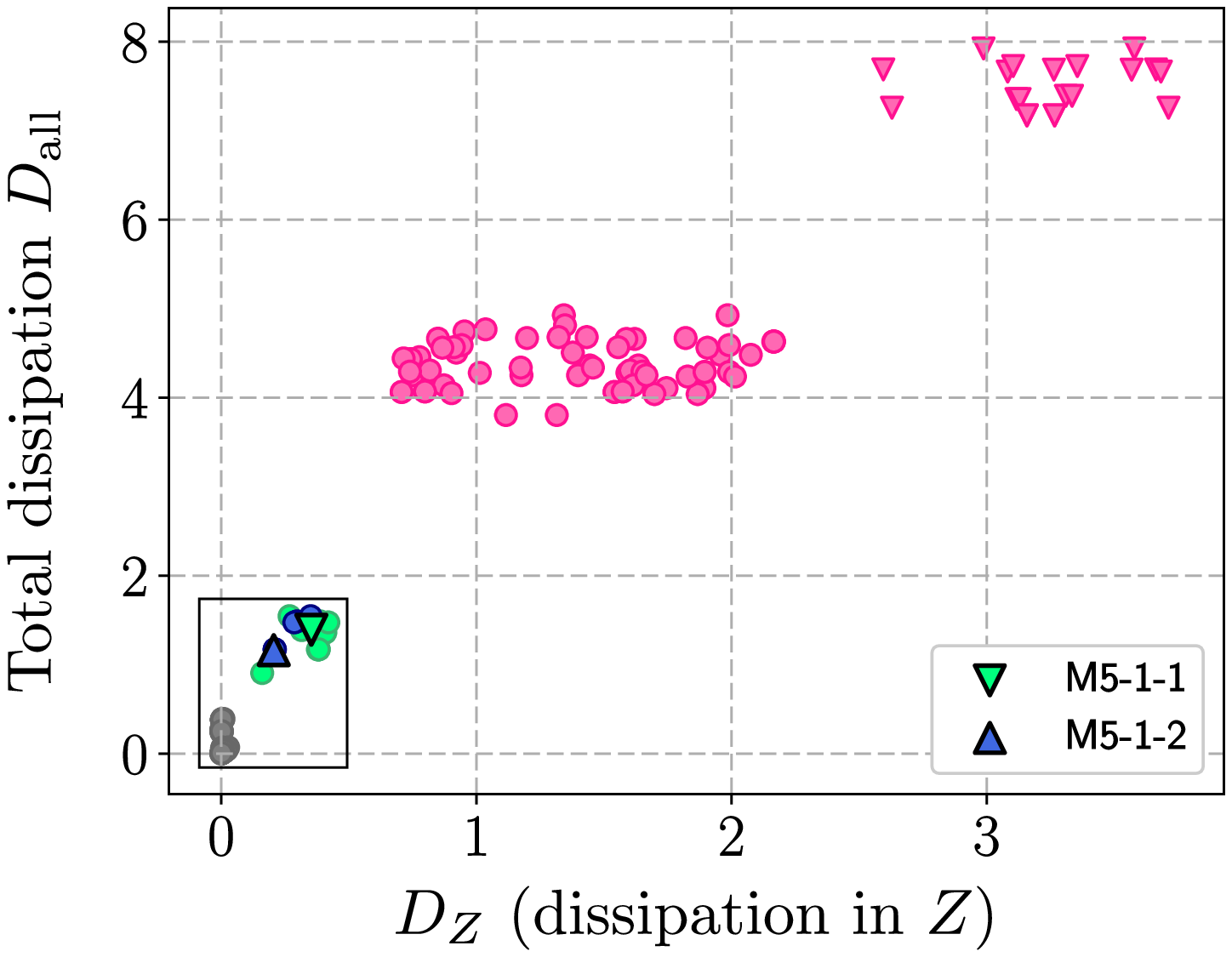}\label{fig: DAll}}&
	\subfigure[]{
		\includegraphics[width = 0.48\linewidth]{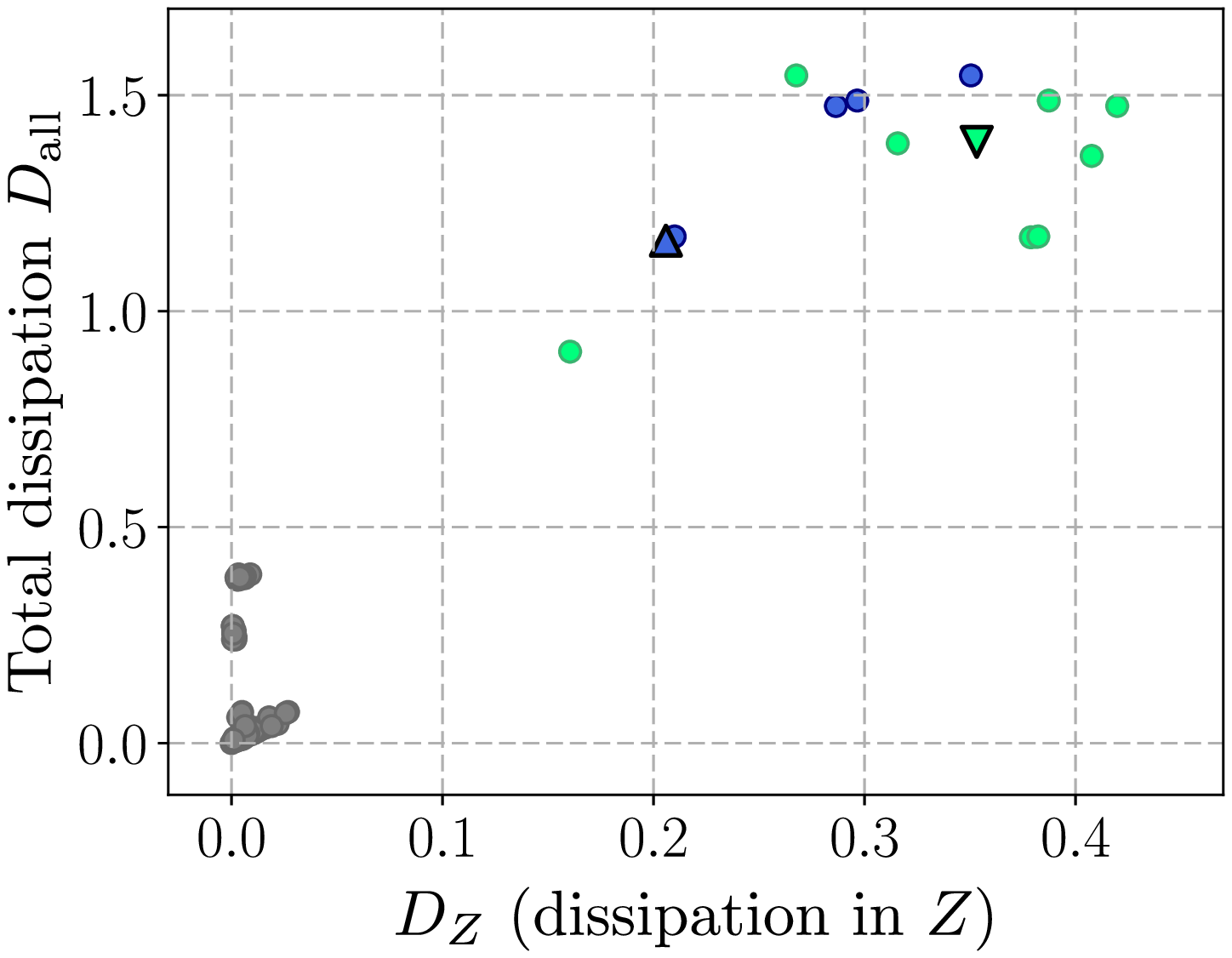}\label{fig: DAll2}}\\
	\subfigure[]{
		\includegraphics[width = 0.48\linewidth]{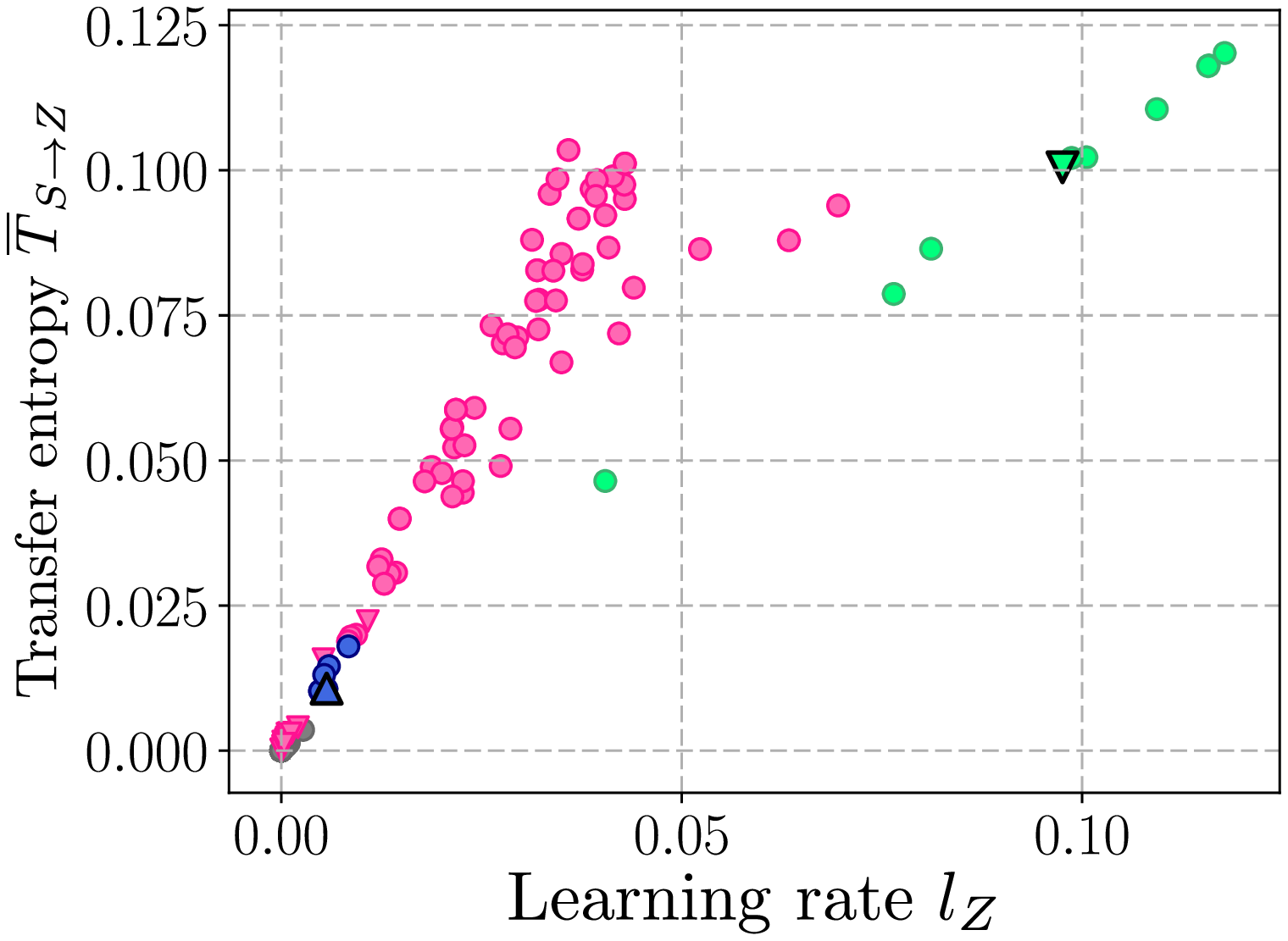}\label{fig: FlowStoZ}}&
	\subfigure[]{
		\includegraphics[width = 0.48\linewidth]{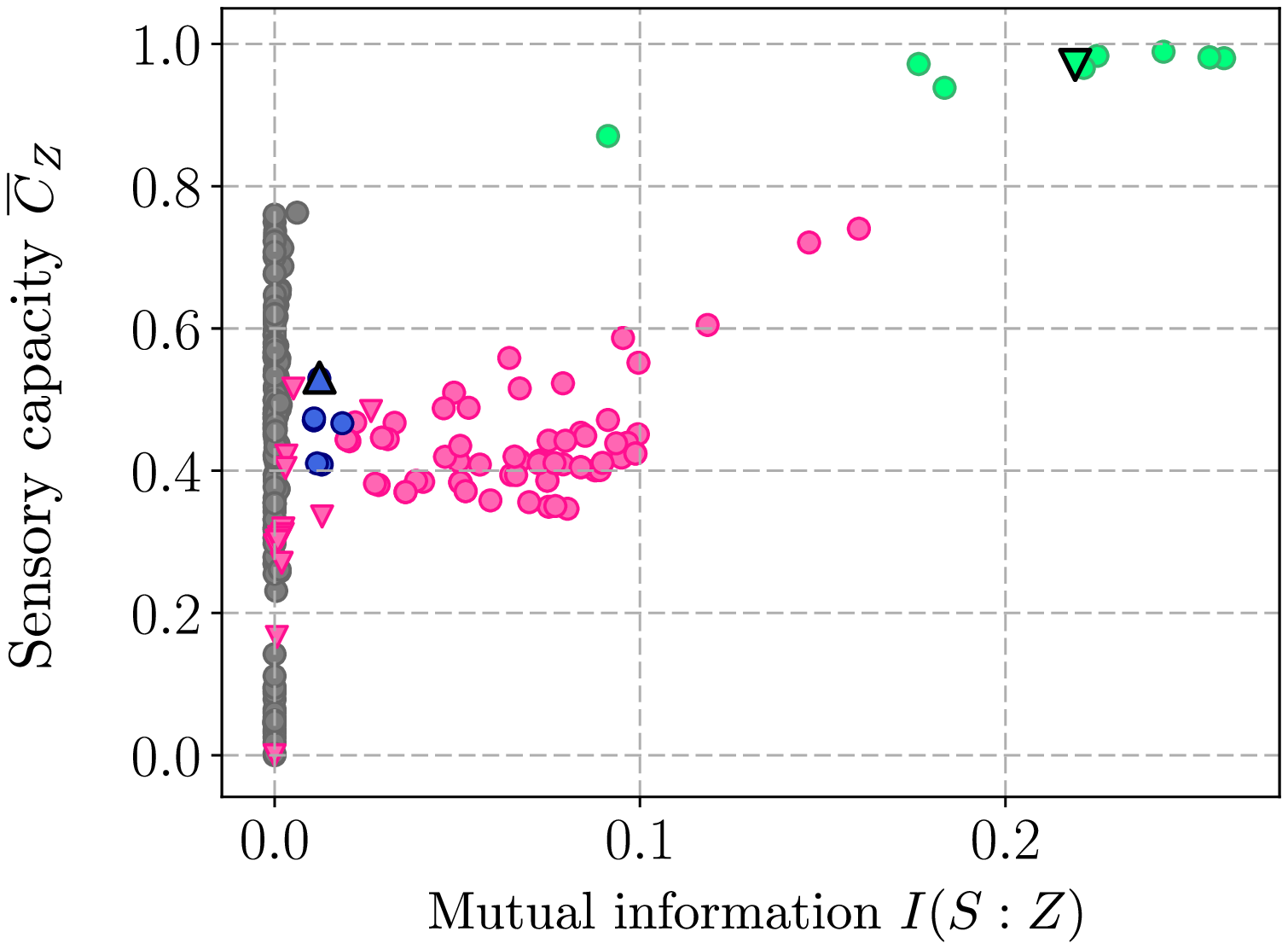}\label{fig: CSZ}}\\		
\end{tabular}
\end{center}
\caption{Information-thermodynamic characterization of three-node patterns: The parameters are set to $\gamma = 1, ~\gamma_S = 0.1, ~e = 0.001$. Each point represents a pattern, and triangles with black edges represent the FFLs. M5-1-1 represents all the CFFLs and M5-1-2 represents all the IFFLs, because all the CFFL (IFFL) patterns become equivalent (see Appendix \ref{sec: app3-1}). The difference in colors and shapes corresponds to the difference in the types (see Table 1). (a) Information-thermodynamic dissipation $D_Z$ and $D_{\rm all}$. (b) Enlarged view of (a). (c) Information flow from S to Z, i.e., $l_Z$ and $\overline{T}_{S\rightarrow Z}$, which shows the details of the sensory capacity $\overline{C}_Z$: If a pattern is located near the line passing through the origin with slope 1, the sensory capacity is close to unity. This figure clarifies how large the numerator and the denominator of the sensory capacity are. (d) $I(S:Z)$ and $\overline{C}_Z$. The parameter dependence of these plots is discussed in Appendix \ref{sec: app7}.
}
\label{fig: main1}
\end{figure*}
\begin{table*}
\begin{center}
\begin{tabular}{| c | c | c | c |}\hline
\begin{minipage}{0.9cm}\includegraphics[width = 0.4cm]{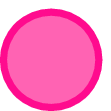} \includegraphics[width = 0.4cm]{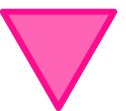}\end{minipage}& dissipative &
\begin{minipage}{0.45cm}\includegraphics[width = 0.4cm]{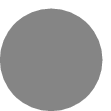}\end{minipage} & static\\ \hline
\begin{minipage}{0.45cm}\includegraphics[width = 0.4cm]{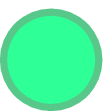}\end{minipage} & informative &
\begin{minipage}{0.45cm}\includegraphics[width = 0.4cm]{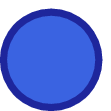} \end{minipage}& adaptive\\\hline
\end{tabular}
\end{center}
\caption{Classification of three-node patterns: On the basis of Fig.~\ref{fig: main1}, all patterns are divided into five groups. In terms of their characteristics, they are classified into four types: dissipative, static, informative, and adaptive.}
\label{table: 1}
\end{table*}
\begin{figure}
\begin{tabular}{c}
	\subfigure[]{
		\includegraphics[width = 0.9\linewidth]{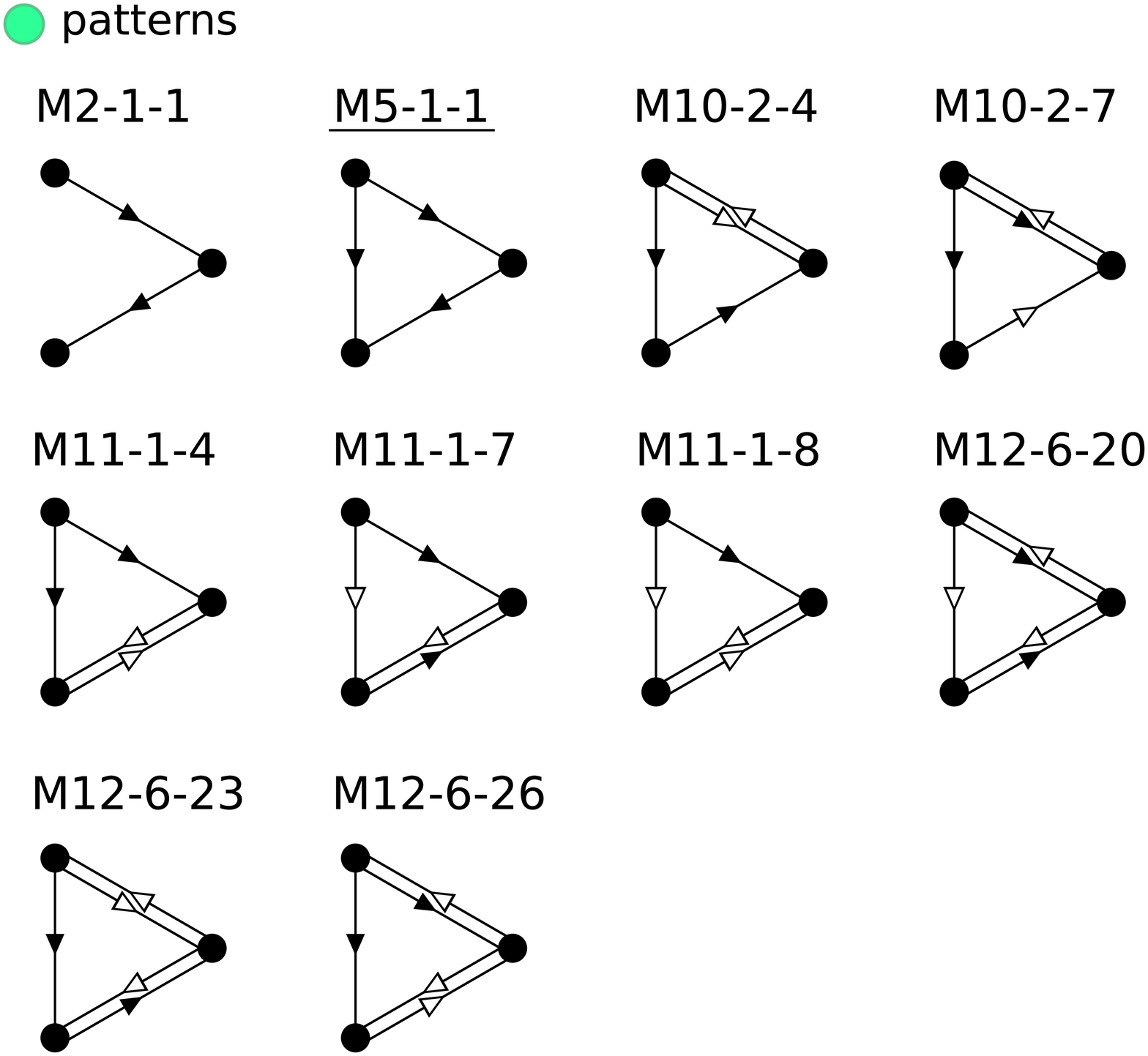}\label{fig: green}}\\
	\subfigure[]{
		\includegraphics[width = 0.9\linewidth]{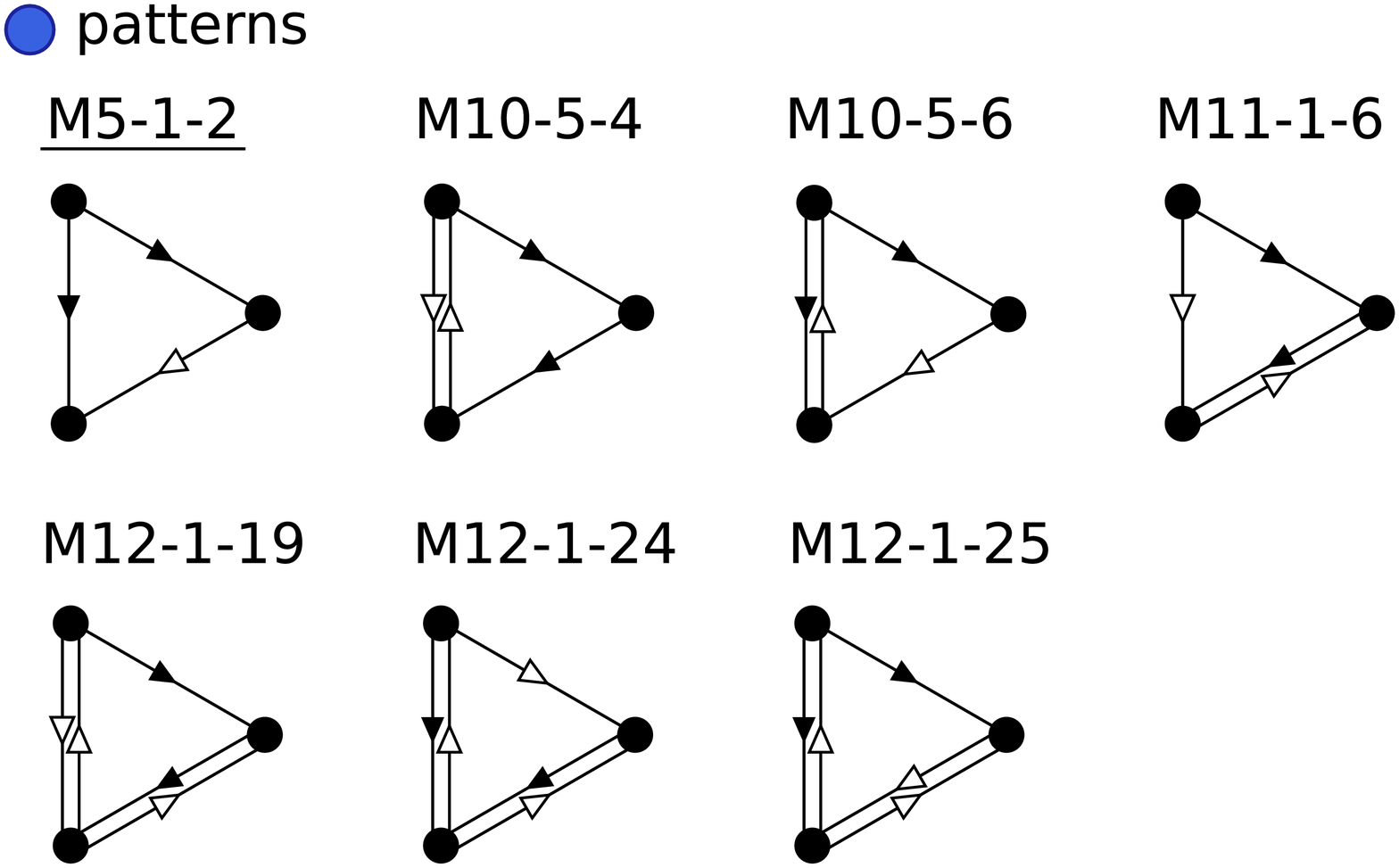}\label{fig: blue}}
\end{tabular}
\caption{Patterns in the green and blue types. In particular, M5-1-1 and M5-1-2 are network motifs.}
\label{fig: gb_list}
\end{figure}

\section{Main results}
\label{sec: main}
We now show our main results on the characterization of three-node networks: the single input case in Sec.~\ref{sec: single} and the double inputs case in Sec.~\ref{sec: double}.
\subsection{Single input: Information propagation}
\label{sec: single}
In the setup of Fig.~\ref{fig: setup}, we classify all the three-node patterns on the basis of information-thermodynamic quantities. We reveal that patterns are classified into four types and to which type each network motif belongs.\\ \indent
We show scatter plots with respect to several information-thermodynamic quantities in Fig.~\ref{fig: main1}. Fig.~\ref{fig: DAll} and Fig.~\ref{fig: DAll2} show information-thermodynamic dissipation $D_Z$ and $D_{\rm all}$. Figure~\ref{fig: FlowStoZ} and Fig.~\ref{fig: CSZ} show four information quantities. All the points split into four groups in Fig.~\ref{fig: DAll} and \subref{fig: DAll2}. We also realize that the group with the second smallest dissipation split into two groups in Fig.~\ref{fig: FlowStoZ} and \subref{fig: CSZ}. 
From these observations, we assign different colors and shapes to the patterns in Fig.~\ref{fig: main1}. These five groups can be rearranged into four types on the basis of their characteristics as shown in Table.~\ref{table: 1}.
Here, the sensory capacity plays a role in the separation of the groups. For example, the mutual information of M11-1-4 (which belongs to \includegraphics[width = 0.25cm]{greenc.eps} and $I(S:Z)\sim0.1$) is small, but its sensory capacity is big. We discuss the characteristic of each type below. Also, the concrete behavior of the patterns in each type is shown by simulation in Appendix \ref{sec: app5}.\\ \indent 
First, \includegraphics[width = 0.25cm]{pinkdt.eps} and \includegraphics[width = 0.25cm]{pinkc.eps} are dissipative groups. We confirmed that all the patterns in  \includegraphics[width = 0.25cm]{pinkdt.eps} and \includegraphics[width = 0.25cm]{pinkc.eps} include negative feedback loops. The reason why they are dissipative is that they show oscillatory behavior due to such negative feedback loops.\\ \indent
On the other hand, the patterns in \includegraphics[width = 0.25cm]{grayc.eps} dissipate much less. These patterns also take small values in terms of the information quantities.
We confirmed that all of them include positive feedback loops. The behavior of these patterns is due to the fact that dynamics of these patterns are static because of positive feedback loops, and they do not react well to the signals from $S$.\\ \indent
The remaining 17 patterns that belong to \includegraphics[width = 0.25cm]{greenc.eps} and \includegraphics[width = 0.25cm]{bluec.eps} are shown in Fig.~\ref{fig: gb_list}.
The patterns in \includegraphics[width = 0.25cm]{bluec.eps} take small values in terms of the information quantities, and the other way around for the patterns in \includegraphics[width = 0.25cm]{greenc.eps}. 
We find that the patterns in \includegraphics[width = 0.25cm]{bluec.eps} include incoherent feedforward loops. According to the simulation results shown in Appendix~\ref{sec: app5}, they show adaptive behavior such that they pass information from $S$ to $Z$ only in a short time when the state of $S$ changes. For example, in the case of M5-1-2 (I1), $Z$ changes from 0 to 1 temporarily when $X$ changes from 0 to 1, but after that, $Z$ returns to 0 due to the inhibition by $Y$. 
This is the reason why the patterns in \includegraphics[width = 0.25cm]{bluec.eps} take small values in terms of the information quantities.
We finally confirm that the patterns in \includegraphics[width = 0.25cm]{greenc.eps} propagate signals from $S$ to $Z$ quite well. 
Thus, these patterns take large values in terms of the information quantities.\\ \indent
The CFFL and IFFL network motifs are classified into \includegraphics[width = 0.25cm]{greenc.eps} and \includegraphics[width = 0.25cm]{bluec.eps} respectively, and the PFBL network motifs belong to the \includegraphics[width = 0.25cm]{grayc.eps} or \includegraphics[width = 0.25cm]{greenc.eps}  type (see Appendix \ref{sec: pfbl} for the details of the classification of the PFBL network motifs). Interestingly, there are only two types other than \includegraphics[width = 0.25cm]{grayc.eps}, \includegraphics[width = 0.25cm]{pinkc.eps} and \includegraphics[width = 0.25cm]{pinkdt.eps}, and the CFFLs and IFFLs are simple patterns in \includegraphics[width = 0.25cm]{greenc.eps} and \includegraphics[width = 0.25cm]{bluec.eps}.
No network motifs belong to the dissipative type \includegraphics[width = 0.25cm]{pinkc.eps}. In terms of information thermodynamics, the dissipative type might waste too much chemical fuels, which may be one of the reasons why no network motifs belong to the dissipative type.\\ \indent
The \includegraphics[width = 0.25cm]{greenc.eps} and \includegraphics[width = 0.25cm]{bluec.eps} types are interesting because they operate with small dissipation and have the capability of propagating information. The patterns in \includegraphics[width = 0.25cm]{greenc.eps} form a stable correlation, and those in \includegraphics[width = 0.25cm]{bluec.eps} propagate information temporarily. The patterns that propagate information would be preferable in gene regulatory networks because the upstream gene can control the state of the pattern. We will discuss the reason why feedforward loops are distinguished from the other patterns in \includegraphics[width = 0.25cm]{greenc.eps} and \includegraphics[width = 0.25cm]{bluec.eps} in the next subsection.\\ \indent

\subsection{Two inputs: Logical operation}
\label{sec: double}
\begin{figure}		
	\includegraphics[width = 0.8\linewidth]{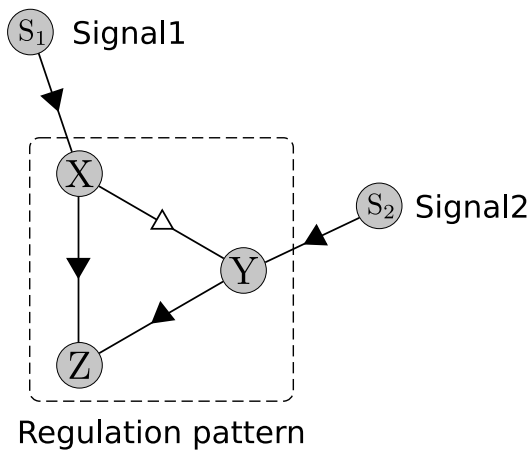}
\caption{Our setup with the two input signals, which focuses on the logical operation performed by a three-node pattern. We investigate how nontrivially the two input signals are processed to an output $Z$.}
\label{fig: setup2}
\end{figure}

\begin{figure}
\begin{tabular}{cc}
	\multicolumn{2}{c}{
	\subfigure[]{
		\includegraphics[width = \linewidth]{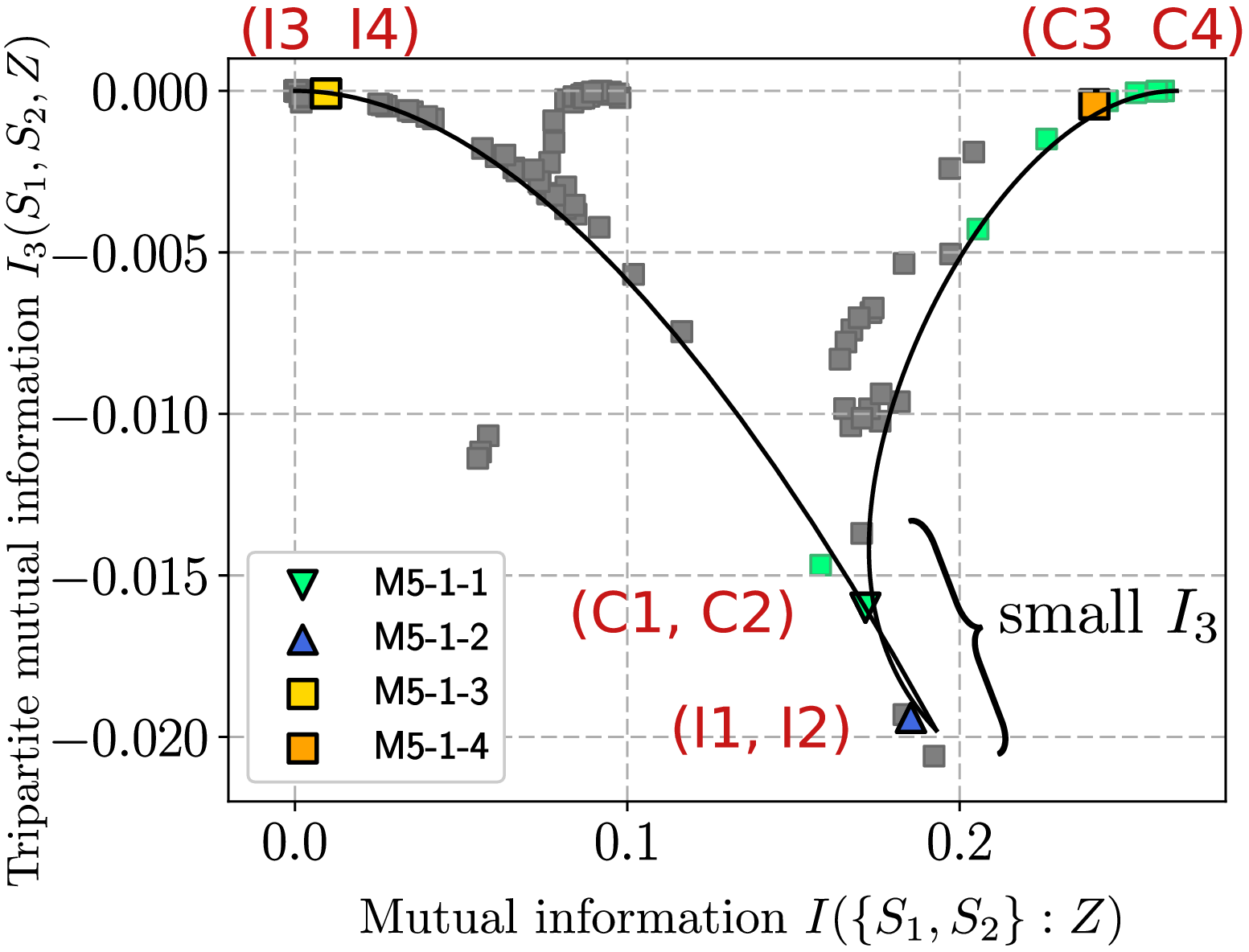}\label{fig: I3}}}\\[0.5cm]
	\subfigure[]{
		\includegraphics[width = 0.4\linewidth]{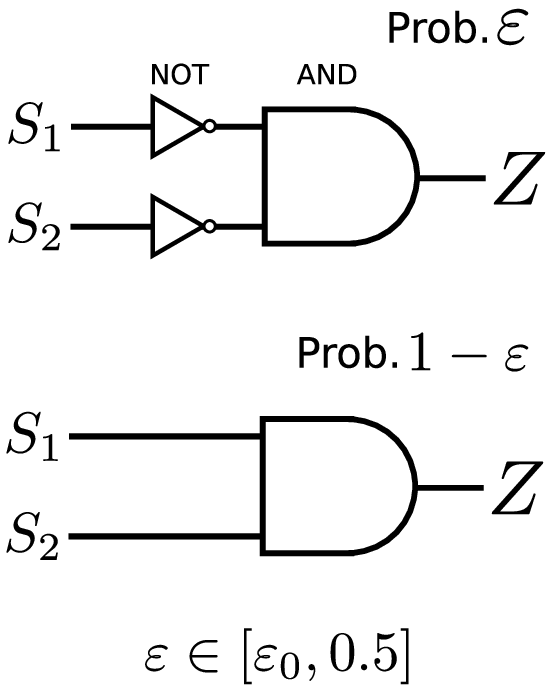}\label{fig: model1}} &
	\subfigure[]{
		\includegraphics[width = 0.4\linewidth]{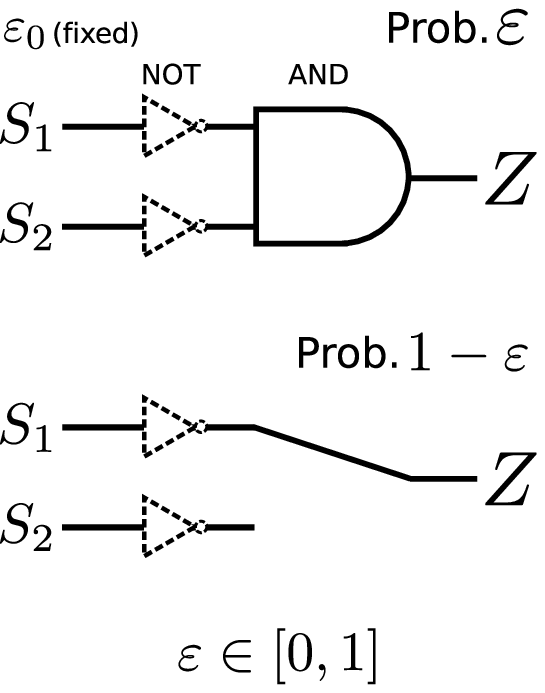}\label{fig: model2}} \\		
\end{tabular}
\caption{Tripartite mutual information of three-node patterns and the fitting models: (a) The scatter plot of $I(\{S_1, S_2\}:Z)$ versus $I_3(S_1, S_2, Z)$. The parameters are set to $\gamma = 1, ~\gamma_S = 0.1, ~e = 0.001$. As shown in the legend, the four feedforward loops are highlighted with black edges, and the other patterns are represented without edges. The commonly used names of the feedforward loops are shown with red letters. Also, the patterns in Fig.~\ref{fig: gb_list} are highlighted with the corresponding colors (green and blue). 
Blue patterns are, however, hidden behind M5-1-3 except for M5-1-2. The black curves are obtained by the fitting with the following models. 
(b) Fitting model for the left curve. (c) Fitting model for the right curve. There is a single fitting parameter, which is given by $\epsilon_0 = 0.153$ in our fitting.}
\label{fig: main2}
\end{figure}

\begin{figure}
\includegraphics[width = 1.0\linewidth]{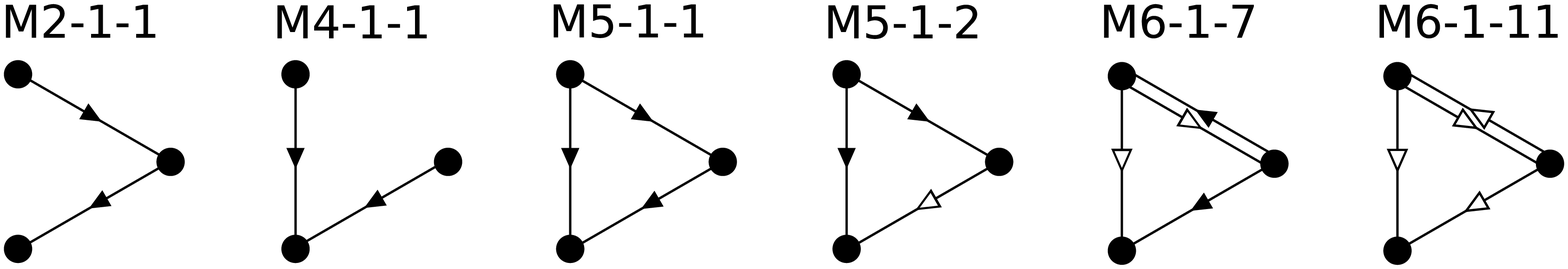}
\caption{Patterns that take small values of the tripartite mutual information, which are shown in Fig.~\ref{fig: I3} with a brace.}
\label{fig: lowI3}
\end{figure}

We now show our second result.
Besides the input to $X$, in real gene regulatory networks, a three-node pattern often takes another input to $Y$.
Therefore, patterns that can nontrivially operate on both the signals would be preferable in natural selection.
Here we evaluate the capacity of logical operation performed by three-node patterns by using an information quantity called tripartite mutual information, and discuss the reason why patterns with fewer edges such as feedforward loops are preferred.\\ \indent
We consider the situation that two signal sources $S_1$ and $S_2$ activate $X$ and $Y$ independently (Fig.~\ref{fig: setup2}), and that
$S_1$ and $S_2$ flip between 0 and 1 randomly (i.e., $e_{S_1} = e_{S_2} = 1$). The parameters are set with the following conditions:
\begin{eqnarray}
&& \gamma_S:= \gamma_{S_1} = \gamma_{S_2},\\
&&\gamma := \gamma_X = \gamma_Y = \gamma_Z,\\
&&\gamma_S \ll \gamma,\\
&& e:= e_X = e_Y = e_Z \ll1.
\end{eqnarray}
We exclude irrelevant patterns from calculation with the same rule as in Sec.~\ref{sec: single}, and additionally exclude patterns that become equivalent by swapping $X$ and $Y$.
As a result, we perform calculation for 204 patterns (see Appendix \ref{sec: app4} for the details). \\ \indent
We discuss the definition and the basic properties of tripartite mutual information \cite{Cerf1996}, which quantifies how nontrivially the output $Z$ depends on the inputs $S_1$ and $S_2$.
The tripartite mutual information between these three nodes is defined as
\begin{eqnarray}
I_3(S_1, S_2, Z) &:=& H(S_1) + H(S_2) + H(Z) - H(S_1, S_2)\nonumber\\
&&- H(S_1, Z) - H(S_2, Z) + H(S_1, S_2, Z)\nonumber\\
&=& I(S_1:Z) + I(S_2:Z) - I(\{S_1, S_2\}:Z).\nonumber\\ 
\end{eqnarray}
\indent
To illustrate the informational meaning of tripartite mutual information, we show two examples with three binary bits $x, y, z$: (i) $I_3(x, y, z) = -\ln2$ if $x = y\oplus z$ (XOR), and $y$ and $z$ are independent and random. In this case, neither $y$ nor $z$ is correlated with $x$, while composite $yz$ has maximum correlation with $x$. This means that information of $y$ and that of $z$ are mixed up in $x$. (ii) $I_3(x, y, z) = \ln2$ if $x = y = z$ and $x$ is random. In this case, three bits are maximally correlated. These examples show that the negative tripartite mutual information quantifies mixing of signals from two inputs ($y$ and $z$ in this case).\\ \indent
In the present setup, it does not take a positive value, as shown from simple calculation with the assumption $I(S_1:S_2) = H(S_1) + H(S_2) - H(S_1, S_2) = 0$. In fact, we have
\begin{eqnarray}
I_3(S_1, S_2, Z) &=& H(Z) + H(S_2) - H(S_2, Z) - H(S_2)\nonumber\\
&&- H(S_1, Z) + H(S_1, S_2, Z)\nonumber\\
&=& I(S_2:Z) - I(S_2:\{S_1, Z\})\nonumber\\
&\leq& 0.
\end{eqnarray}
\\ \indent
The smaller the tripartite mutual information is, the more non-trivial the logical operation is.
In fact, if $Z$ depends on both $S_1$ and $S_2$, $I(S_1:Z) + I(S_2:Z)$ becomes smaller than $I(\{S_1, S_2\}:Z)$, which implies a small negative value of $I_3(S_1, S_2, Z)$ (or equivalently, the absolute value $|I_3(S_1, S_2, Z)|$ becomes large). The logical operation is non-trivial in this case. On the other hand, if $Z$ depends only on $S_1$ and is independent of $S_2$, $I(S_1:Z)=I(\{S_1, S_2\}:Z)$ and $I(S_2:Z)=0$ hold, resulting in $I_3(S_1, S_2, Z) = 0$, where the logical operation is trivial.\\ \indent
We argue that a pattern with small tripartite mutual information is preferable in gene regulatory networks because of the following reason. If the tripartite mutual information is small, both information from $S_1$ and $S_2$ propagate to $Z$, which enables the pattern to process information in a complex way. In fact, a previous study shows that information propagates more widely in gene regulatory networks than in random networks \cite{Kim2015}.\\ \indent
We show the result of our calculation in Fig.~\ref{fig: I3}. The horizontal axis is the mutual information $I(\{S_1, S_2\}:Z)$, and the vertical axis is the tripartite mutual information $I_3(S_1, S_2, Z)$.
Network motifs M5-1-1 and M5-1-2 take small values of the tripartite mutual information. There are only six patterns that take small values of the tripartite mutual information, which are shown with a brace in Fig.~\ref{fig: main2}. These six patterns are listed in Fig.~\ref{fig: lowI3}. Interestingly, these six patterns have only a few edges, which may explain the reason why gene regulatory networks consist of patterns with fewer edges.\\ \indent
There seems to be a bifurcation structure in Fig.~\ref{fig: I3}.
The upper left patterns do not propagate information at all, i.e., $I(\{S_1, S_2\}:Z)\simeq0$, and the upper right patterns make $Z$ only react to one of the two signals $S_1$ or $S_2$. On the basis of these observations, we consider two simple models (Fig.~\ref{fig: main2}(b)(c)) that explain the bifurcation structure.
Intuitively speaking, these two models capture the following properties: (i) how much information propagates from $S_1$ or $S_2$ to $Z$, and (ii) how equally information propagates in the two paths. \\ \indent
In the model of Fig.~\ref{fig: model1}, the NOT gates are inserted into the two inputs with probability $\varepsilon$ independently. In this model, the NOT gates represent the errors that independently occur on the input signals during the logical operation. In the model of Fig.~\ref{fig: model2}, the probability of inserting the NOT gates is fixed with $\varepsilon_0$, and one of the AND gate or a straight pass from $S_1$ to $Z$ is chosen with probabilities $\varepsilon$ and $1-\varepsilon$, respectively. In this model, the non-triviality of the logical operation is represented by the probability $\varepsilon$. The left curve in Fig.~\ref{fig: I3} is drawn with model \subref{fig: model1} by varying $\varepsilon$ in the range of $\varepsilon_0\leq\varepsilon\leq0.5$, and the right curve is drawn with model \subref{fig: model2} by varying $\varepsilon$ in $0\leq\varepsilon\leq1$. 
Here, there is only a single fitting parameter $\varepsilon_0$. The good agreement with the data means that these models capture the essence of the behavior of the three-node patterns.\\ \indent
In particular, network motifs M5-1-1 and M5-1-2 perform non-trivial logical operation with relatively small errors. On the other hand, most of the patterns in \includegraphics[width = 0.25cm]{greenc.eps} perform trivial logical operation, and the patterns in \includegraphics[width = 0.25cm]{bluec.eps} other than M5-1-2 do not propagate information at all. We can thus consider that these patterns are unfavorable in terms of information processing. \\ \indent
We note that M4-1-1 is not a network motif. It should be remarked, however, that M4-1-1 occurs most frequently in many gene regulatory networks \cite{Gerstein2012}, and it also occurs frequently in random networks (and thus is not regarded as a network motif). In this way, tripartite mutual information gives us an intuitive account for not only network motifs but also frequent patterns in gene regulatory networks.\\ \indent
The above result may give a reasonable explanation for the difference in the occurrence frequencies among different types of feedforward loops. 
As shown in Fig.~\ref{fig: motif}, there are eight types of feedforward loops, but only the C1 and I1 FFLs occur most frequently with around 40 percent and 30 percent each, and the other six types occur with around five percent each. This has been confirmed in the gene regulatory networks of {\it E. coli} and {\it S. cerevisiae} \cite{Mangan2006}.
The tripartite mutual information of C3, C4, I3 and I4 is around zero in Fig.~\ref{fig: I3}. Since a pattern whose tripartite mutual information is around 0 is trivial in terms of logical operation, this may explain the low occurrence frequencies of C3, C4, I3 and I4 FFLs, which suggests that the value of tripartite mutual information indeed quantifies the capability of logical operation of three-node patterns.\\ \indent
It should be noted that we assumed each regulatory function as the AND gate in the foregoing argument. If the regulatory function of a three-node pattern is given by a combination of AND and OR, the conclusion becomes opposite to the above (see Appendix \ref{sec: app3} for the details). Therefore, it is a future issue to check the consistency between our result and real experimental data. Qualitatively the same argument has also been mentioned in Refs.~\cite{Mangan2003, Alon2006}. In Refs.~\cite{Mangan2003, Alon2006}, the authors argued that the eight types of FFLs are different with each other, in that stationary $Z$ shows different dependence on the two signal molecules (which correspond to $S_1$ and $S_2$ in this study). In the present work, we clarify the different dependence by using tripartite mutual information.

\section{Concluding remarks}
\label{sec: conclusion}
In a system like a gene regulatory network with information propagation, it is crucial to take into account information flow to analyze dissipation. In this study, we have adopted information thermodynamics to quantify such dissipation in a three-node pattern which is only a part of a large-scale network.\\ \indent
We first considered the case where there is a single input, and characterized all the possible three-node patterns with information-thermodynamic quantities. We found that these patterns are classified into four types. We discussed the characteristics of network motifs, by considering to which types they are categorized.\\ \indent
We next considered the case where there are two inputs. By quantifying how the two inputs affect the output by tripartite mutual information, we argued the reason why feedforward-loop network motifs occur frequently in an intuitive manner. This result is consistent with a previous study which claims that information propagates more widely in gene regulatory networks than in random networks \cite{Kim2015}. 
In addition, we found that the different occurrence frequencies among the eight types of feedforward loops might correspond to the difference in the amounts of tripartite mutual information.\\ \indent
We have partly succeeded to explain in what respect network motifs have advantages compared to other patterns in terms of information-thermodynamic quantities. We note that previous studies \cite{Alon2006, Alon2007} explain that IFFL network motifs would be preferable because they show adaptive behavior and accelerate the response time. On the other hand, we argue that IFFLs are preferable because they show adaptive behavior with small dissipation and they take small tripartite mutual information. We believe that these two approaches are complimentary to each other.\\ \indent
However, these results might involve some uncertainty due to the fact that the stochastic Boolean model might be too much simplified. 
Although this model is suitable for understanding the basic behavior of gene regulatory patterns, the oscillatory or static solution often deviates from the actual behavior. In addition, since the stochastic Boolean model is a coarse-grained model, some dissipative processes such as protein productions are not included. Therefore, it is a future issue to investigate dissipation by using more detailed models. For example, the difference in the occurrence frequencies between C1 (I1) and C2 (I2) might be accounted for by such detailed analysis of dissipation.\\ \indent
Meanwhile, our study suggests that tripartite mutual information is important in terms of the network formation of gene regulatory networks. The symmetry between the occurrence frequencies of the C1 to C4 and the I1 to I4 FFLs \cite{Mangan2006} may arise from the symmetry in the tripartite mutual information between them. It is also worth investigating whether the same tendencies can be found in other gene regulatory networks than those of {\it E.coli} and {\it S.cerevisiae}. \\ \indent
Our results suggest that information thermodynamics gives us a useful methodology for a systematic analysis of biochemical reaction networks. We note that tripartite mutual information has not received much attention in this context so far, while information quantities such as mutual information and transfer entropy have often been used in a wide context\cite{Honey2007, Tostevin2009}. Further application of our approach to a broader class of networks, including both biological and artificial ones, is a future issue.
\begin{acknowledgments}
We thank Tetsuya J. Kobayashi for fruitful discussion. T. S. is supported by JSPS KAKENHI Grant No. JP16H02211 and No. 25103003.
\end{acknowledgments}

\appendix
\section{Master equation}
\label{sec: app1}
In this appendix, we present an example of the master equation for the stochastic Boolean model. 
We consider the pattern described in Fig.~\ref{fig: setup} as an example.
The joint probability distribution $p(s_t, x_t, y_t, z_t)$ is represented by the probability vector $\bm{p}$ defined by 
\begin{eqnarray}
&&\bm{p} = (p(0, 0, 0, 0), p(0, 0, 0, 1), p(0, 0, 1, 0), p(0, 0, 1, 1), \nonumber\\
&&p(0, 1, 0, 0), p(0, 1, 0, 1), p(0, 1, 1, 0), p(0, 1, 1, 1),\nonumber\\
&&p(1, 0, 0, 0), p(1, 0, 0, 1), p(1, 0, 1, 0), p(1, 0, 1, 1), \nonumber\\
&&p(1, 1, 0, 0), p(1, 1, 0, 1), p(1, 1, 1, 0), p(1, 1, 1, 1))^{\mathrm{T}},
\end{eqnarray}
where T means the transpose of a matrix. Then the transition matrix $H$ is given by
\begin{eqnarray}
H = \left(\begin{matrix}
A & B\\
C & D\\
\end{matrix}
\right),
\end{eqnarray}
where
\begin{eqnarray}
&&A = \nonumber\\&&\left(\begin{matrix}
* & \gamma_Z & e_Y\gamma_Y & 0 & \gamma_X & 0 & 0 & 0\\
e_Z\gamma_Z & * & 0 & e_Y\gamma_Y & 0 & \gamma_X & 0 & 0\\
\gamma_Y & 0 & * & \gamma_Z & 0 & 0 & \gamma_X & 0\\
0 & \gamma_Y & e_Z\gamma_Z & * & 0 & 0 & 0 & \gamma_X\\
e_X\gamma_X & 0 & 0 & 0 & * & \gamma_Z & \gamma_Y & 0\\
0 & e_X\gamma_X & 0 & 0 & e_Z\gamma_Z & * & 0 & \gamma_Y\\
0 & 0 & e_X\gamma_X & 0 & e_Y\gamma_Y & 0 & * & e_Z\gamma_Z\\
0 & 0 & 0 & e_X\gamma_X & 0 & e_Y\gamma_Y & \gamma_Z & *\\
\end{matrix}\right),\nonumber\\
\end{eqnarray}
\begin{eqnarray}
&&D = \nonumber\\ &&\left(\begin{matrix}
* & \gamma_Z & e_Y\gamma_Y & 0 & e_X\gamma_X & 0 & 0 & 0\\
e_Z\gamma_Z & * & 0 & e_Y\gamma_Y & 0 & e_X\gamma_X & 0 & 0\\
\gamma_Y & 0 & * & \gamma_Z & 0 & 0 & e_X\gamma_X & 0\\
0 & \gamma_Y & e_Z\gamma_Z & * & 0 & 0 & 0 & e_X\gamma_X\\
\gamma_X & 0 & 0 & 0 & * & \gamma_Z & \gamma_Y & 0\\
0 & \gamma_X & 0 & 0 & e_Z\gamma_Z & * & 0 & \gamma_Y\\
0 & 0 & \gamma_X & 0 & e_Y\gamma_Y & 0 & * & e_Z\gamma_Z\\
0 & 0 & 0 & \gamma_X & 0 & e_Y\gamma_Y & \gamma_Z & *\\
\end{matrix}\right),\nonumber\\
\end{eqnarray}
and $B = C = \gamma_S I$ with $I$ being the identity matrix.
Here, an element represented by $*$ above is determined so that the sum of each column becomes zero.
For example, the $(1, 1)$ element of $H$ is given by $H_{11} = - e_X\gamma_X - \gamma_Y - e_Z\gamma_Z - \gamma_S$.
The time evolution of the total system is described by the master equation
\begin{eqnarray}
\frac{d\bm{p}}{dt} = H\bm{p}.
\end{eqnarray}
For example, the first row of the above equation is given by
\begin{eqnarray}
\frac{dp(0, 0, 0, 0)}{dt} &=& H_{11} p(0, 0, 0, 0) + \gamma_Z p(0, 0, 0, 1) \nonumber\\
&+& e_Y\gamma_Y p(0, 0, 1, 0)+ \gamma_X p(0, 1, 0, 0)\nonumber\\
&+& \gamma_S p(1, 0, 0, 0).
\end{eqnarray}

\section{Equivalence of network patterns}
\label{sec: app3}
In this appendix, we first formulate the equivalence of network patterns and then discuss the relation between the OR-logic patterns and the AND-logic ones.
\subsection{Definition of the equivalence}
\label{sec: app3-1}
\begin{figure}[H]
\begin{center}
\includegraphics[width = 0.8\linewidth]{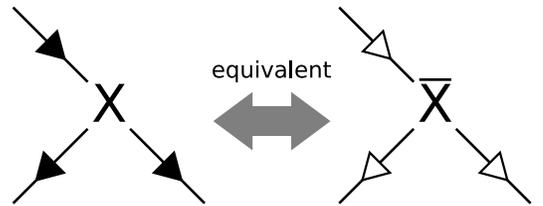}
\caption{Equivalent transformation of patterns: The NOT transformations on all the input and output edges give an equivalent pattern.}
\label{fig: equiv}
\end{center}
\end{figure}

We consider the stationary state of the setup in Fig.~\ref{fig: setup}. We show the definition of the equivalence of two patterns by an example. If the join probability distributions $p_1(s_t, x_t, y_t, z_t, s_{t+dt}, x_{t+dt}, y_{t+dt}, z_{t+dt})$ for a pattern 1 and $p_2(s_t, x_t, y_t, z_t, s_{t+dt}, x_{t+dt}, y_{t+dt}, z_{t+dt})$ for another pattern 2 satisfy the following relation
for all $s_t, x_t, y_t, z_t, s_{t+dt}, x_{t+dt}, y_{t+dt}, z_{t+dt}\in\{0, 1\}$, then patterns 1 and 2 are said to be equivalent:
\begin{eqnarray}
&&p_1(s_t, x_t, y_t, z_t, s_{t+dt}, x_{t+dt}, y_{t+dt}, z_{t+dt}) = \nonumber\\
&&p_2(s_t, x_t, y_t, \overline{z_t}, s_{t+dt},x_{t+dt}, y_{t+dt}, \overline{z_{t+dt}})\label{eq: equiv},
\end{eqnarray}
where the bar on a letter represents the inversion $0\mapsto 1$ and $1\mapsto 0$.
In general, the inversion is allowed on multiple nodes, but it should be performed on both variables at time $t$ and at time $t+dt$ simultaneously.
If pattern 1 and 2 are equivalent, their information quantities take the same values.
For example, when Eq.~(\ref{eq: equiv}) holds, the mutual information $I_1(S_t, Z_t)$ for pattern 1 and $I_2(S_t, Z_t)$ for pattern 2 become the same:
\begin{eqnarray}
I_1(S_t:Z_t) &=& \sum_{s_t, z_t}p_1(s_t, z_t)\ln\frac{p_1(s_t, z_t)}{p_1(s_t)p_1(z_t)}\nonumber\\
&=&  \sum_{s_t, z_t}p_2(s_t, \overline{z_t})\ln\frac{p_2(s_t, \overline{z_t})}{p_2(s_t)p_2(\overline{z_t})}\nonumber\\
&=& \sum_{s_t, z_t}p_2(s_t, z_t)\ln\frac{p_2(s_t, z_t)}{p_2(s_t)p_2(z_t)}\nonumber\\
&=& I_2(S_t:Z_t).
\end{eqnarray}
\indent
This equivalence of patterns can be shown graphically. 
For example, Fig.~\ref{fig: equiv} shows a transformation from one pattern to another equivalent pattern.

\begin{figure}
\begin{center}
\includegraphics[width = 0.9\linewidth]{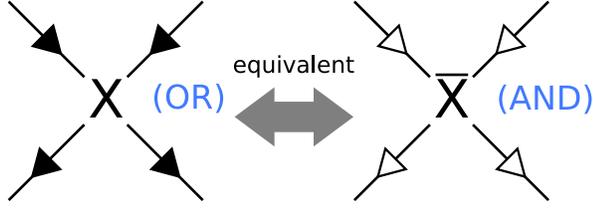}
\caption{Equivalent transformation from the OR-logic pattern to the AND-logic one.}
\label{fig: equiv2}
\end{center}
\end{figure}

\noindent
This transformation is based on the following equations: $f^{\overline{X}} = {\rm NOT}(S_t)$ (if $f^X = S_t$ for example), and $f^i(X, ...) = f^i(\overline{\overline{X}}, ...) ~(i = Y ~{\rm or} ~Z)$.  \\ \indent
We discuss the equivalence among feedforward loops. In the setup of Fig.~\ref{fig: setup}, the equivalent transformation on $X$ and $Y$ is possible. Thus, the patterns from the C1 to C4 and those from the I1 to I4 FFLs become equivalent, respectively.
On the other hand, in the setup of Fig.~\ref{fig: setup2}, the transformation is possible only on $X$, because if we make the transformation on $Y$, the regulatory function of $Y$ changes accordingly (see Fig.~\ref{fig: equiv2}). In this case, the C1 and C2, the C3 and C4, the I1 and I2, the I3 and I4 FFLs become equivalent, respectively.\\ \indent

\subsection{Equivalence between the OR-logic patterns and the AND-logic ones}

\begin{figure}
\begin{center}
\includegraphics[width = 1\linewidth]{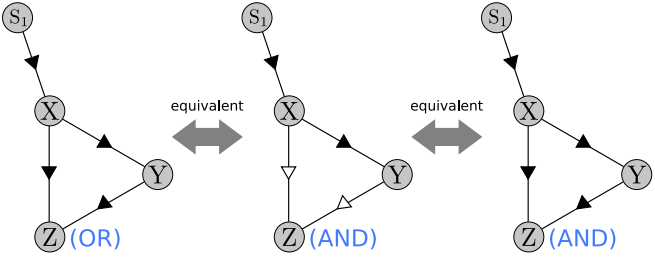}
\caption{Equivalent transformation with the setup of Fig.~2: An OR-logic FFL is equivalent to the same pattern with the AND gate.
In the second transformation, we conduct the equivalent transformation on $S_1$, $X$ and $Y$.
}
\label{fig: equiv3}
\end{center}
\end{figure}

\begin{figure}
\begin{center}
\includegraphics[width = 1\linewidth]{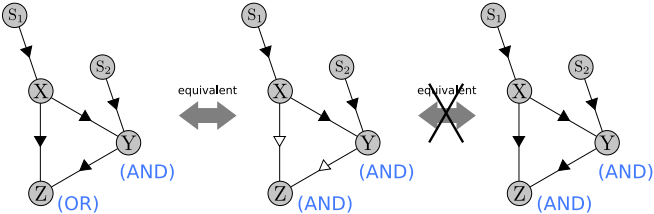}
\caption{Equivalent transformation with the setup of Fig.~5: Unlike Fig.~\ref{fig: equiv3}, FFLs with different regulatory functions are generally not equivalent each other.}
\label{fig: equiv4}
\end{center}
\end{figure}
In this study, we have assumed that the regulatory functions are given by the AND gates, but there is other type of gene whose regulatory function can be described by the OR gate. Even if we assume the regulatory functions of a pattern to be a combination of the AND and OR gates, the scatter plots of Fig.~\ref{fig: main1} and Fig.~\ref{fig: main2} themselves do not change, while a pattern that corresponds to each point becomes different from that of Fig.~\ref{fig: main1} and Fig.~\ref{fig: main2}. 
Therefore, it is enough to consider the mapping from a pattern with the AND and OR gates to a pattern with only the AND gates on the basis of the equivalent transformation.\\ \indent
We first consider the results in Fig.~\ref{fig: main1} that are independent of the regulatory functions. (i) The points of the FFLs are independent of the choice of the regulatory functions in Fig.~\ref{fig: main1} (see Fig.~\ref{fig: equiv3}). Therefore, the CFFLs belong to the informative type and the IFFLs belong to the adaptive type independently of the regulatory functions. (ii) No network motifs belong to the dissipative type. This is because a negative feedback loop is mapped to a negative feedback loop by the equivalent transformation, and it is still a necessary condition to belong to the dissipative type that a pattern includes negative feedback loops. \\ \indent
We next consider the results in Fig.~\ref{fig: main1} that are dependent on the regulatory functions. The classification of the positive feedback loop network motifs are dependent on the choice of regulatory functions. This is because a positive feedback loop shows two distinct behavior ``static" or ``bistable" depending on the regulatory functions (see Appendix~\ref{sec: pfbl} and Fig.~\ref{fig: pfblnm}).  \\ \indent
We also consider the results in Fig.~\ref{fig: main2} that are dependent on the regulatory functions. The tripartite mutual information of the FFLs are dependent on the regulatory functions. In the setup of Fig.~\ref{fig: setup2}, for example, let us consider a situation that the regulatory functions of Y and Z nodes of the C1 FFL are AND and OR respectively. If we express this as C1(AND, OR), C1(AND, OR) is equivalent to C3(AND, AND) as shown in Fig.~\ref{fig: equiv4}. By considering the other cases in the same manner, we can summarize the dependence of the results on the regulatory functions as Table \ref{table: 2}.\\ \indent
It mat be possible to explain the difference in the occurrence frequencies among the FFLs by the statistical tendency in the regulatory functions. This is because the FFLs which have similar occurrence frequencies are categorized into the same subgroups C1, I1, C2, I2 and C3, I3, C4, I4. If the combination of (AND, AND) or (OR, OR) occurs more frequently than (AND, OR) or (OR, AND), for example, we can expect that the C1, I1, C2, I2 FFLs occur more frequently than the C3, I3, C4, I4 FFLs. It is a remaining question why the occurrence frequencies of the C2 and I2 FFLs are low comparing to the C1 and I1 FFLs.

\begin{table}
\begin{center}
\begin{tabular}{| c | c | c |}\hline
& C1, I1, C2, I2 & C3, I3, C4, I4 \\ \hline
(AND, AND), (OR, OR) & small & around 0 \\ \hline
(AND, OR), (OR, AND) & around 0 & small \\ \hline
\end{tabular}
\end{center}
\caption{The tripartite mutual information of the FFLs are dependent on the regulatory functions. The vertical column represents the regulatory functions of (Y, Z) and the horizontal column represents the two types of the FFLs.}
\label{table: 2}
\end{table}

\section{Irrelevant patterns}
\label{sec: app4}
\begin{figure}[H]
\begin{center}
\includegraphics[width = 0.75\linewidth]{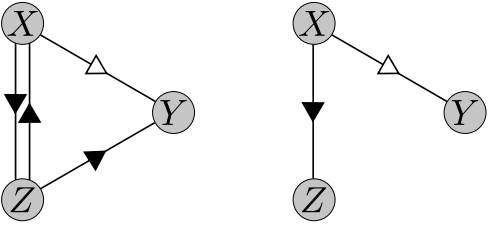}
\caption{Examples of patterns without causal relationship from $Y$ to $Z$.}
\label{fig: abbrev}
\end{center}
\end{figure}

\begin{figure}[H]
\begin{center}
\includegraphics[width = 0.75\linewidth]{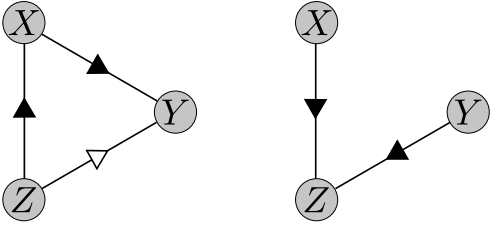}
\caption{Examples of patterns that have unregulated nodes.}
\label{fig: nonregu}
\end{center}
\end{figure}

\begin{figure}[H]
\begin{center}
\includegraphics[width = 0.75\linewidth]{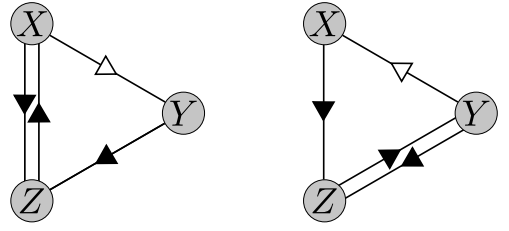}
\caption{Examples of patterns that are symmetric in terms of $X$ and $Y$.}
\label{fig: symmetry}
\end{center}
\end{figure}

There are totally $3^6 - 8\cdot3 - 1 = 704$ three-node patterns, but some of them are irrelevant.
In this study we exclude patterns from calculation with the following rules: (i) patterns without causal relationship from $X$ to $Z$ or $Y$ to $Z$, (ii) patterns that have unregulated nodes. \\ \indent
Two examples of the case (i) are shown in Fig.~\ref{fig: abbrev}, where $Y$ does not regulate $X$ or $Z$ in these patterns. Since such patterns are equivalent to some two-node patterns, we exclude them from our calculation.
Two examples of the case (ii) are shown in Fig.~\ref{fig: nonregu}. Patterns without regulation of $Y$ are equivalent to some two-node patterns in the setup of Fig.~\ref{fig: setup}, while patterns without regulation of $Z$ are meaningless.\\ \indent
Moreover, in the setup of Fig.~\ref{fig: setup2}, the roles of $X$ and $Y$ become equivalent.
We calculated only one of patterns that become equivalent by swapping nodes $X$ and $Y$ (Fig.~\ref{fig: symmetry}).
In summary, we include only a single pattern out of equivalent ones for our calculation on the basis of the above rules. 
As a result, we performed our calculation for 283 patterns for the setup of Fig.~\ref{fig: setup} and 204 patterns for the setup of Fig.~\ref{fig: setup2}
(see Supplemental Material for the list of these patterns).

\section{Dynamics of patterns}
\label{sec: app5}
\begin{figure}
\subfigure[M9-1-2 (pink type)]{\includegraphics[width = 0.9\linewidth]{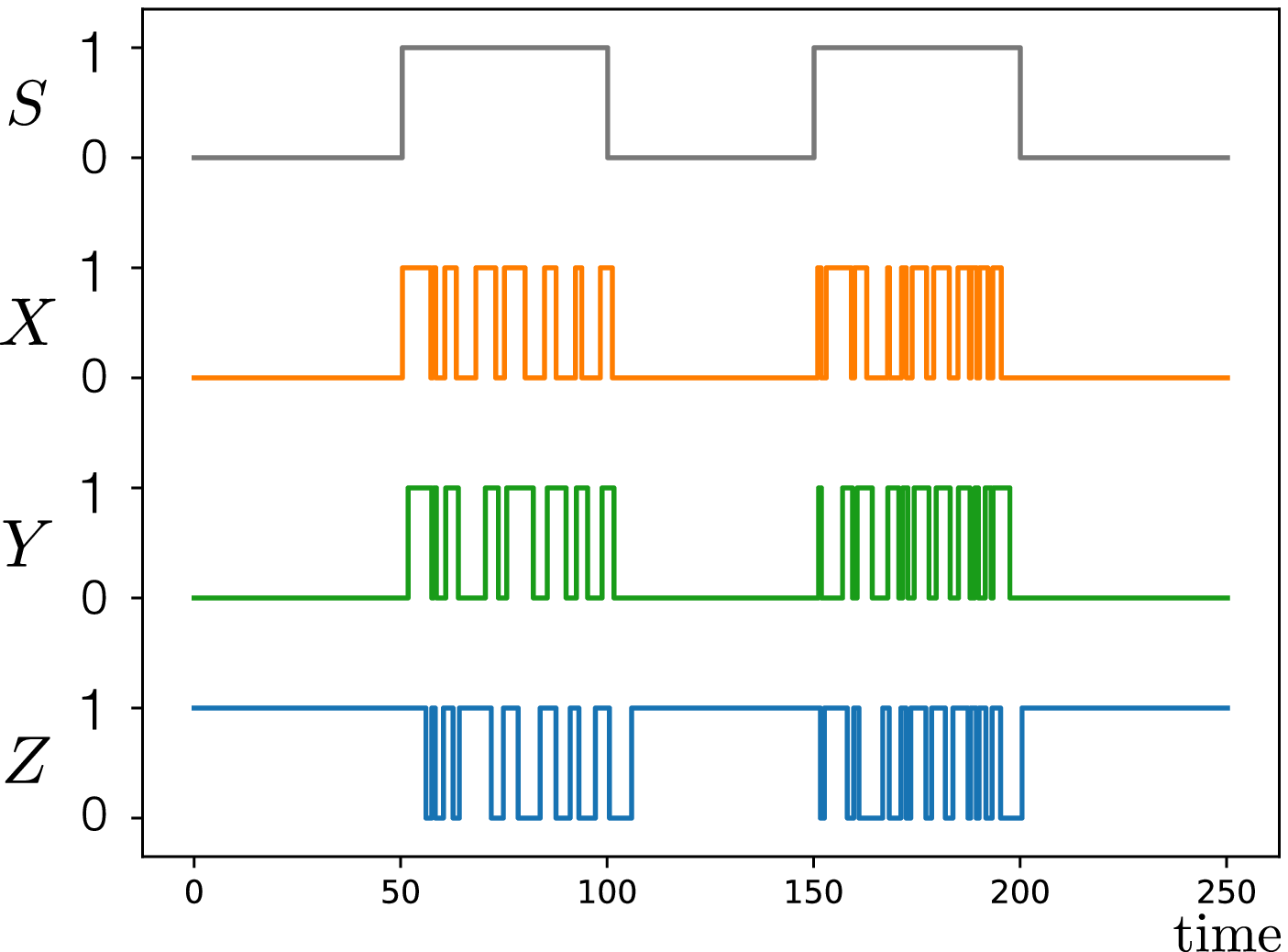}\label{fig: 9-1-2}}
\subfigure[M10-3-4 (pink type)]{\includegraphics[width = 0.9\linewidth]{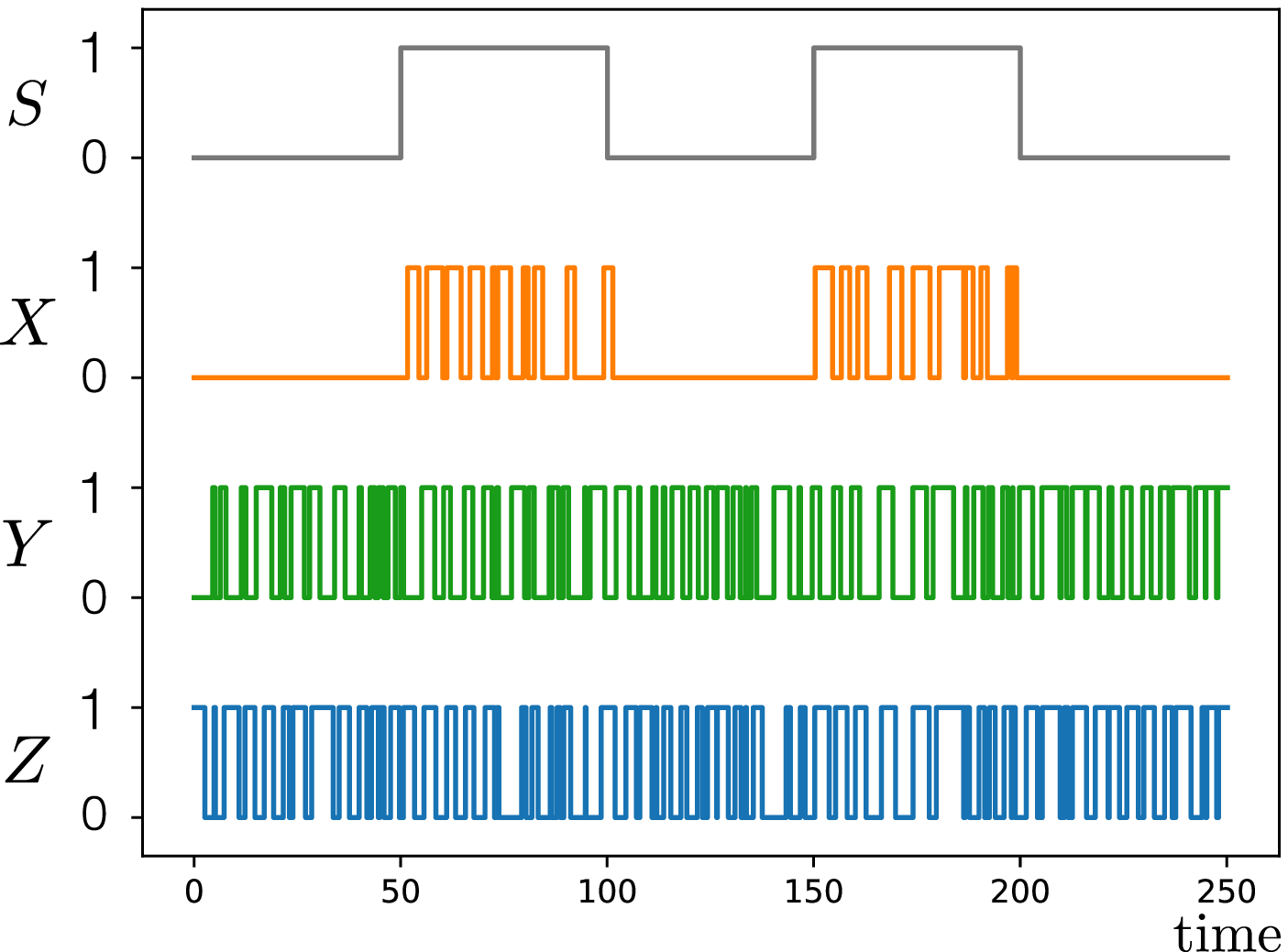}\label{fig: 10-3-4}}
\caption{Examples of dynamics of patterns in the pink type (dissipative). Both of them show oscillatory behavior.}
\label{fig: ex_pink}
\end{figure}

\begin{figure}
\subfigure[M11-1-1 (gray type)]{\includegraphics[width = 0.9\linewidth]{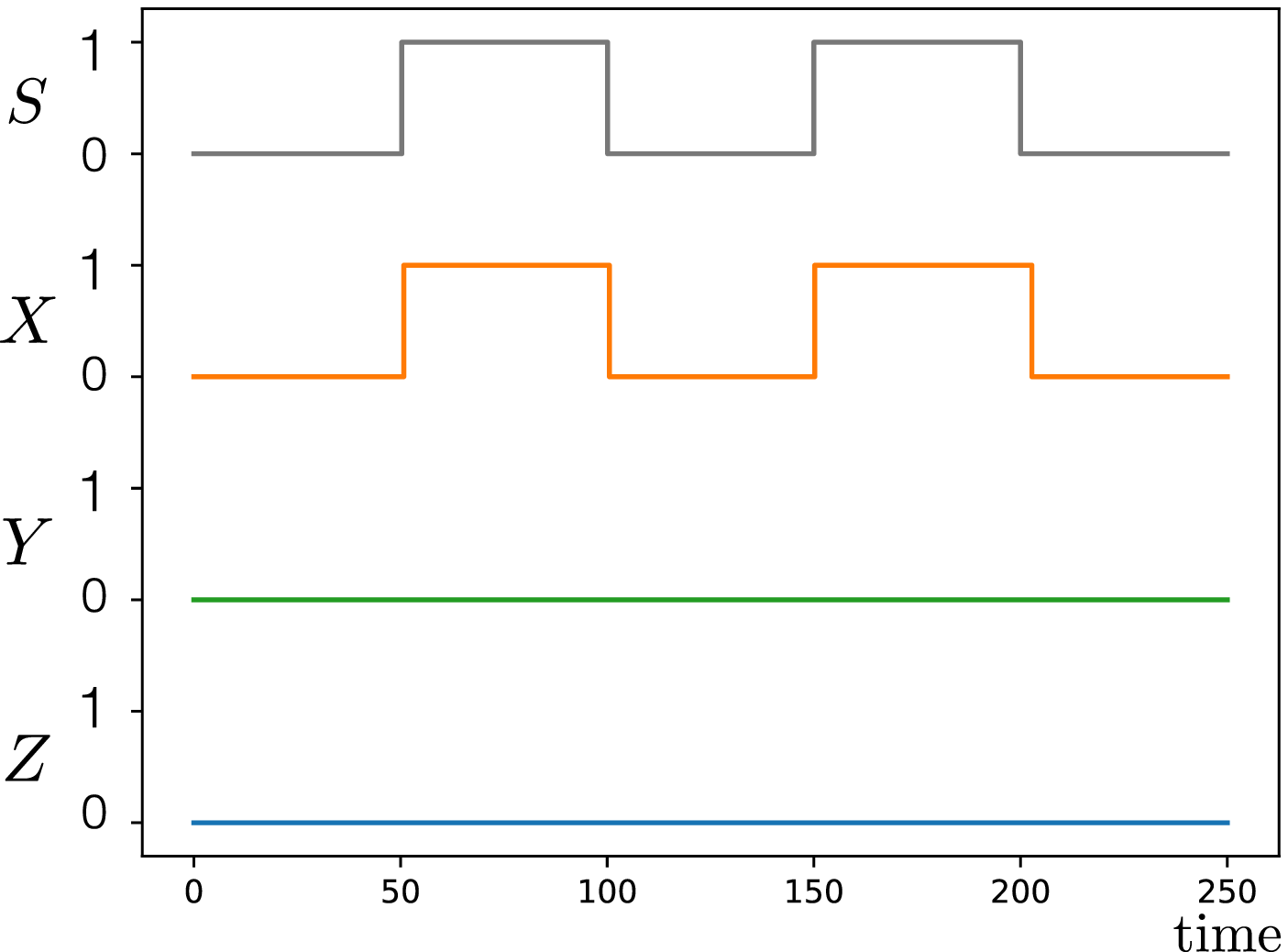}\label{fig: 11-1-1}}
\caption{An example of dynamics of a pattern in the gray type (static). The states of $Y$ and $Z$ become static due to the positive feedback loop between them.}
\label{fig: ex_gray}
\end{figure}

\begin{figure}
\subfigure[M11-1-8 (green type)]{\includegraphics[width = 0.9\linewidth]{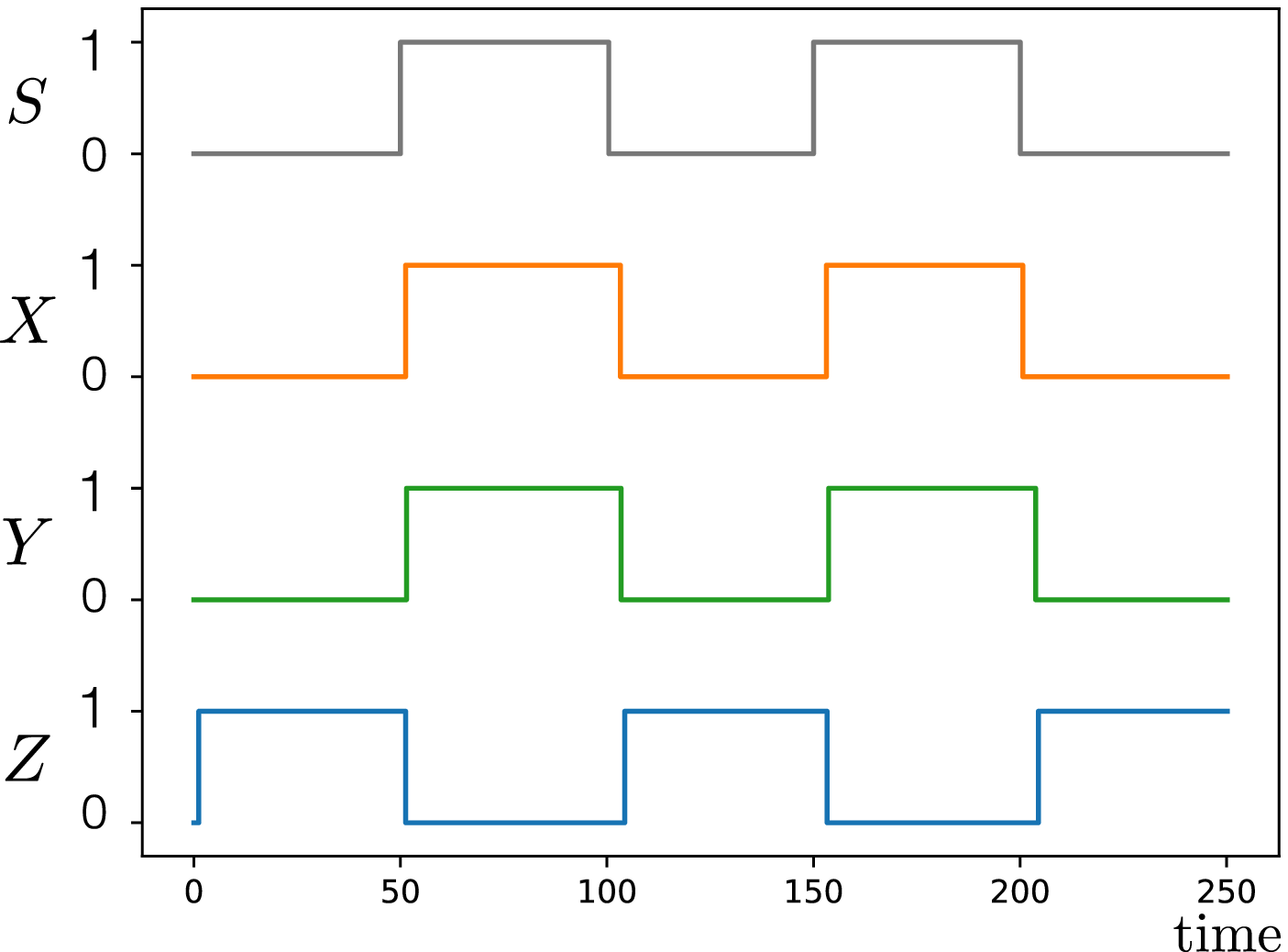}\label{fig: 11-1-8}}
\subfigure[M10-5-6 (blue type)]{\includegraphics[width = 0.9\linewidth]{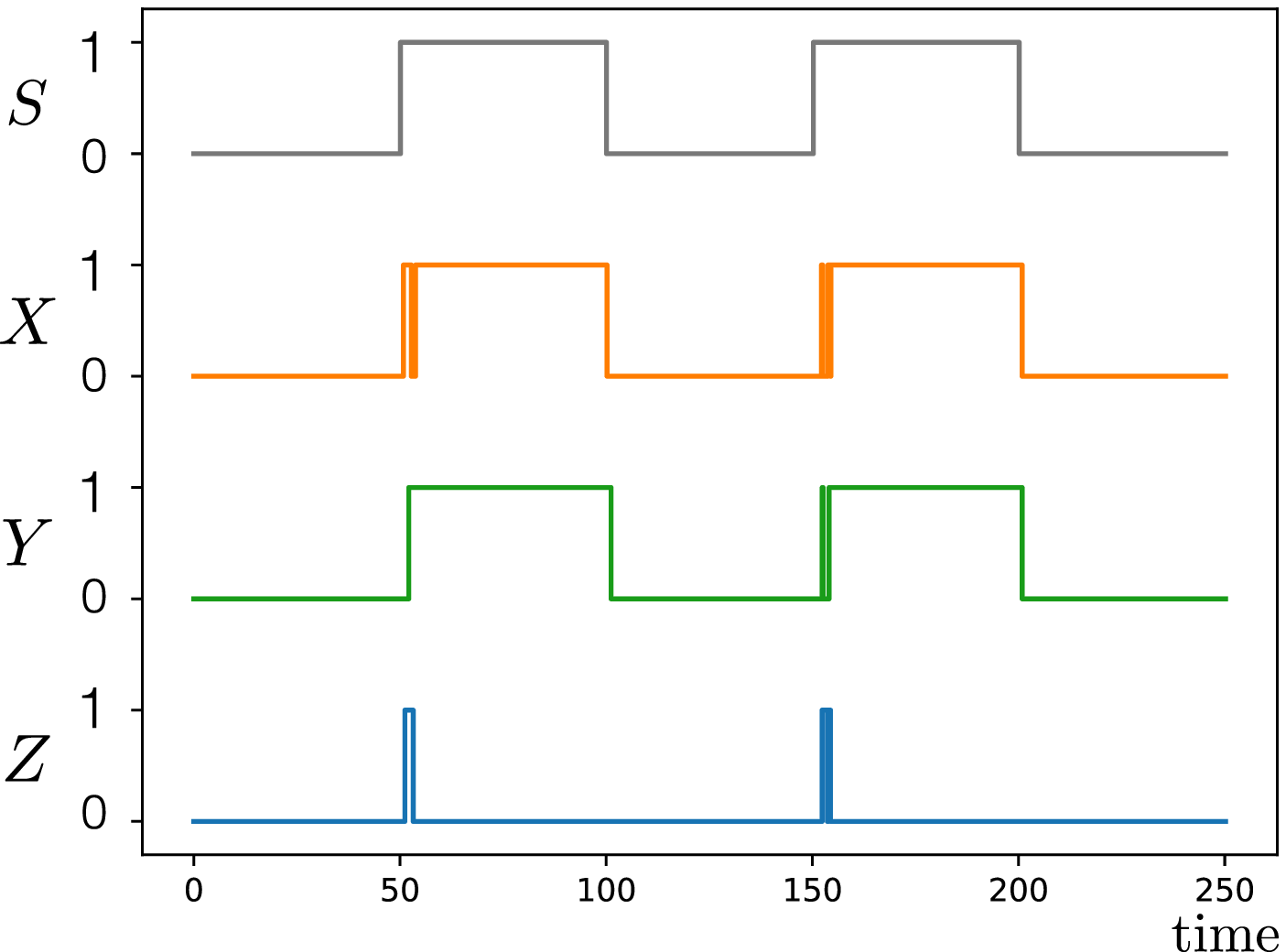}\label{fig: 10-5-6}}
\caption{Examples of dynamics of patterns in the green type (informative) and the blue type (adaptive). M11-1-8 in the green type propagates information efficiently, while M10-5-6 in the blue type shows an adaptive behavior.}
\label{fig: ex_gb}
\end{figure}

In this Appendix, we show the characteristics of the four types shown in Table.~\ref{table: 1} with examples of numerical simulation based on the Gillespie method \cite{Gillespie1977}. In addition, we argue the function of a positive feedback loop by using two examples.
From Fig.~\ref{fig: ex_pink} to Fig.~\ref{fig: ex_gb}, we show the time evolution of $X$, $Y$ and $Z$ when $S$ changes with a constant period.
For simplicity, we set $e = 0$, while $\gamma = 1$ as in Fig.~\ref{fig: main1}.
\subsection{Pink type: Dissipative}
M9-1-2 belongs to \includegraphics[width = 0.25cm]{pinkc.eps}, and M10-3-4 belongs to \includegraphics[width = 0.25cm]{pinkdt.eps}, which are respectively shown in Fig.~\ref{fig: 9-1-2} and \subref{fig: 10-3-4}.
Both of them include negative feedback loops, and are dissipative types. 
We can see oscillatory behavior.
In M10-3-4, $Z$ shows oscillations independently of $S_t$, while in M9-1-2, $Z$ shows oscillations only when $S_t$ is 1.
Thus we argue that this difference leads to the difference in the dissipation of \includegraphics[width = 0.25cm]{pinkc.eps} and \includegraphics[width = 0.25cm]{pinkdt.eps}.

\subsection{Gray type: Static}
M11-1-1 belongs to \includegraphics[width = 0.25cm]{grayc.eps}, which includes a positive feedback loop.
We can see that $Z$ converges to a stationary value.

\subsection{Green and blue types: Informative and adaptive}
M11-1-8 belongs to the informative type \includegraphics[width = 0.25cm]{greenc.eps} and M10-5-6 belongs to the adaptive type \includegraphics[width = 0.25cm]{bluec.eps}, which are shown in Fig.~\ref{fig: 11-1-8} and \subref{fig: 10-5-6}, respectively.
In M11-1-8, the variation in $S$ propagates to $Z$ as it is, while in M10-5-6, $Z$ reacts only when $S$ changes from 0 to 1.

\begin{figure}
\includegraphics[width = 0.7\linewidth]{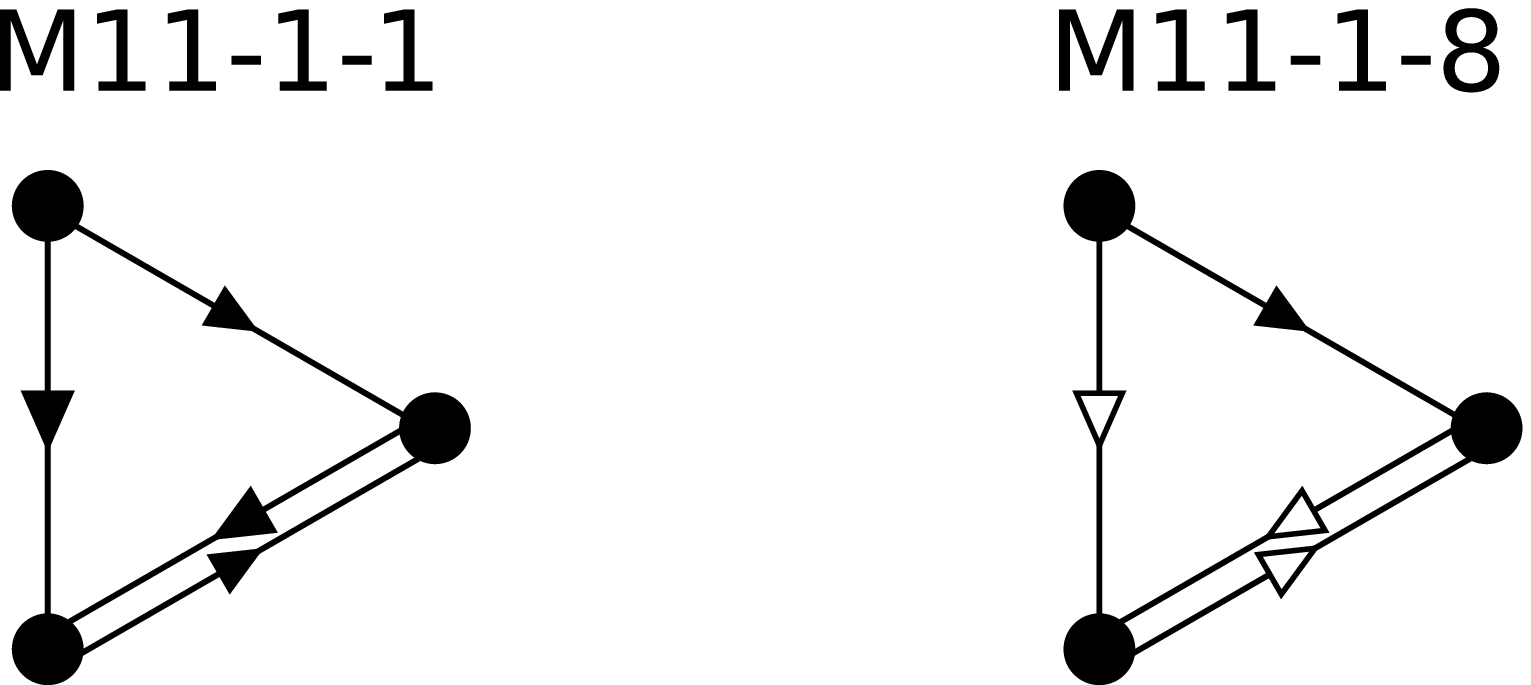}
\caption{Patterns with a positive feedback loop. These are network motifs which show distinct behavior as in Fig.~\ref{fig: 11-1-1} and Fig.~\ref{fig: 11-1-8}}
\label{fig: pfbl}
\end{figure}

\subsection{Positive feedback loop: Static or Bistable}\label{sec: pfbl}
\begin{figure}
\begin{center}
\includegraphics[width = 1\linewidth]{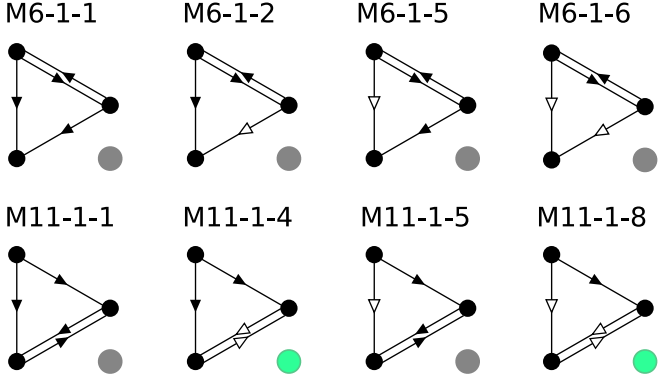}
\caption{The classification of the positive feedback loop network motifs. We list all the positive feedback loop network motifs excluding the equivalent ones. Most of them are classified into the gray type, but some of them belong to the green type.}
\label{fig: pfblnm}
\end{center}
\end{figure}

A positive feedback loop has two functions depending on the signs of the two regulations. A positive feedback loop composed of two positive regulations has a static property, and that composed of two negative regulations is bistable (see Fig.~\ref{fig: 11-1-1}, \ref{fig: 11-1-8} and \ref{fig: pfbl}). Here, we assume that the regulatory functions are the AND gates. Thus, although most of the PFBL network motifs belong to the gray type, some of them are classified into the green type (Fig.~\ref{fig: pfblnm}). Therefore, it is necessary to check the signs and the logics of the regulatory functions when we discuss the role of the PFBL network motifs in gene regulatory networks.

\section{Details of the fitting in Fig.~\ref{fig: main2}}
\label{sec: app6}
We show the details of the fitting curves in Fig.~\ref{fig: main2}.
For example, the probability distribution of the model of Fig.~\ref{fig: model2} is given by
\begin{eqnarray}
P(S_1 = 0, S_2 = 0, Z = 0) &=& \frac{1}{4}\left\{\varepsilon\cdot(1 - \varepsilon_0\cdot\varepsilon_0) \right.\nonumber\\
&&\left.+(1-\varepsilon)\cdot(1-\varepsilon_0)\right\},\\
P(S_1 = 0, S_2 = 0, Z = 1) &=& \frac{1}{4}\left\{\varepsilon\cdot\varepsilon_0\cdot\varepsilon_0 \right.\nonumber\\
&&\left.+ (1-\varepsilon)\cdot\varepsilon_0\right\},\\
P(S_1 = 1, S_2 = 0, Z = 0) &=& \frac{1}{4}\left\{\varepsilon\cdot(1 - (1-\varepsilon_0)\cdot\varepsilon_0) \right.\nonumber\\
&&\left.+ (1-\varepsilon)\cdot\varepsilon_0\right\},\\
P(S_1 = 1, S_2 = 0, Z = 1) &=& \frac{1}{4}\left\{\varepsilon\cdot(1-\varepsilon_0)\cdot\varepsilon_0 \right.\nonumber\\
&&\left.+ (1-\varepsilon)\cdot(1-\varepsilon_0)\right\},\\
P(S_1 = 0, S_2 = 1, Z = 0) &=& \frac{1}{4}\left\{\varepsilon\cdot(1 - (1-\varepsilon_0)\cdot\varepsilon_0) \right.\nonumber\\
&&\left.+ (1-\varepsilon)\cdot(1-\varepsilon_0)\right\},\\
P(S_1 = 0, S_2 = 1, Z = 1) &=& \frac{1}{4}\left\{\varepsilon\cdot(1-\varepsilon_0)\cdot\varepsilon_0 \right.\nonumber\\
&&\left.+ (1-\varepsilon)\cdot\varepsilon_0\right\},
\end{eqnarray}

\begin{eqnarray}
P(S_1 = 1, S_2 = 1, Z = 0) &=& \frac{1}{4}\left\{\varepsilon\cdot(1 - (1-\varepsilon_0)\cdot(1-\varepsilon_0)) \right.\nonumber\\
&&\left.+ (1-\varepsilon)\cdot\varepsilon_0\right\},\\
P(S_1 = 1, S_2 = 1, Z = 1) &=& \frac{1}{4}\left\{\varepsilon\cdot(1-\varepsilon_0)\cdot(1-\varepsilon_0) \right.\nonumber\\
&&\left.+ (1-\varepsilon)\cdot(1-\varepsilon_0)\right\}.
\end{eqnarray}
Then, $I(\{S_1, S_2\}:Z)$ and $I_3(S_1, S_2, Z)$ are determined as functions of $\varepsilon_0$ and $\varepsilon$ by putting the above expressions into their definitions.\\

\section{Parameter dependence of the main results}
\label{sec: app7}
We discuss the parameter dependence of Fig.~\ref{fig: main1} and Fig.~\ref{fig: main2}.
For example, Fig.~\ref{fig: S100_DAll} represents the case where $\gamma_S$ is set to 0.01, while the other parameters are kept in the same values as those of Fig.~\ref{fig: main1}.
We note that the information-thermodynamic dissipation diverges and thus is meaningless for $e = 0$.\\ \indent
We show the cases of various parameters in Fig.~\ref{fig: param1} to Fig.~\ref{fig: param3}. There are some cases where the type classification become ambiguous, but it can be reasonably understood by considering the characteristics of individual types, as discussed below. 
The classification becomes the most ambiguous in the case of $\gamma_S = 1$. This is the case where the signal changes before the system relaxes, which is unrealistic in real biological systems. 
The case where the classification becomes ambiguous the second most is $e = 0.1$. This is the case where stochasticity of the system is too large, and the static nature of positive feedback loops disappears, leading to the small difference between \includegraphics[width = 0.25cm]{grayc.eps} and \includegraphics[width = 0.25cm]{greenc.eps}\includegraphics[width = 0.25cm]{bluec.eps}.\\ \indent
In Fig.~\ref{fig: param4}, we show scatter plots of the tripartite mutual information. The models of Fig.~\ref{fig: model1} and \subref{fig: model2} do not fit well with the data in Fig.~\ref{fig: S1_I3}, \subref{fig: diff_I3}, \subref{fig: ediff_I3}, and \subref{fig: e01_I3}. However, the conclusion does not change from that of Fig.~\ref{fig: main2}. For example, M4-1-1, M5-1-1 and M5-1-2 still take smaller values in terms of the tripartite mutual information in the plots of Fig.~\ref{fig: param4}. The reason why the models in Fig.~\ref{fig: main2} do not fit well with the data in Fig.~\ref{fig: diff_I3} and \subref{fig: ediff_I3} is that these models suppose the symmetric properties between $X$ and $Y$. On the other hand, in Fig.~\ref{fig: param4} \subref{fig: S100_I3} and \subref{fig: e0_I3}, the models fit well with the data, and the fitting parameters are determined as (a) $\varepsilon_0 = 0.022$, (e) $\varepsilon_0 = 0.153$.

\begin{widetext}
\newpage
\begin{figure}[H]
\begin{center}
\begin{tabular}{cc}
	\subfigure[$\gamma_S = 0.01$]{\includegraphics[width = 0.41\linewidth]{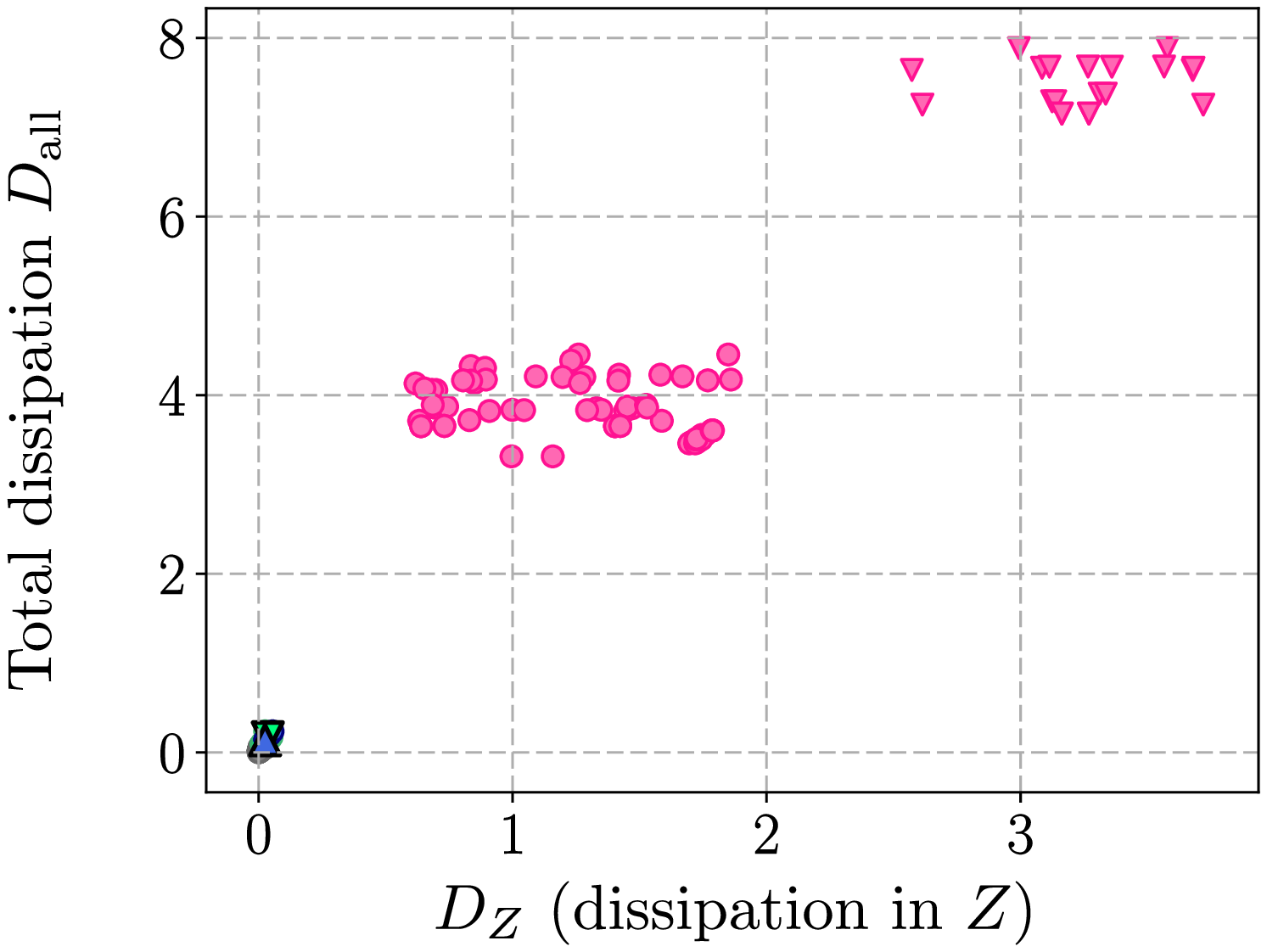}\label{fig: S100_DAll}}& 
	\subfigure[Enlarged view of (a)]{\includegraphics[width = 0.41\linewidth]{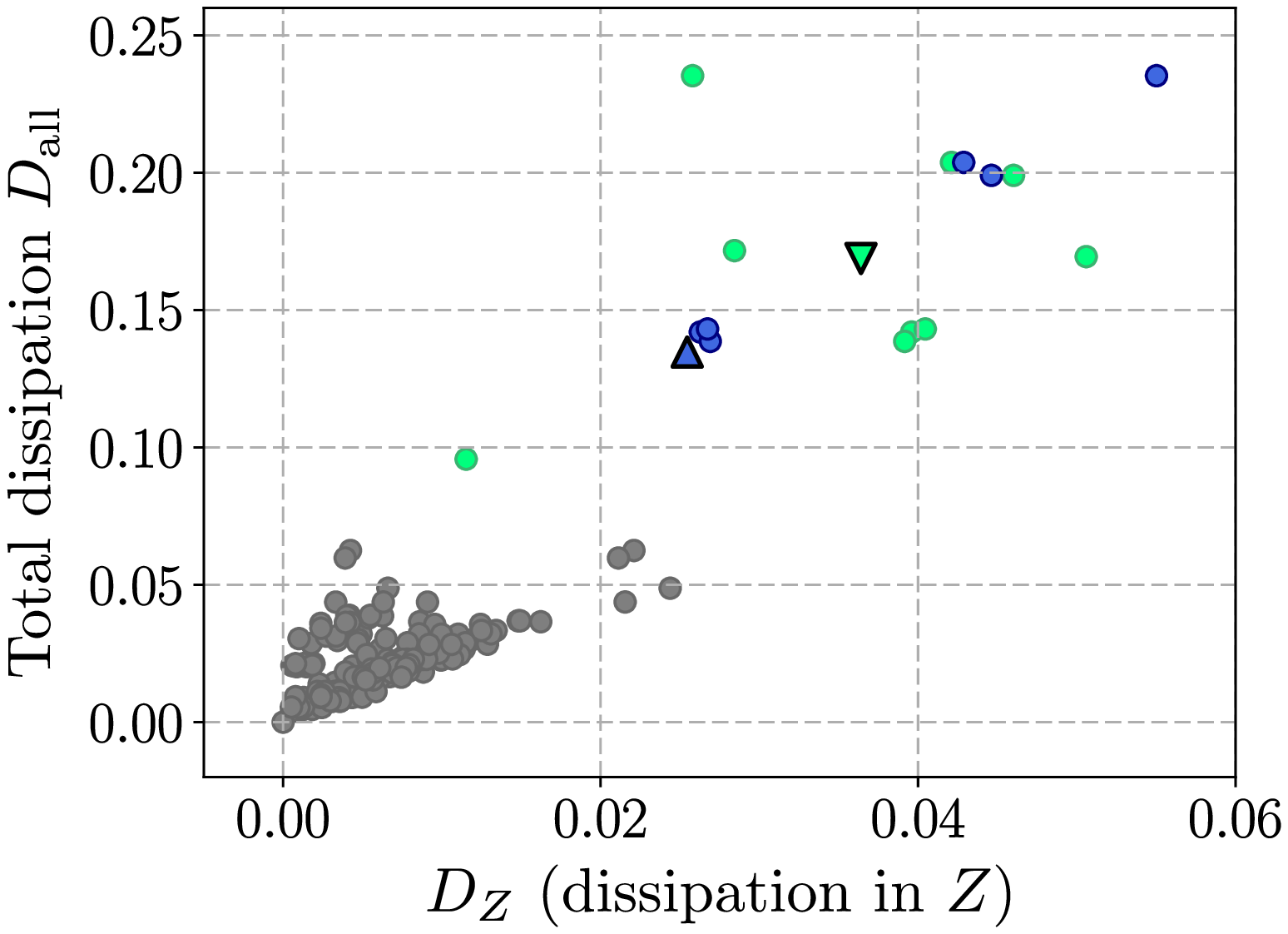}\label{fig: S100_DAll2}}
\end{tabular}
\end{center}
\end{figure}

\begin{figure}[H]
\begin{center}
\begin{tabular}{cc}
	\subfigure[$\gamma_X = 0.5, ~\gamma_Y = 1, ~\gamma_Z = 2$]{\includegraphics[width = 0.41\linewidth]{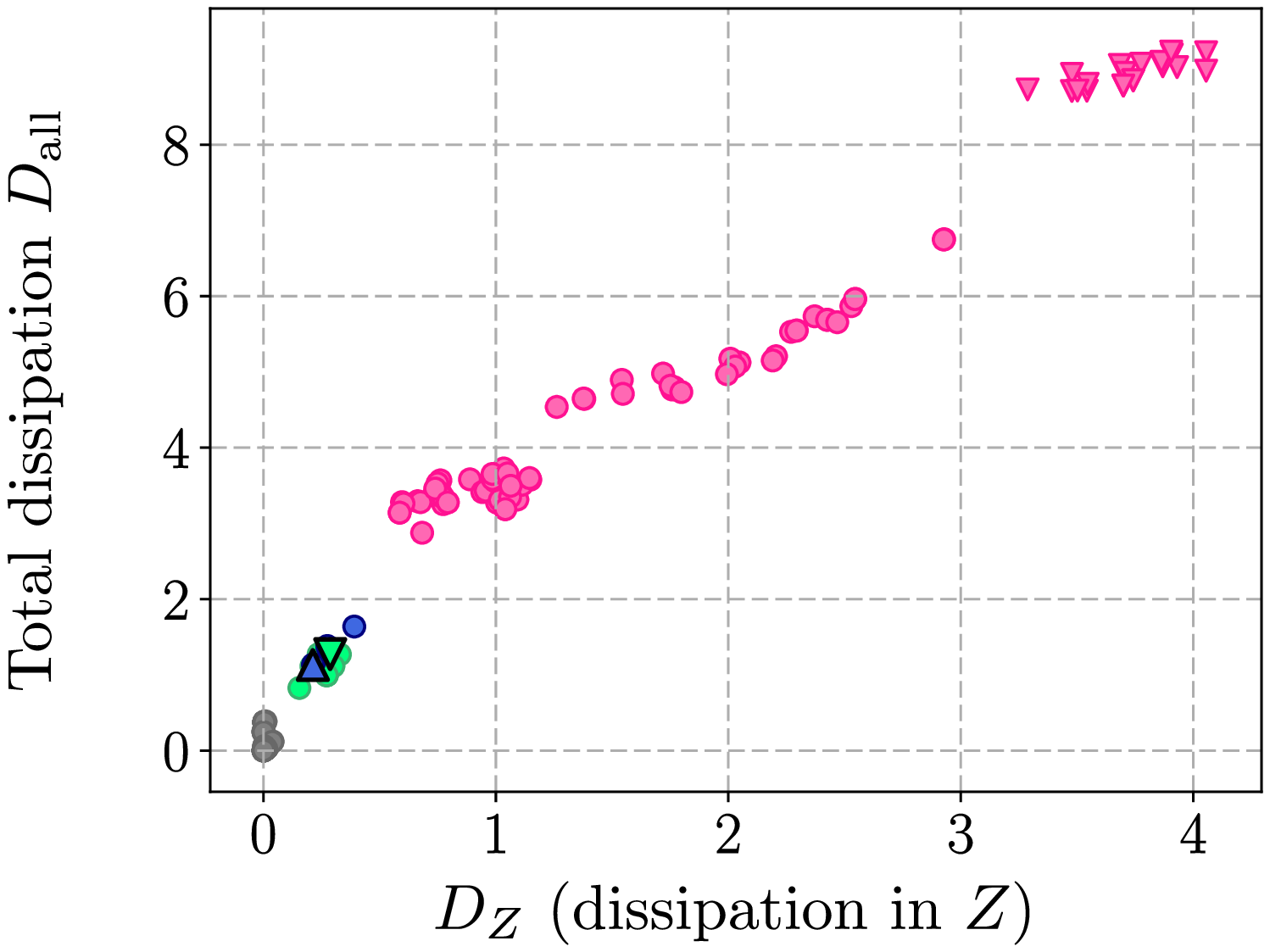}\label{fig: diff_DAll}}&
	\subfigure[$e_X = 0.01, ~e_Y = 0.02, ~e_Z = 0.005$]{\includegraphics[width = 0.41\linewidth]{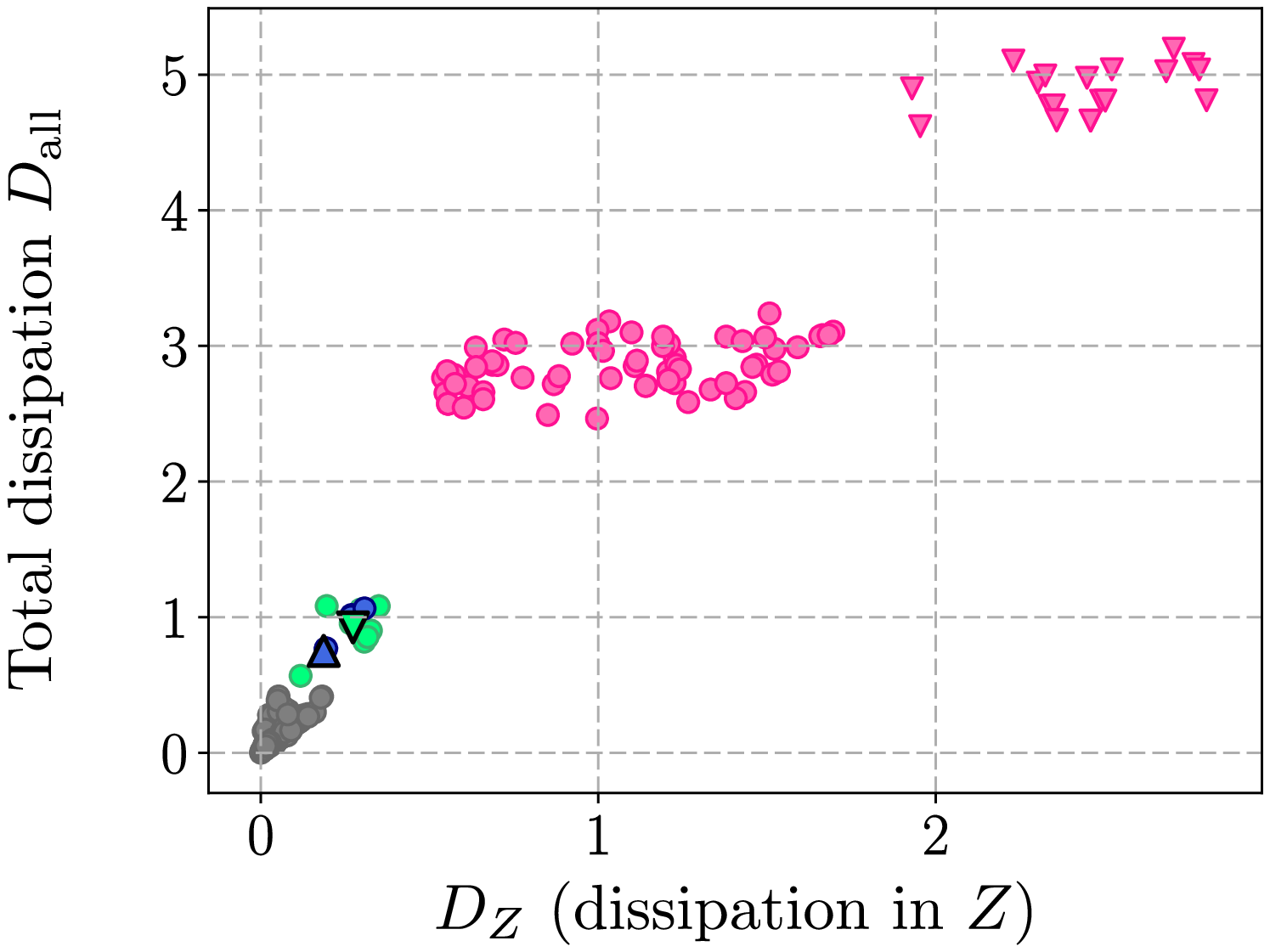}\label{fig: ediff_DAll}}\\
\end{tabular}
\end{center}
\end{figure}

\begin{figure}[H]
\begin{center}
\begin{tabular}{cc}
	\subfigure[$\gamma_S = 1$]{\includegraphics[width = 0.41\linewidth]{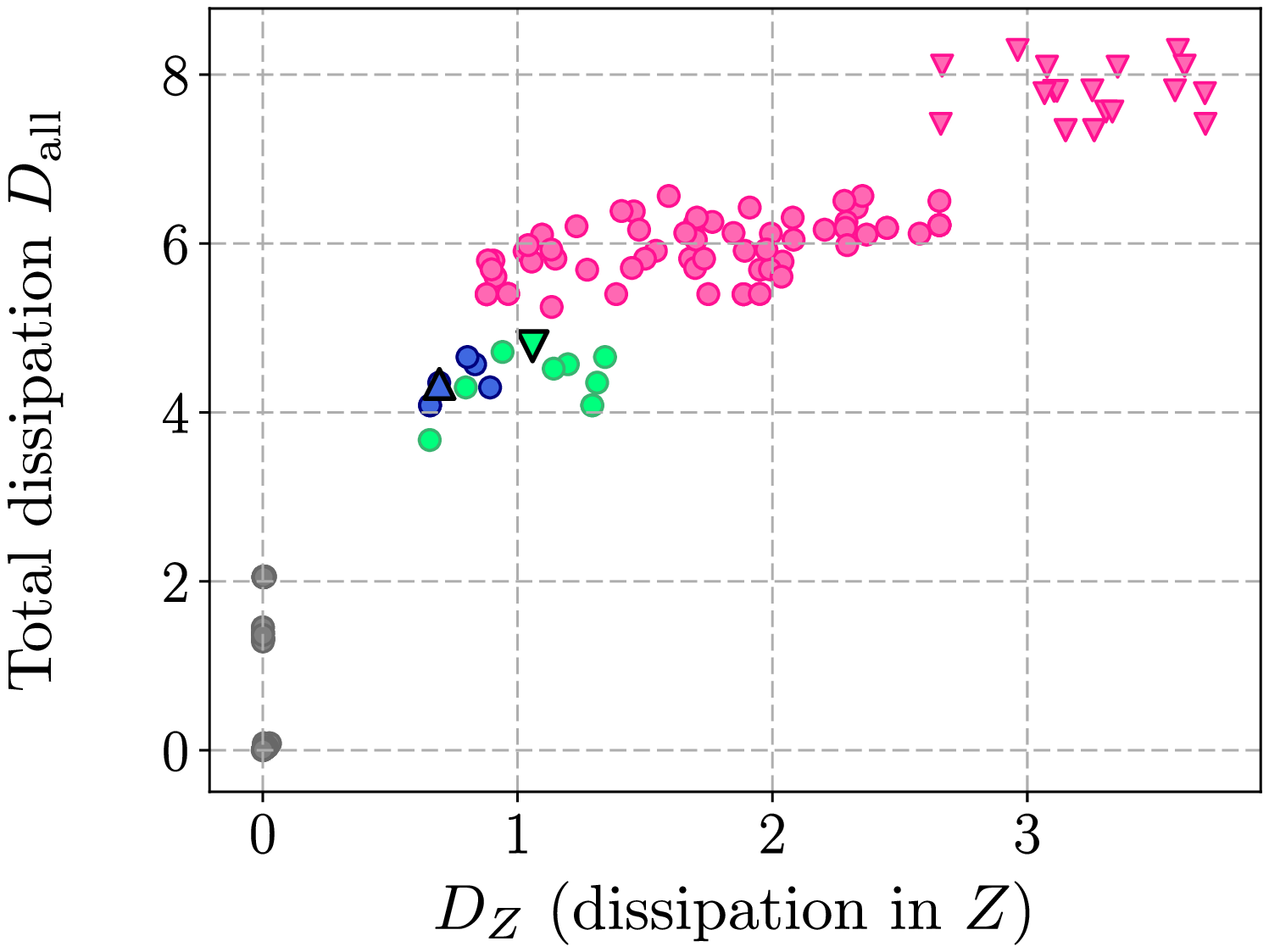}\label{fig:  S1_DAll}}&
	\subfigure[$e = 0.1$]{\includegraphics[width = 0.41\linewidth]{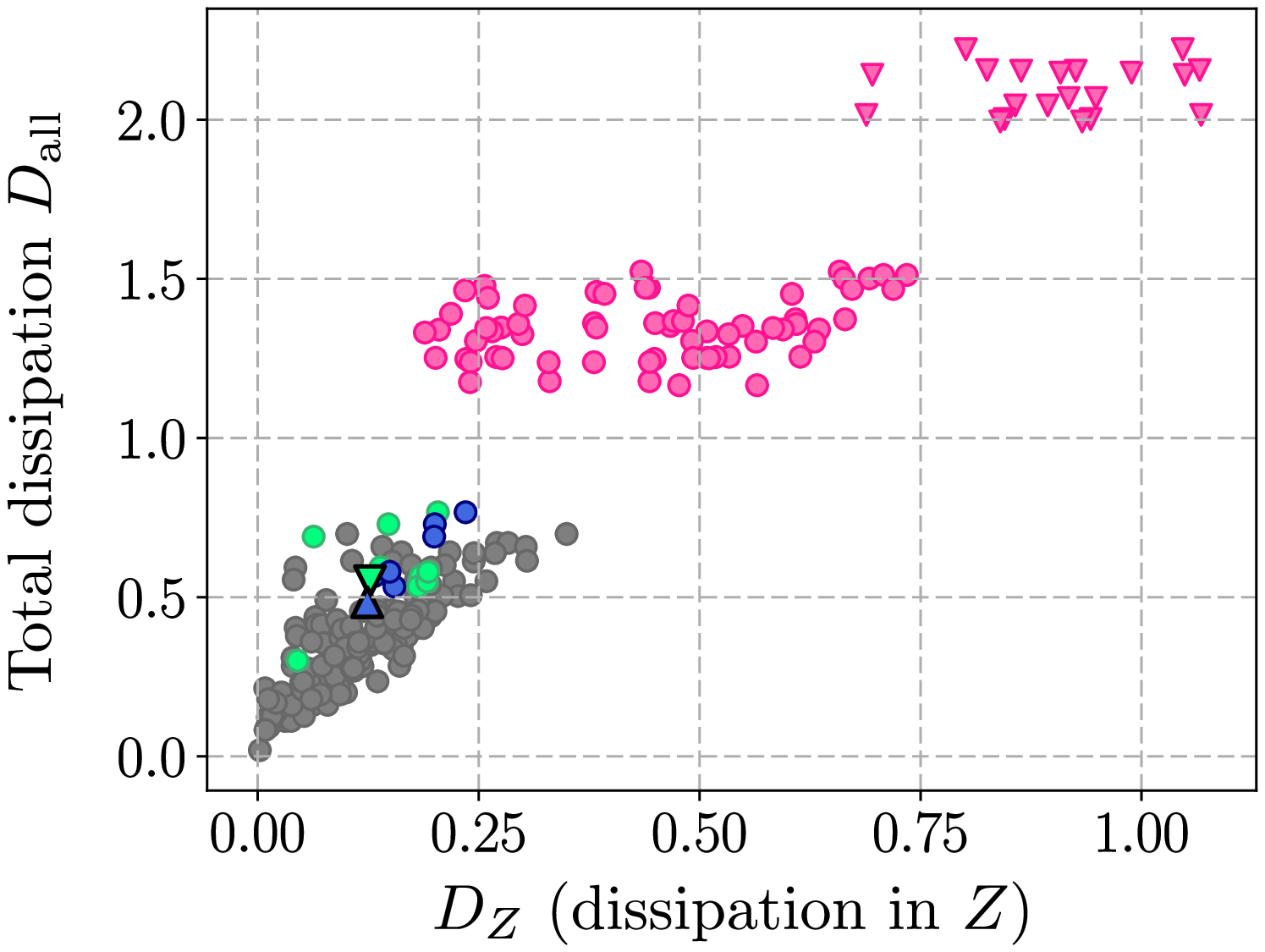}\label{fig: e01_DAll}}\\
\end{tabular}
\end{center}
\caption{Parameter dependence of the information-thermodynamic dissipation: (a) The case where the signal changes quite slowly compared to the characteristic times of the three nodes. (b) Enlarged view of (a). (c) The case where the transition rates are different in the three nodes. (d) The case where the reverse transition ratios are different in each node. (e) The case where the signal changes as fast as the characteristic times of the three nodes. (f) The case where the reverse transition ratios are large. In any case, the other parameters are set to the same values as those of Fig.~\ref{fig: main1}.}
\label{fig: param1}
\end{figure}
	
\begin{figure}[H]
\begin{center}
\begin{tabular}{cc}
	\subfigure[$\gamma_S = 0.01$]{\includegraphics[width = 0.41\linewidth]{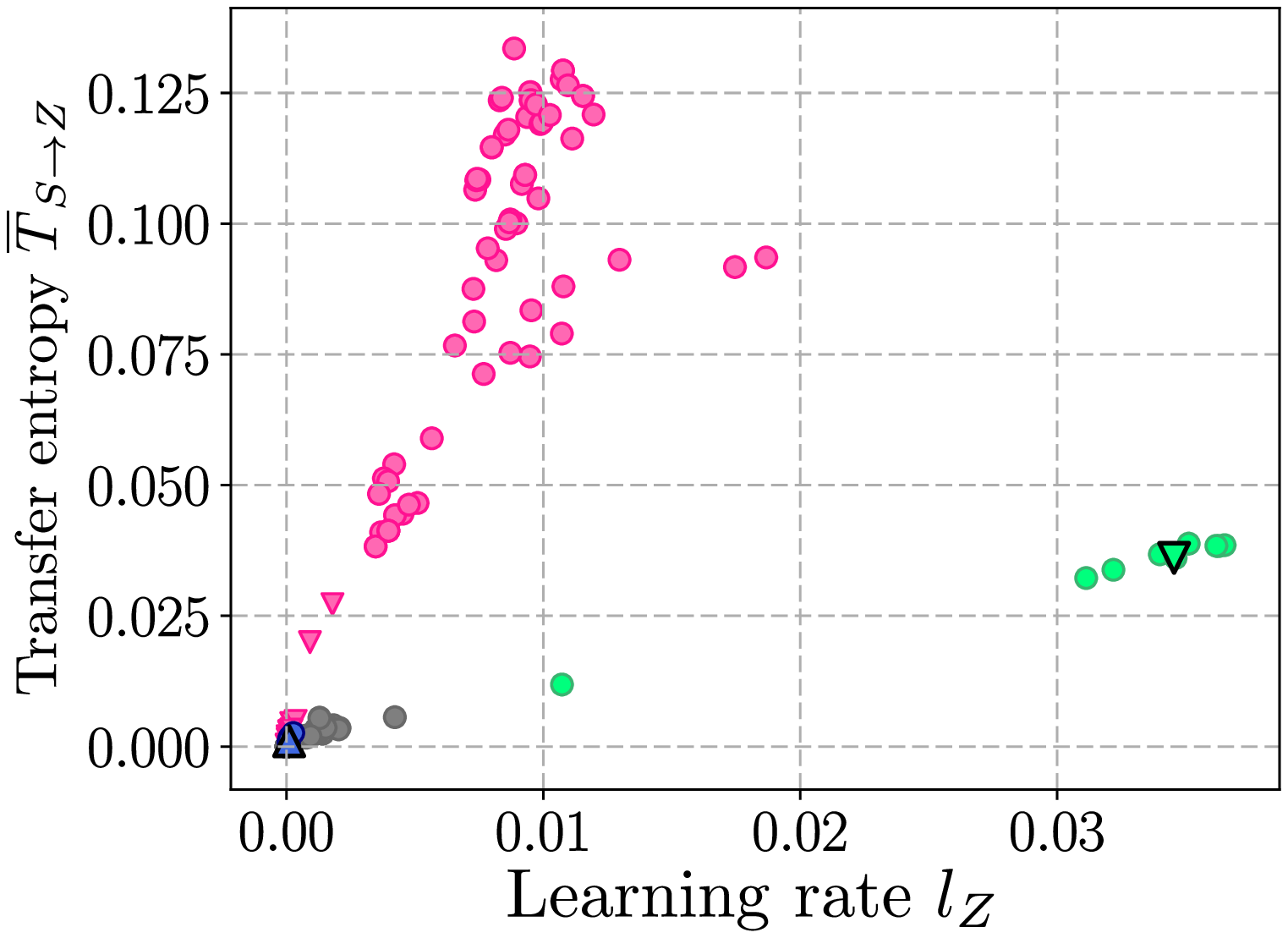}\label{fig: S100_FlowStoZ}}&
	\subfigure[$\gamma_S = 1$]{\includegraphics[width = 0.41\linewidth]{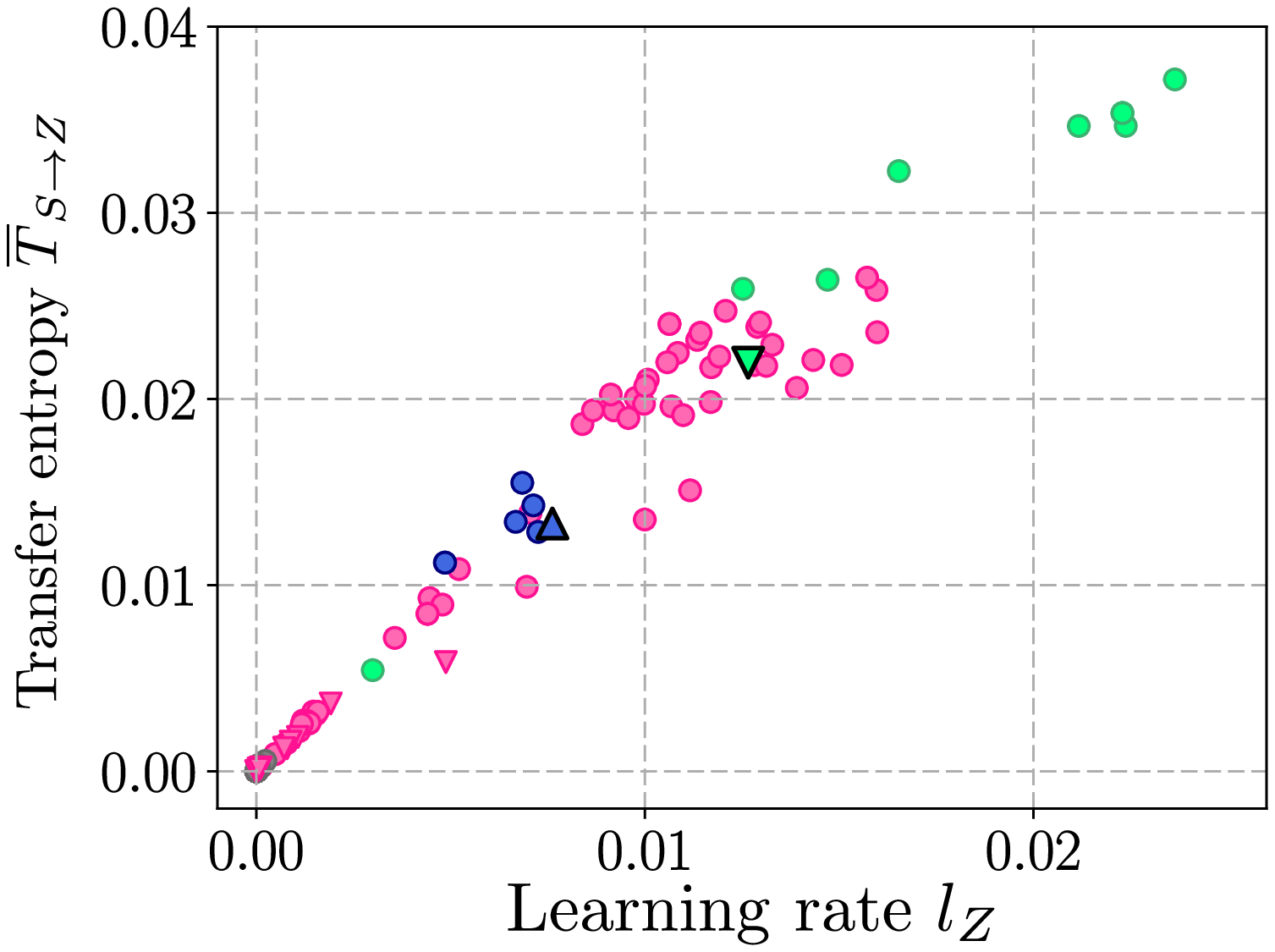}\label{fig: S1_FlowStoZ}}\\
\end{tabular}
\end{center}
\end{figure}

\begin{figure}[H]
\begin{center}
\begin{tabular}{cc}
	\subfigure[$\gamma_X = 0.5, ~\gamma_Y = 1, ~\gamma_Z = 2$]{\includegraphics[width = 0.41\linewidth]{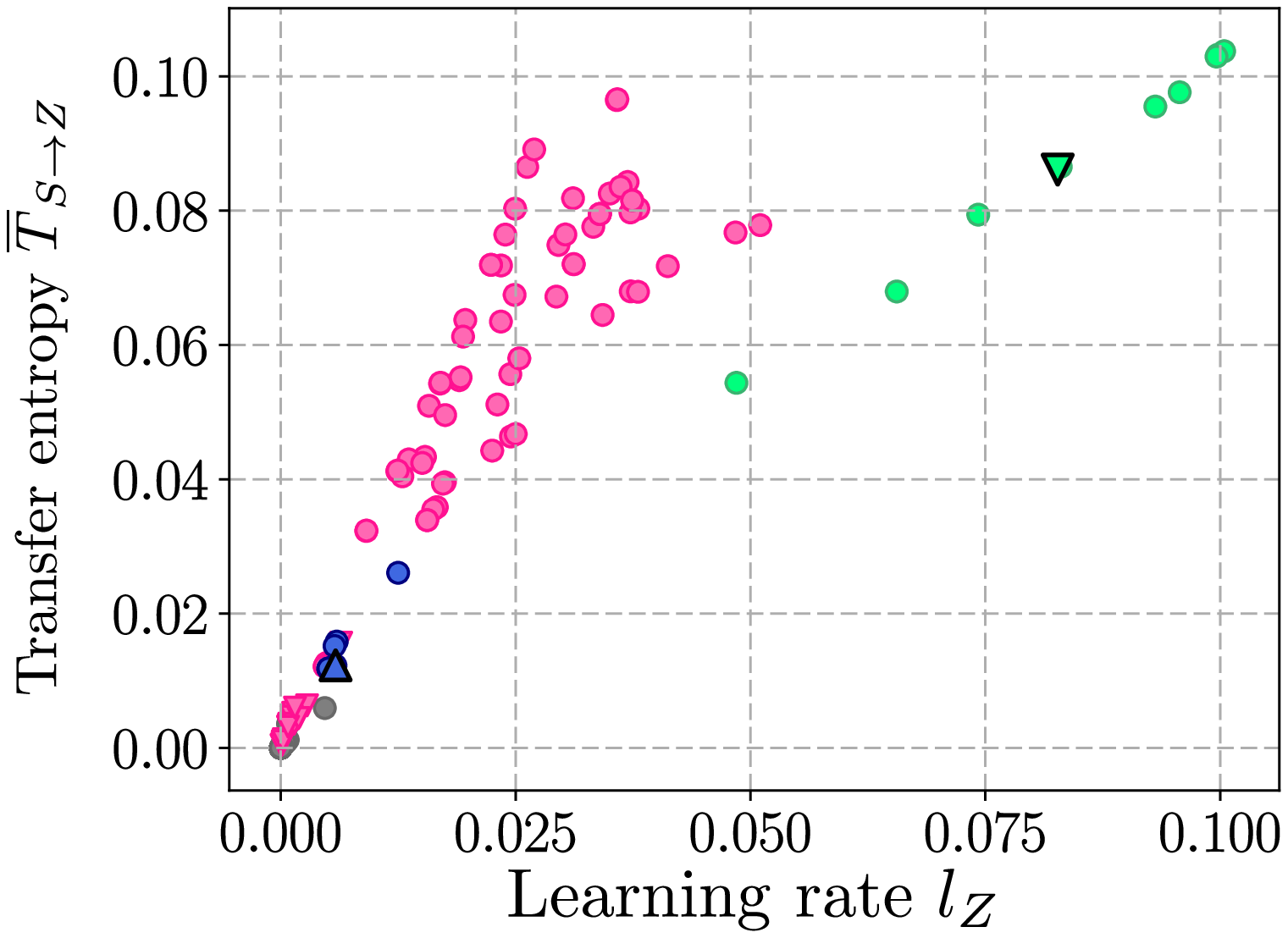}\label{fig: diff_FlowStoZ}}&
	\subfigure[$e_X = 0.01, ~e_Y = 0.02, ~e_Z = 0.005$]{\includegraphics[width = 0.41\linewidth]{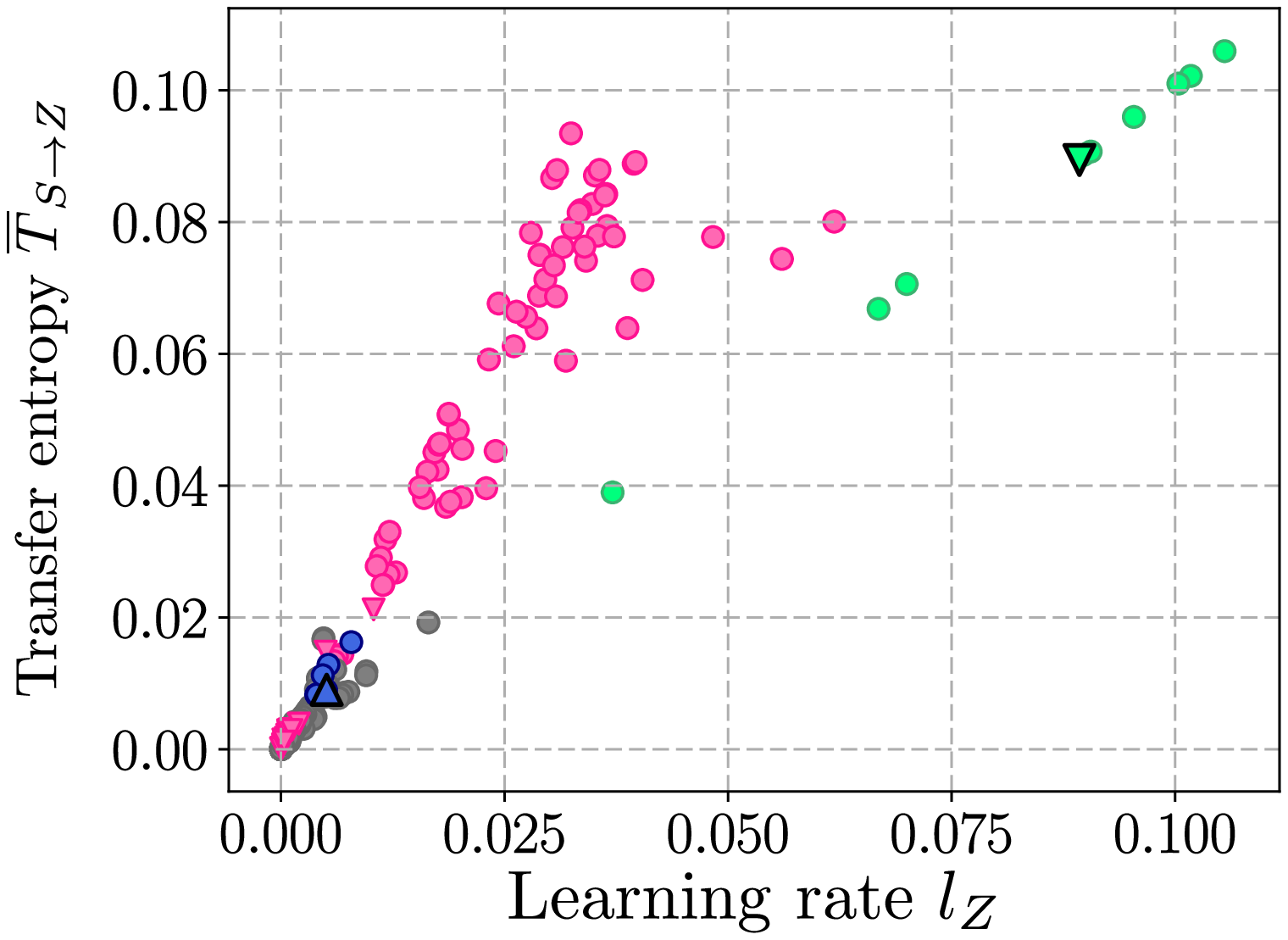}\label{fig: ediff_FlowStoZ}}\\
\end{tabular}
\end{center}
\end{figure}

\begin{figure}[H]
\begin{center}
\begin{tabular}{cc}
	\subfigure[$e = 0$]{\includegraphics[width = 0.41\linewidth]{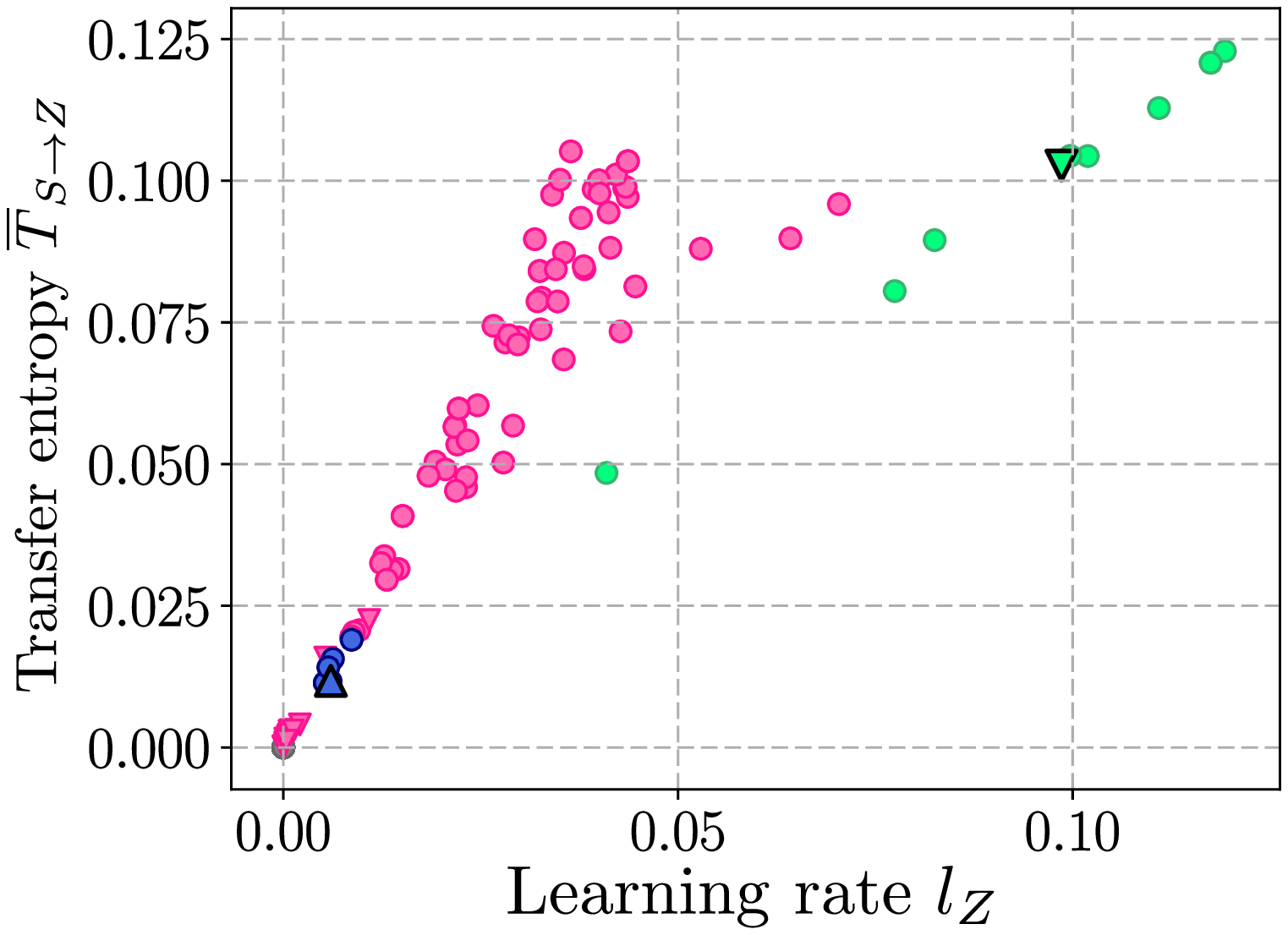}\label{fig: e0_FlowStoZ}}&
	\subfigure[$e = 0.1$]{\includegraphics[width = 0.41\linewidth]{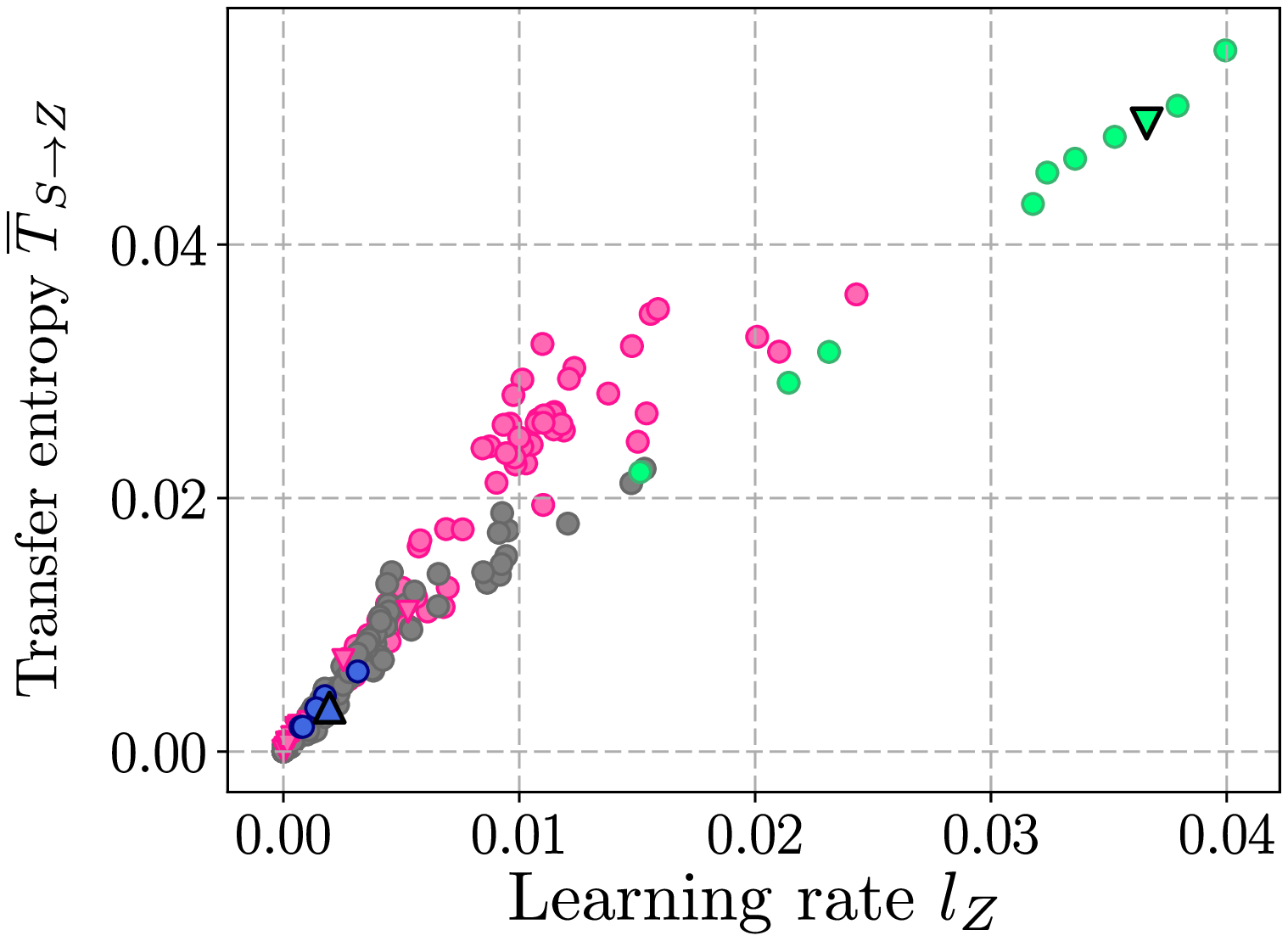}\label{fig: e01_FlowStoZ}}\\
\end{tabular}
\end{center}
\caption{Parameter dependence of the informational quantities: (a) The case where the signal changes quite slowly compared to the characteristic times of the three nodes. (b) The case where the signals changes as fast as the characteristic times of the three nodes. (c) The case where the transition rates are different in three nodes. (d) The case where the reverse transition ratios are different in the three nodes. (e) The case where the reverse transition ratios are zero. (f) The case where the reverse transition ratios are large. In any case, the other parameters are set to the same values as those of Fig.~\ref{fig: main1}.}
\label{fig: param2}
\end{figure}

\begin{figure}[H]
\begin{center}
\begin{tabular}{cc}
	\subfigure[$\gamma_S = 0.01$]{\includegraphics[width = 0.41\linewidth]{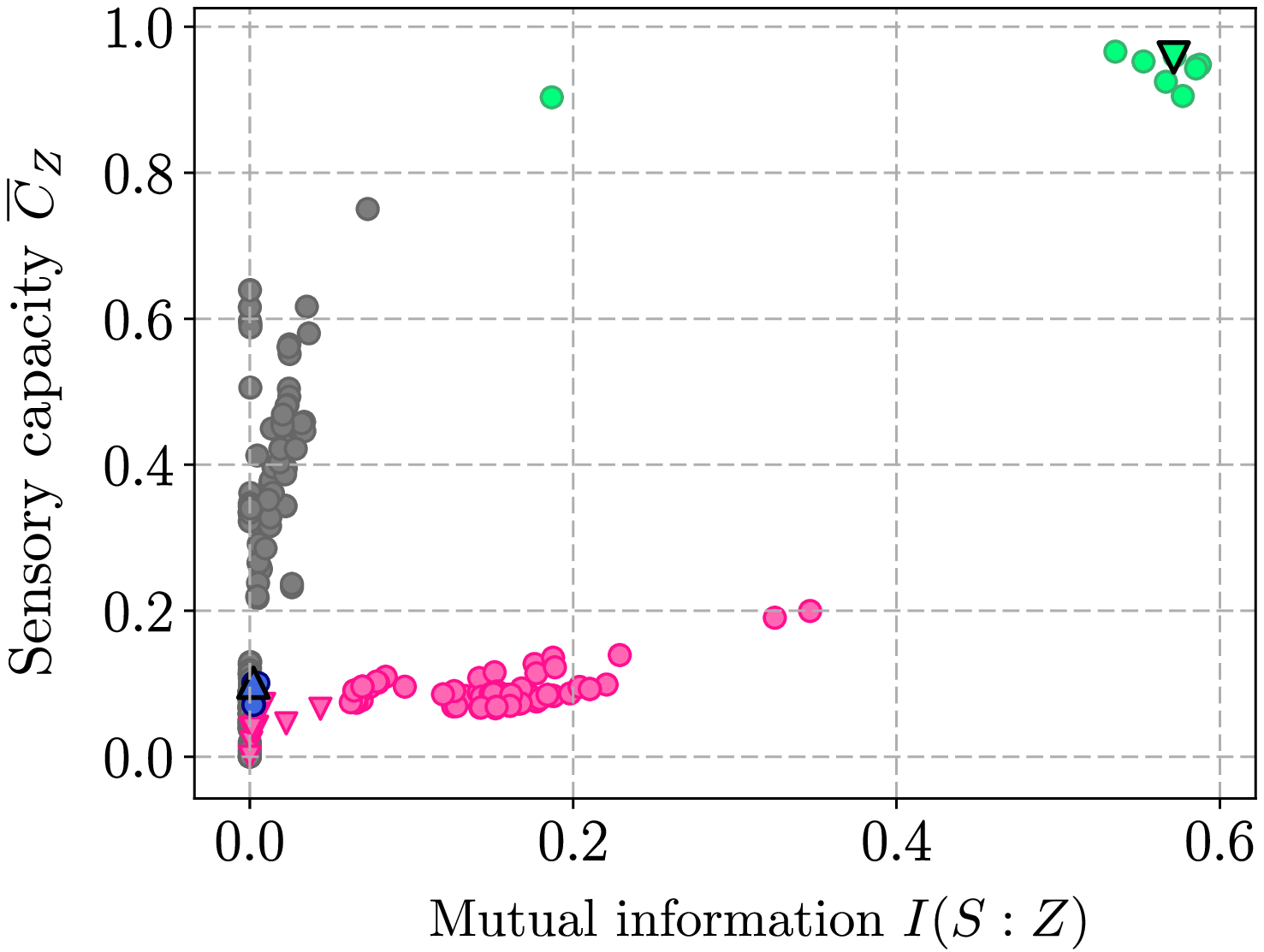}\label{fig: S100_CSZ}}&
	\subfigure[$\gamma_S = 1$]{\includegraphics[width = 0.41\linewidth]{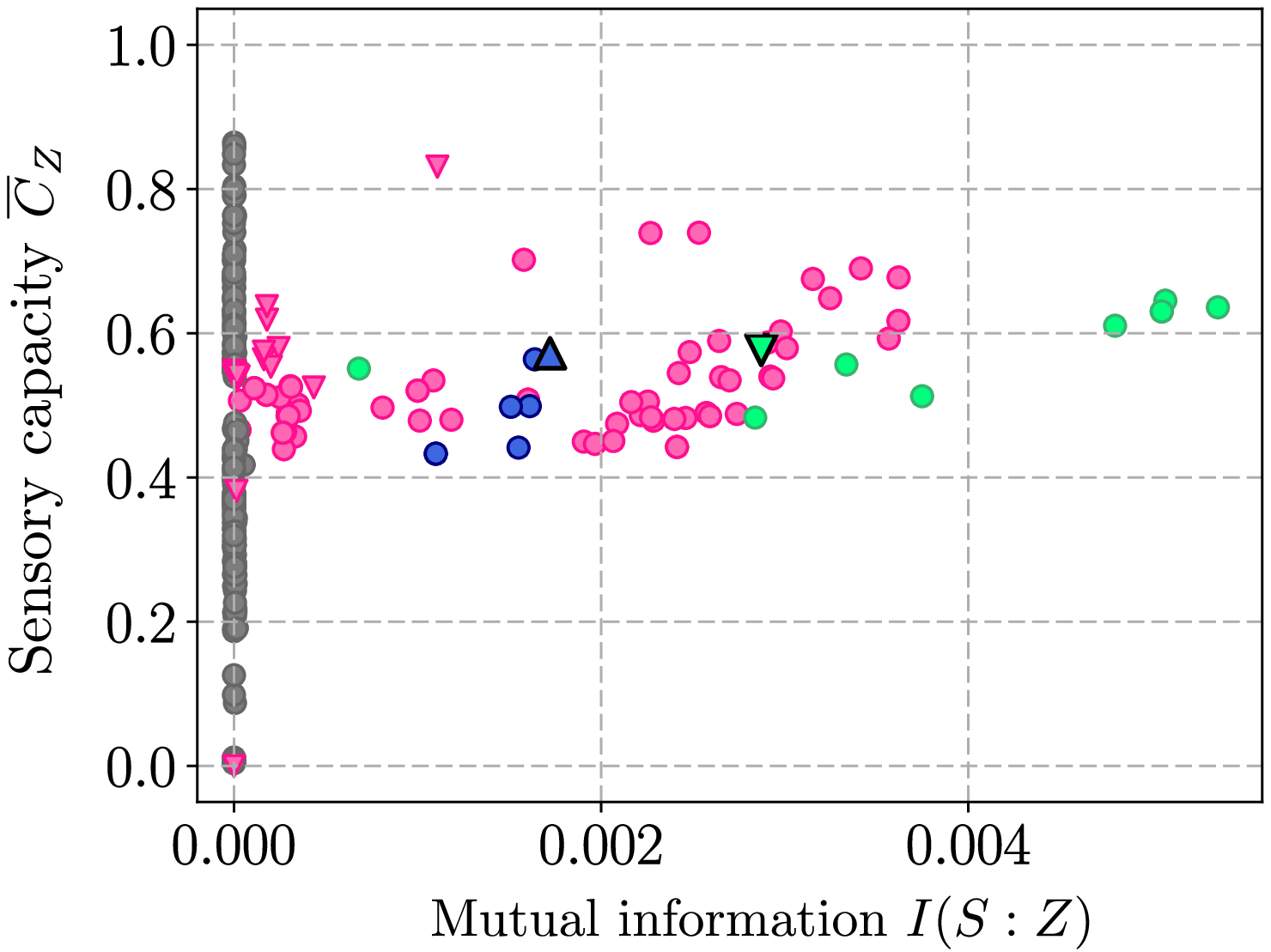}\label{fig: S1_CSZ}}\\
\end{tabular}
\end{center}
\end{figure}

\begin{figure}[H]
\begin{center}
\begin{tabular}{cc}
	\subfigure[$\gamma_X = 0.5, ~\gamma_Y = 1, ~\gamma_Z = 2$]{\includegraphics[width = 0.41\linewidth]{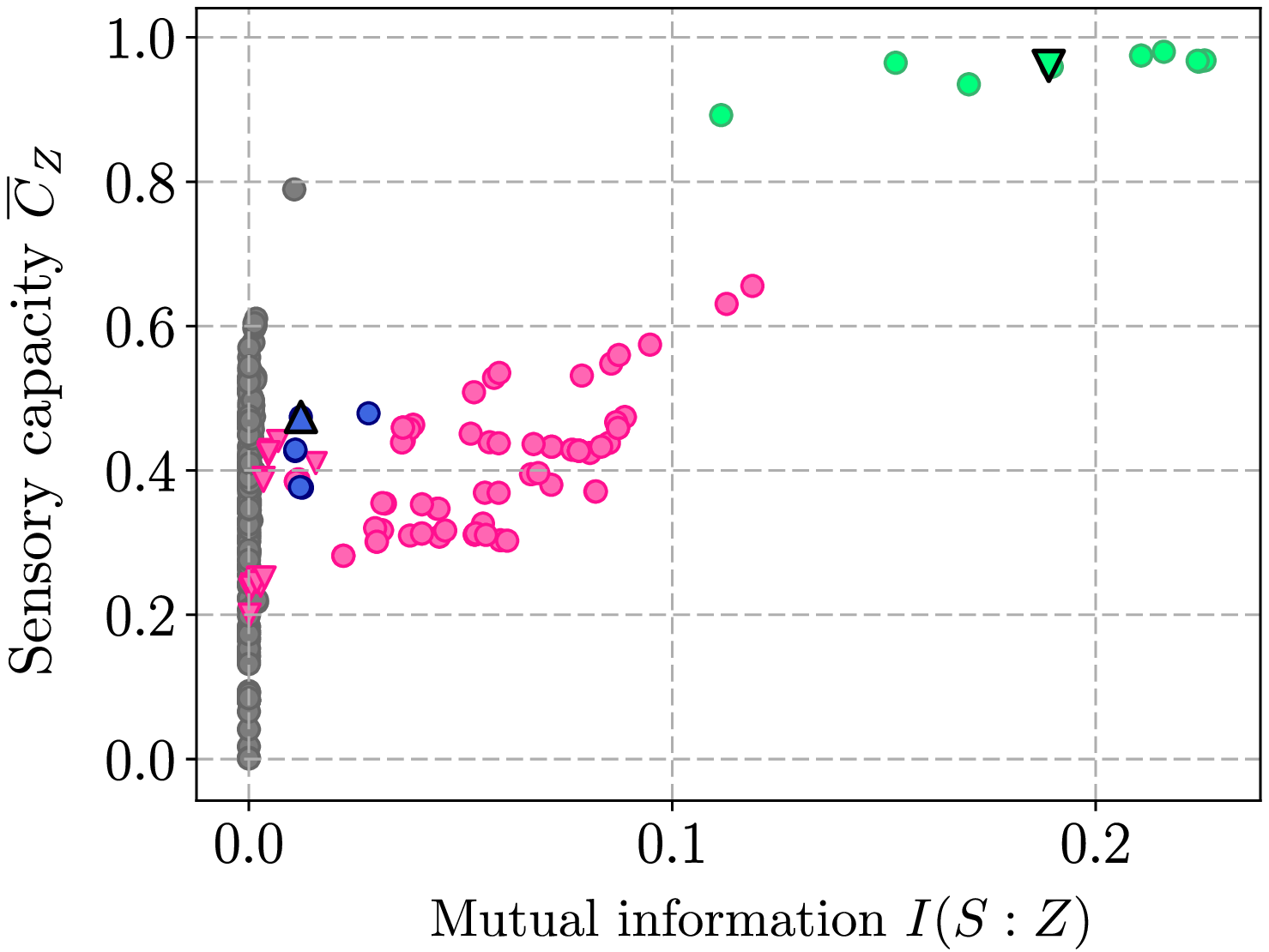}\label{fig: diff_CSZ}}&
	\subfigure[$e_X = 0.01, ~e_Y = 0.02, ~e_Z = 0.005$]{\includegraphics[width = 0.41\linewidth]{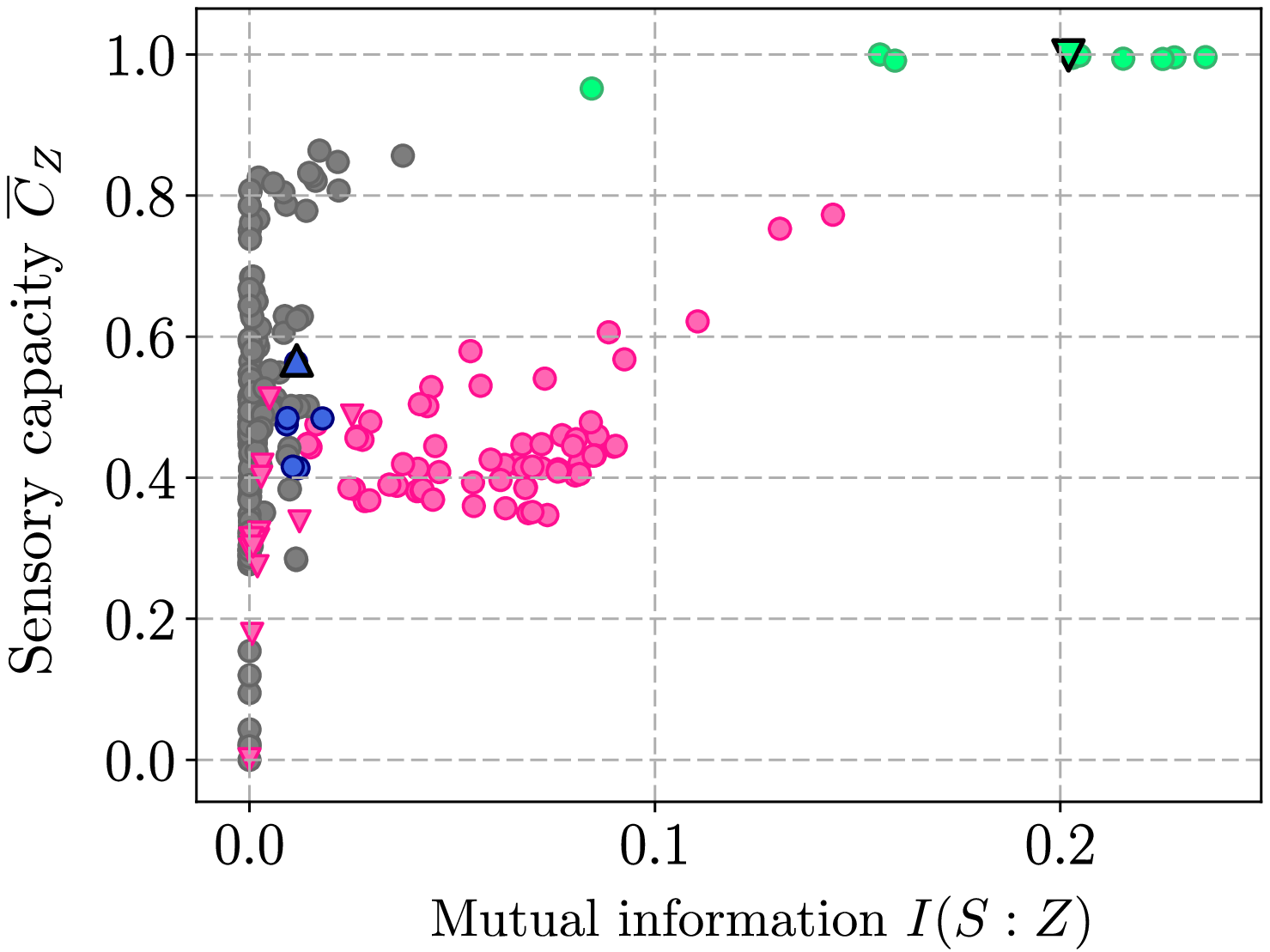}\label{fig: ediff_CSZ}}\\
\end{tabular}
\end{center}
\end{figure}

\begin{figure}[H]
\begin{center}
\begin{tabular}{cc}
	\subfigure[$e = 0$]{\includegraphics[width = 0.41\linewidth]{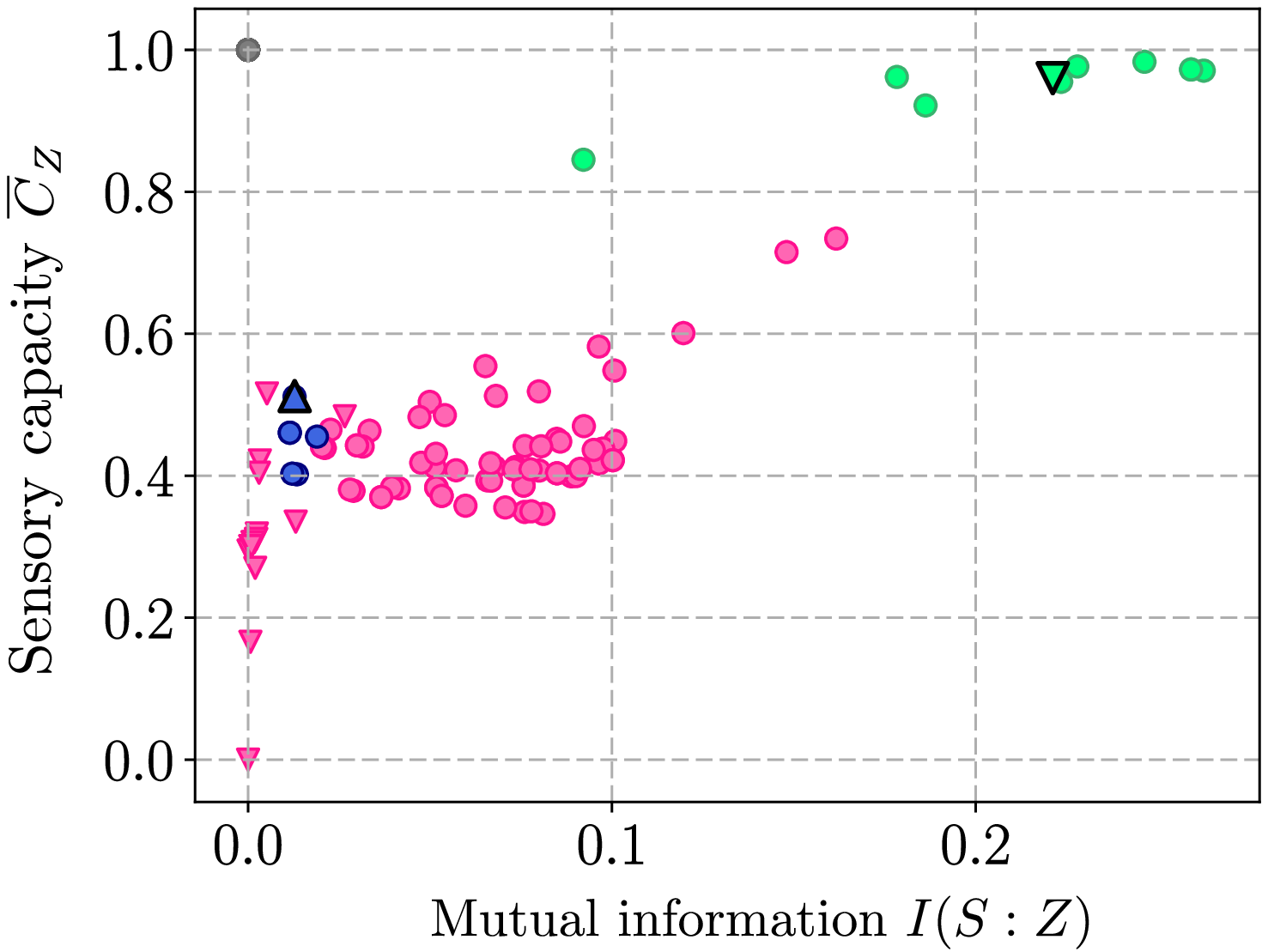}\label{fig: e0_CSZ}}&
	\subfigure[$e = 0.1$]{\includegraphics[width = 0.41\linewidth]{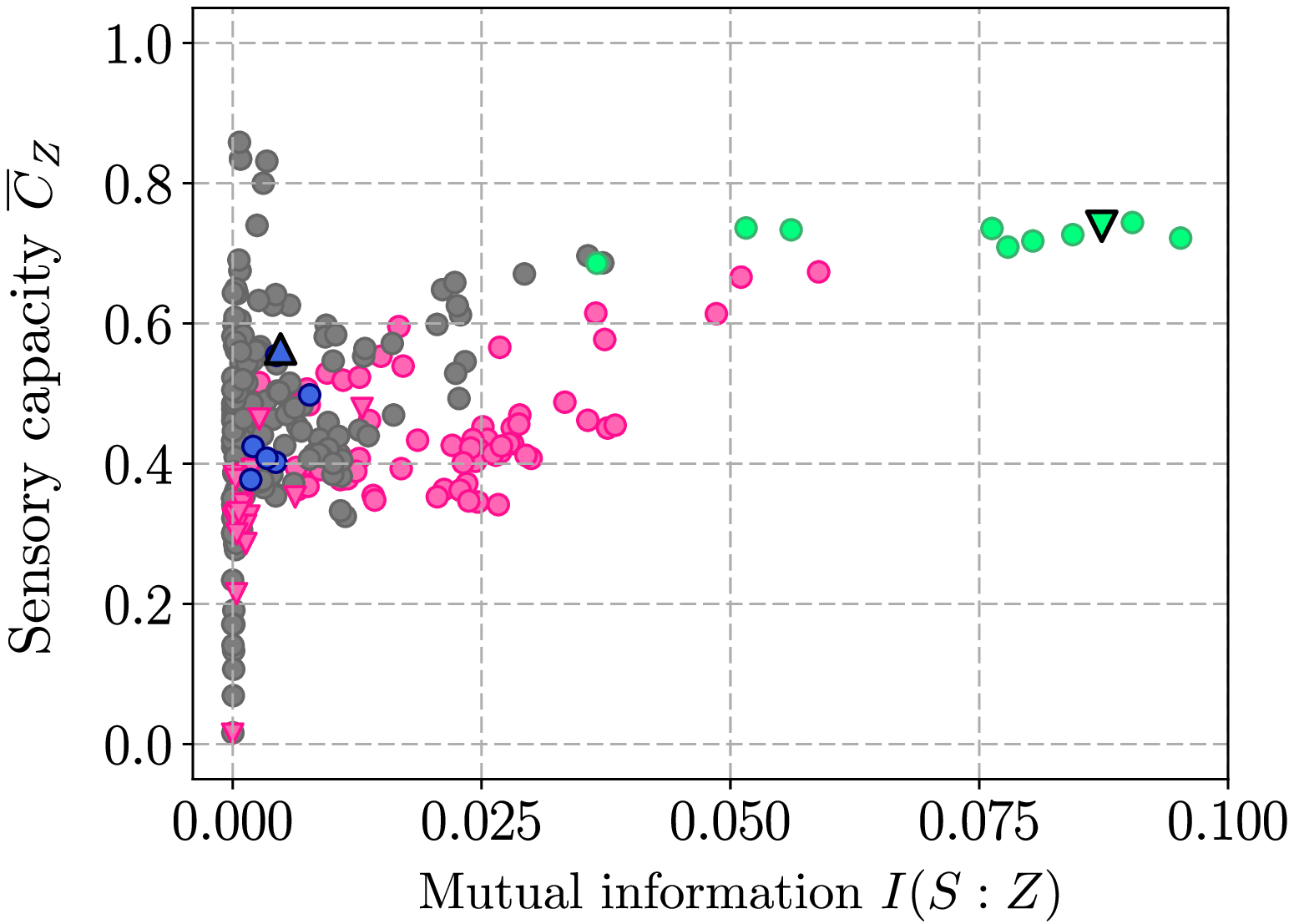}\label{fig: e01_CSZ}}\\
\end{tabular}
\end{center}
\caption{Parameter dependence of the mutual information and the sensory capacity: (a) The case where the signal changes quite slowly compared to the characteristic times of the three nodes. (b) The case where the signals changes as fast as the characteristic times of the three nodes. (c) The case where the transition rates are different in three nodes. (d) The case where the reverse transition ratios are different in the three nodes. (e) The case where the reverse transition ratios are zero. (f) The case where the reverse transition ratios are large. In any case, the other parameters are set to the same values as those of Fig.~\ref{fig: main1}.}
\label{fig: param3}
\end{figure}

\begin{figure}[H]
\begin{center}
\begin{tabular}{cc}
	\subfigure[$\gamma_S = 0.01$]{\includegraphics[width = 0.41\linewidth]{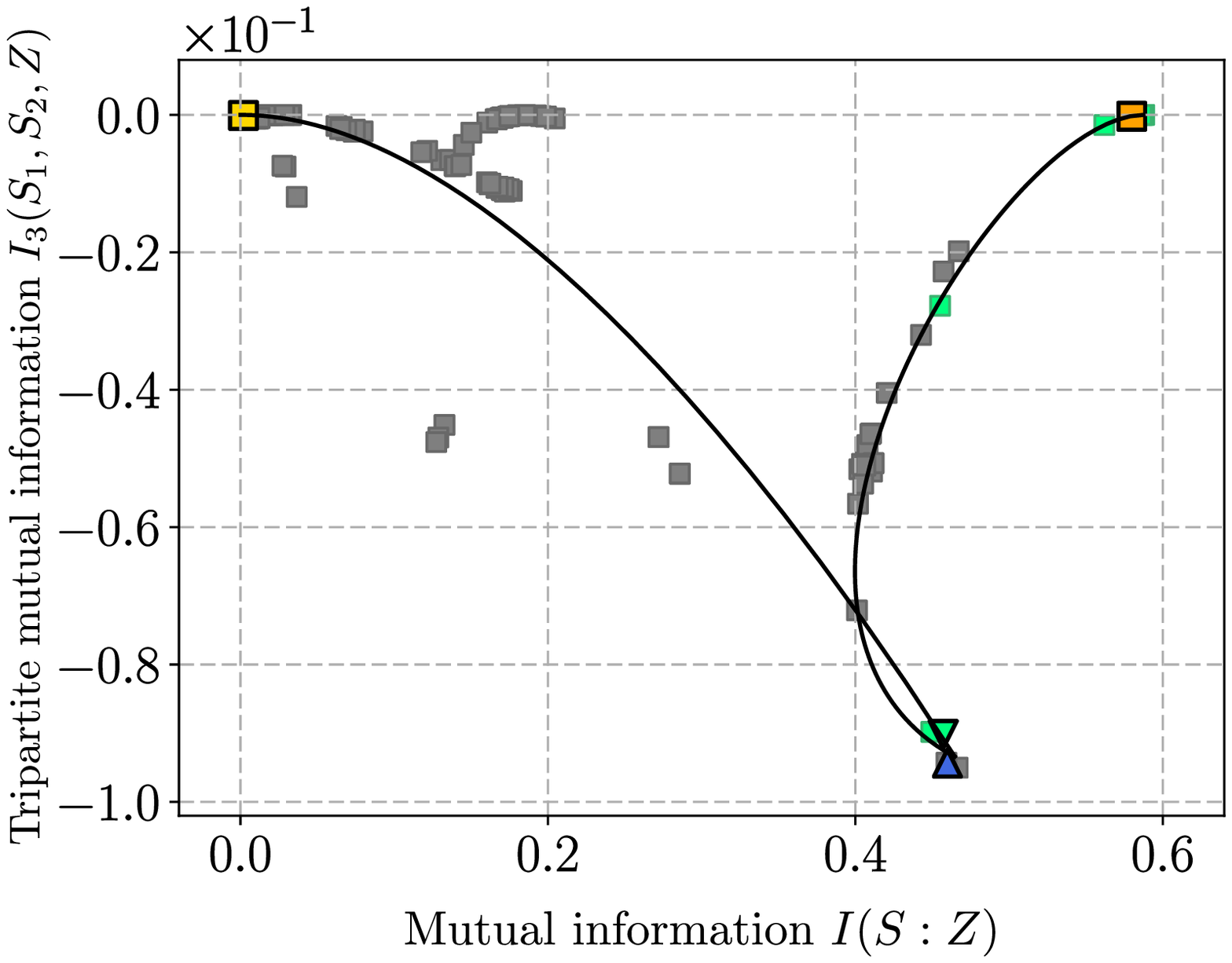}\label{fig: S100_I3}}&
	\subfigure[$\gamma_S = 1$]{\includegraphics[width = 0.41\linewidth]{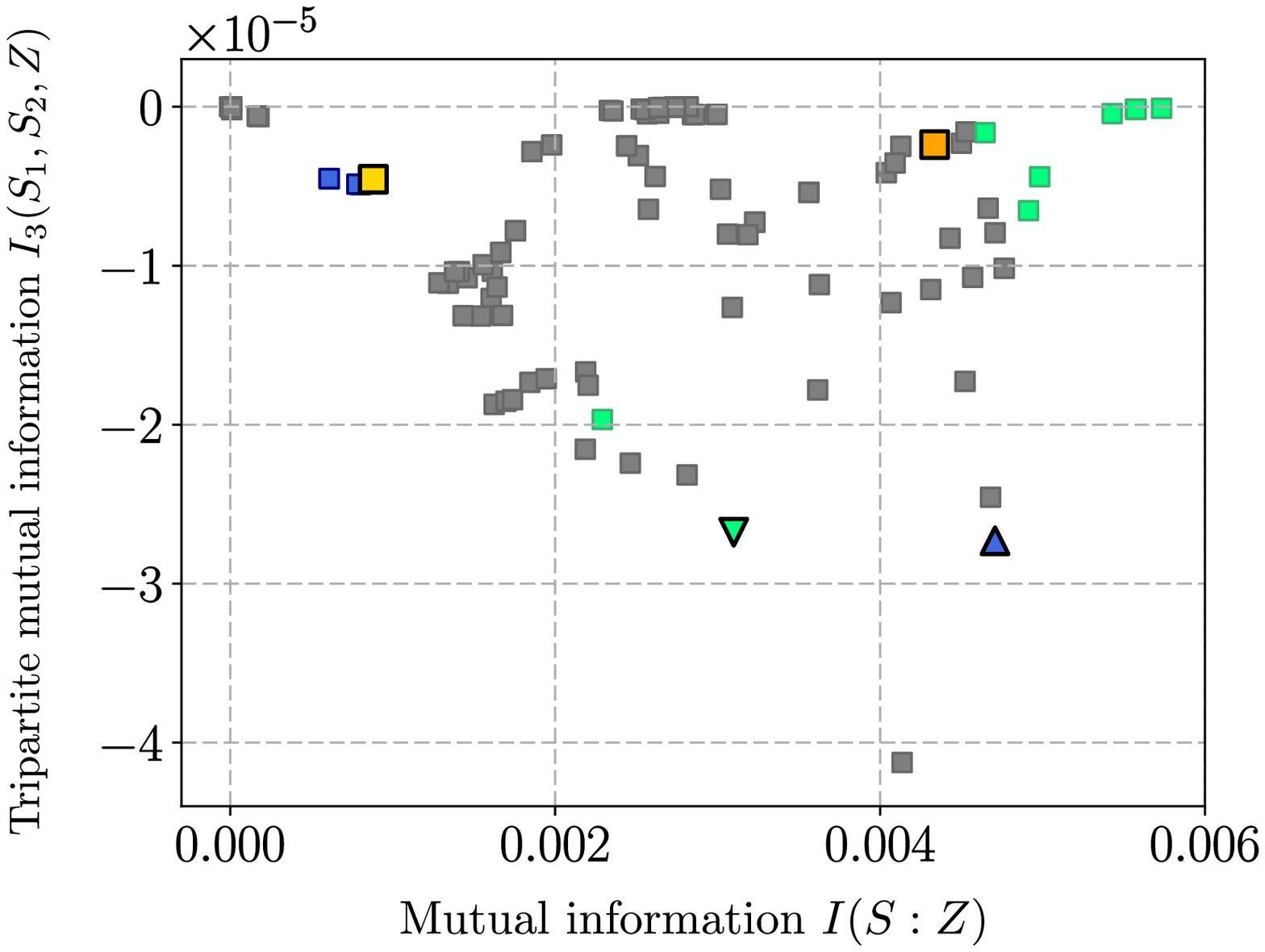}\label{fig: S1_I3}}\\
\end{tabular}
\end{center}
\end{figure}

\begin{figure}[H]
\begin{center}
\begin{tabular}{cc}
	\subfigure[$\gamma_X = 0.5, ~\gamma_Y = 1, ~\gamma_Z = 2$]{\includegraphics[width = 0.41\linewidth]{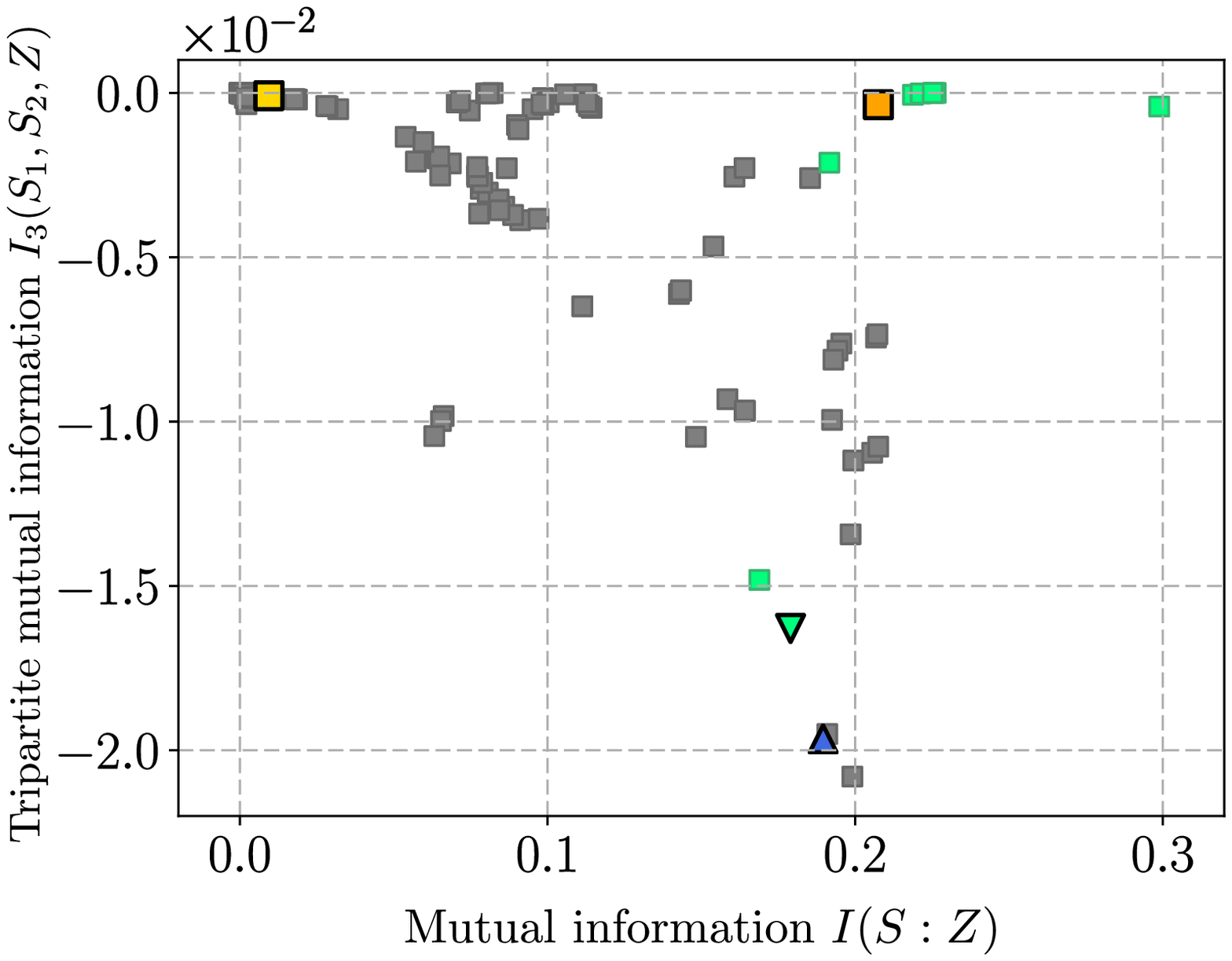}\label{fig: diff_I3}}&
	\subfigure[$e_X = 0.01, ~e_Y = 0.02, ~e_Z = 0.005$]{\includegraphics[width = 0.41\linewidth]{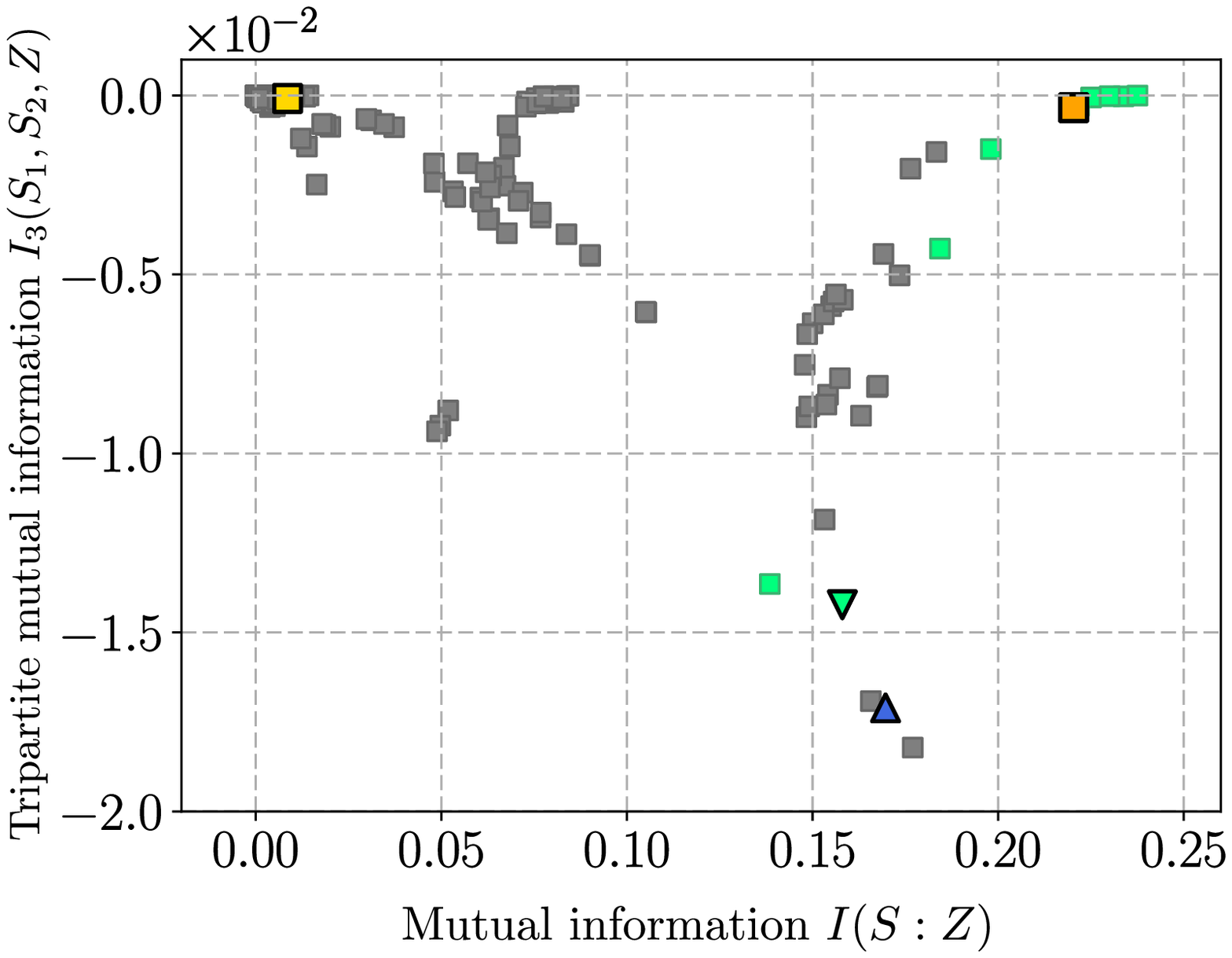}\label{fig: ediff_I3}}\\
\end{tabular}
\end{center}
\end{figure}

\begin{figure}[H]
\begin{center}
\begin{tabular}{cc}
	\subfigure[$e = 0$]{\includegraphics[width = 0.41\linewidth]{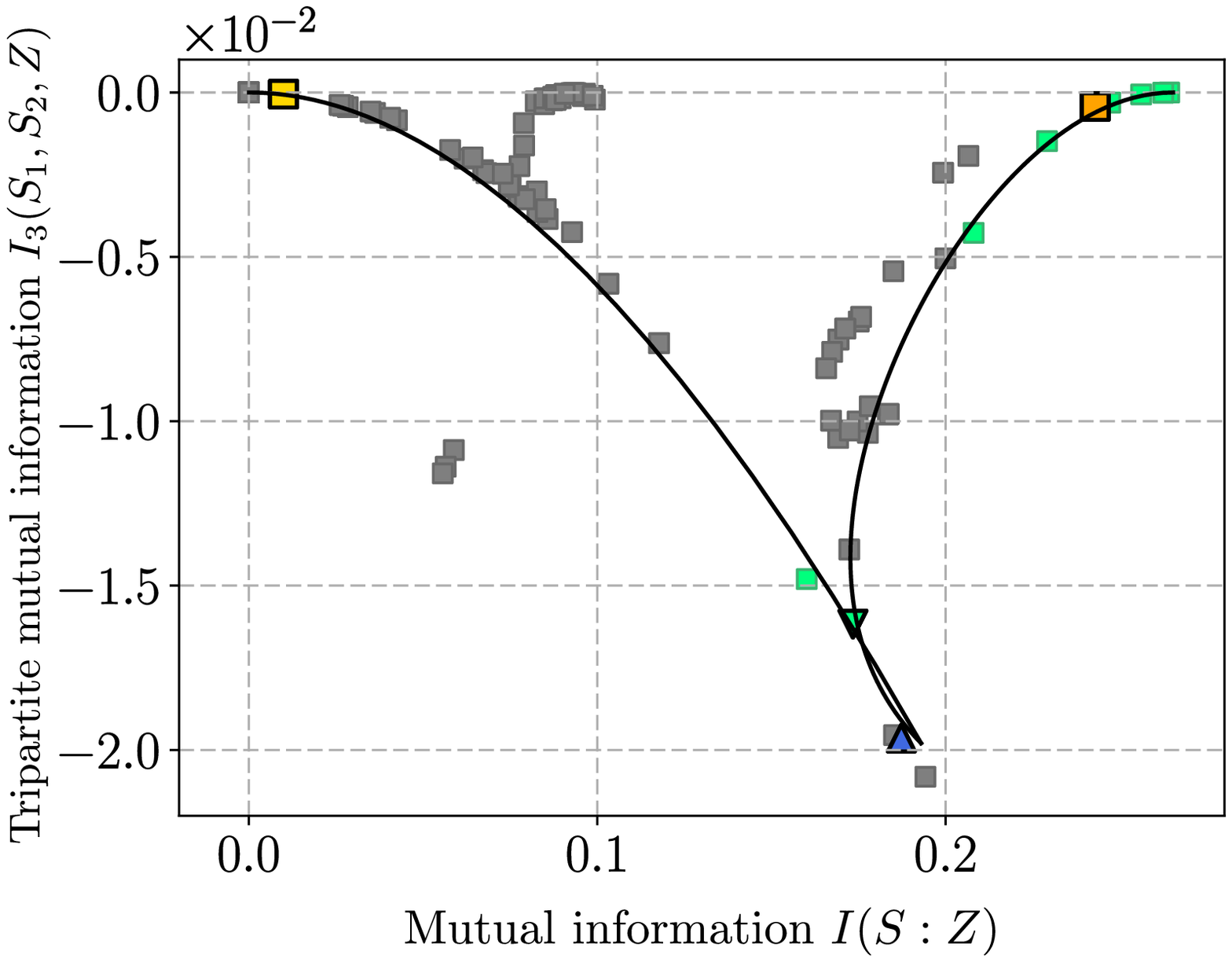}\label{fig: e0_I3}}&
	\subfigure[$e = 0.1$]{\includegraphics[width = 0.41\linewidth]{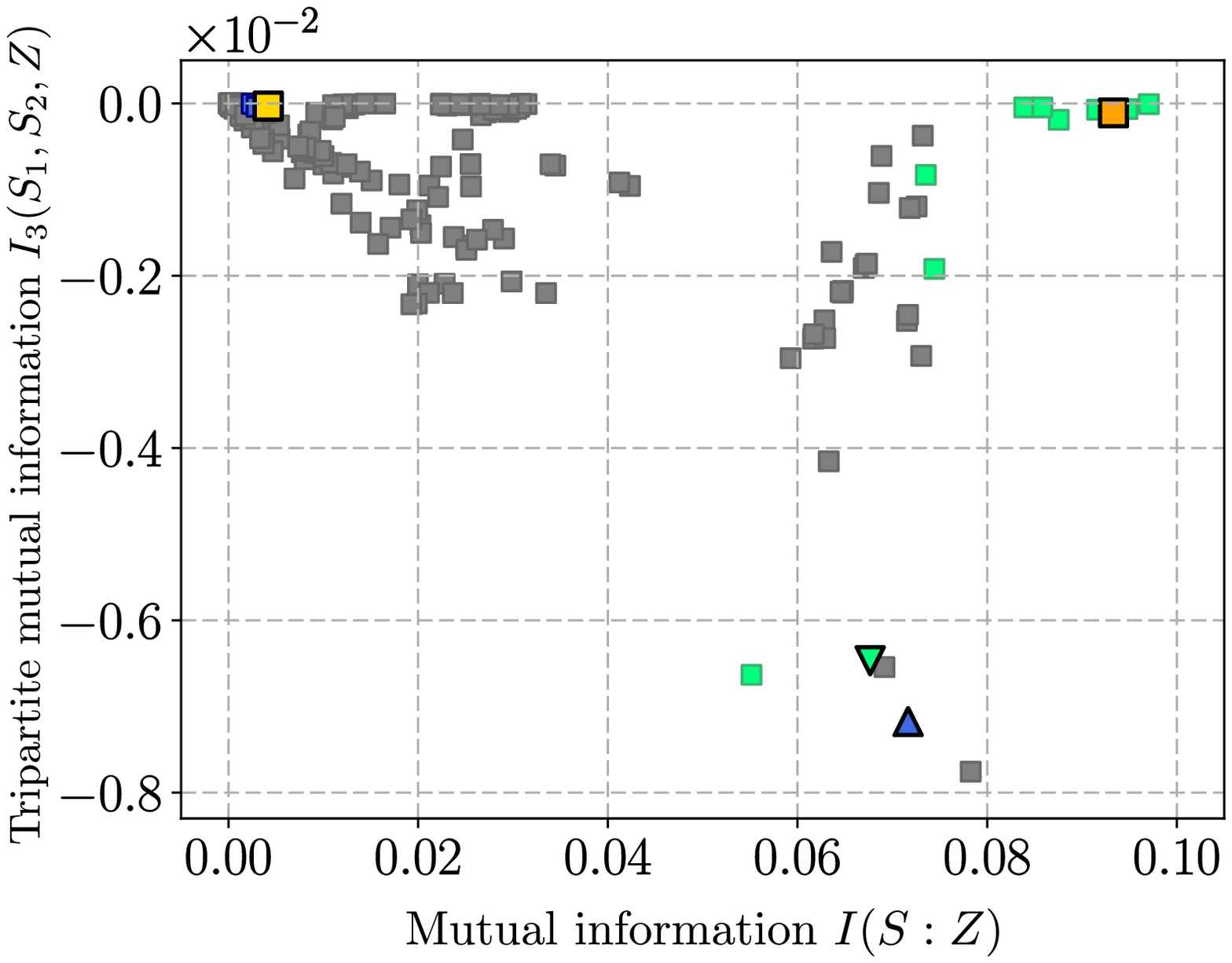}\label{fig: e01_I3}}\\
\end{tabular}
\end{center}
\caption{Parameter dependence of the tripartite mutual information: (a) The case where the signal changes quite slowly compared to the characteristic times of the three nodes. (b) The case where the signals changes as fast as the characteristic times of the three nodes. (c) The case where the transition rates are different in three nodes. (d) The case where the reverse transition ratios are different in the three nodes. (e) The case where the reverse transition ratios are zero. (f) The case where the reverse transition ratios are large. In any case, the other parameters are set to the same values as those of Fig.~\ref{fig: main2}. The fitting parameter are given by (a) $\varepsilon_0 = 0.022$, (e) $\varepsilon_0 = 0.153$.}
\label{fig: param4}
\end{figure}
\end{widetext}



\newpage
\widetext
\begin{center}
\textbf{\large Supplemental Material}
\end{center}

\setcounter{equation}{0}
\setcounter{figure}{0}
\setcounter{table}{0}
\setcounter{page}{1}
\makeatletter
\renewcommand{\theequation}{S\arabic{equation}}
\renewcommand{\thefigure}{S\arabic{figure}}
\renewcommand{\bibnumfmt}[1]{[S#1]}

\begin{figure}[H]
\begin{center}
\includegraphics[width = 12.5cm]{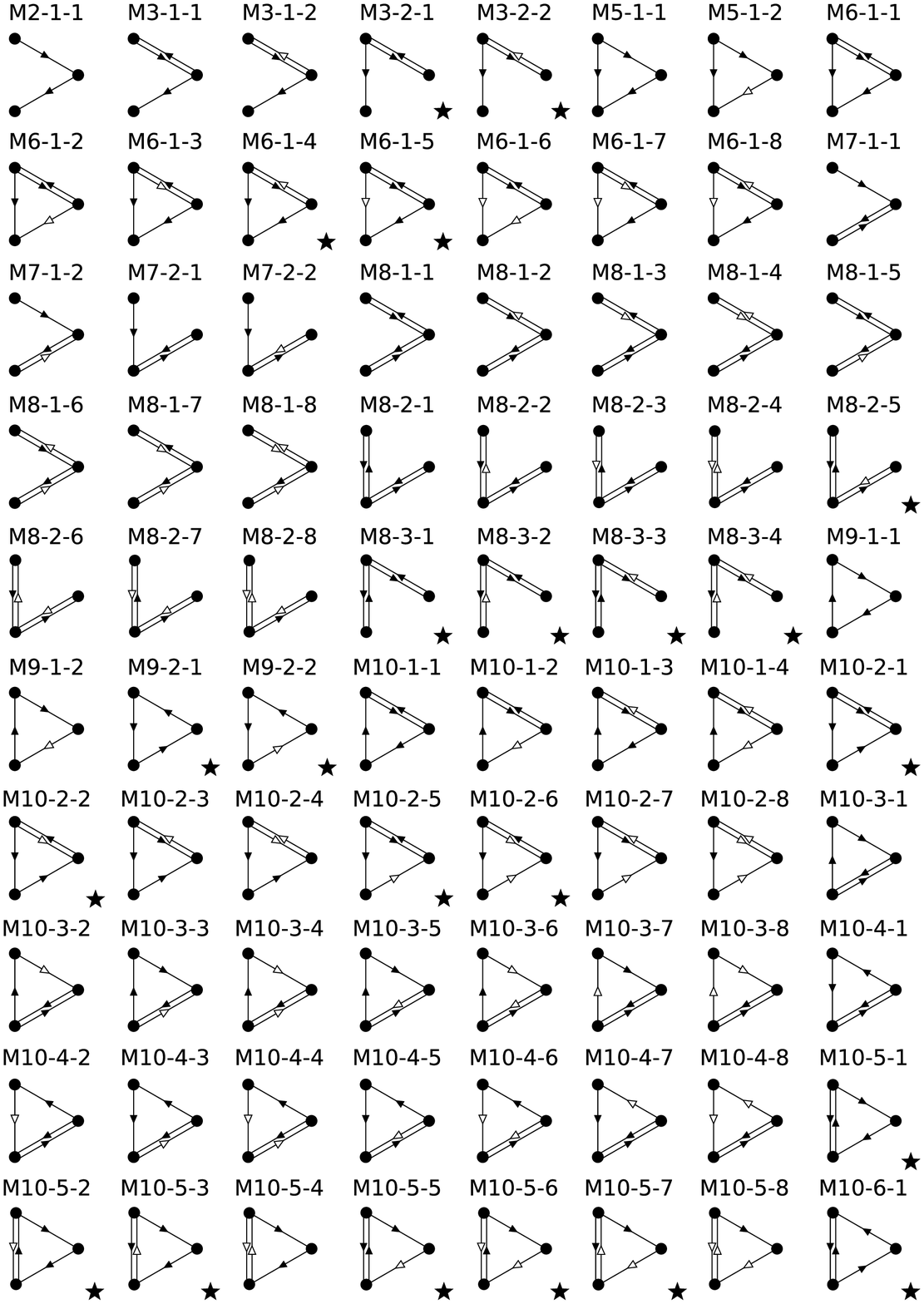}
\caption{List 1: These patterns are included for calculation in the setup of Fig.~\ref{fig: setup}, while those with ★ marks at lower right are excluded from calculation in the setup of Fig.~\ref{fig: setup2}.}
\label{fig: regu1}
\end{center}
\end{figure}

\begin{figure}[H]
\vspace{2cm}
\begin{center}
\includegraphics[width = 14cm]{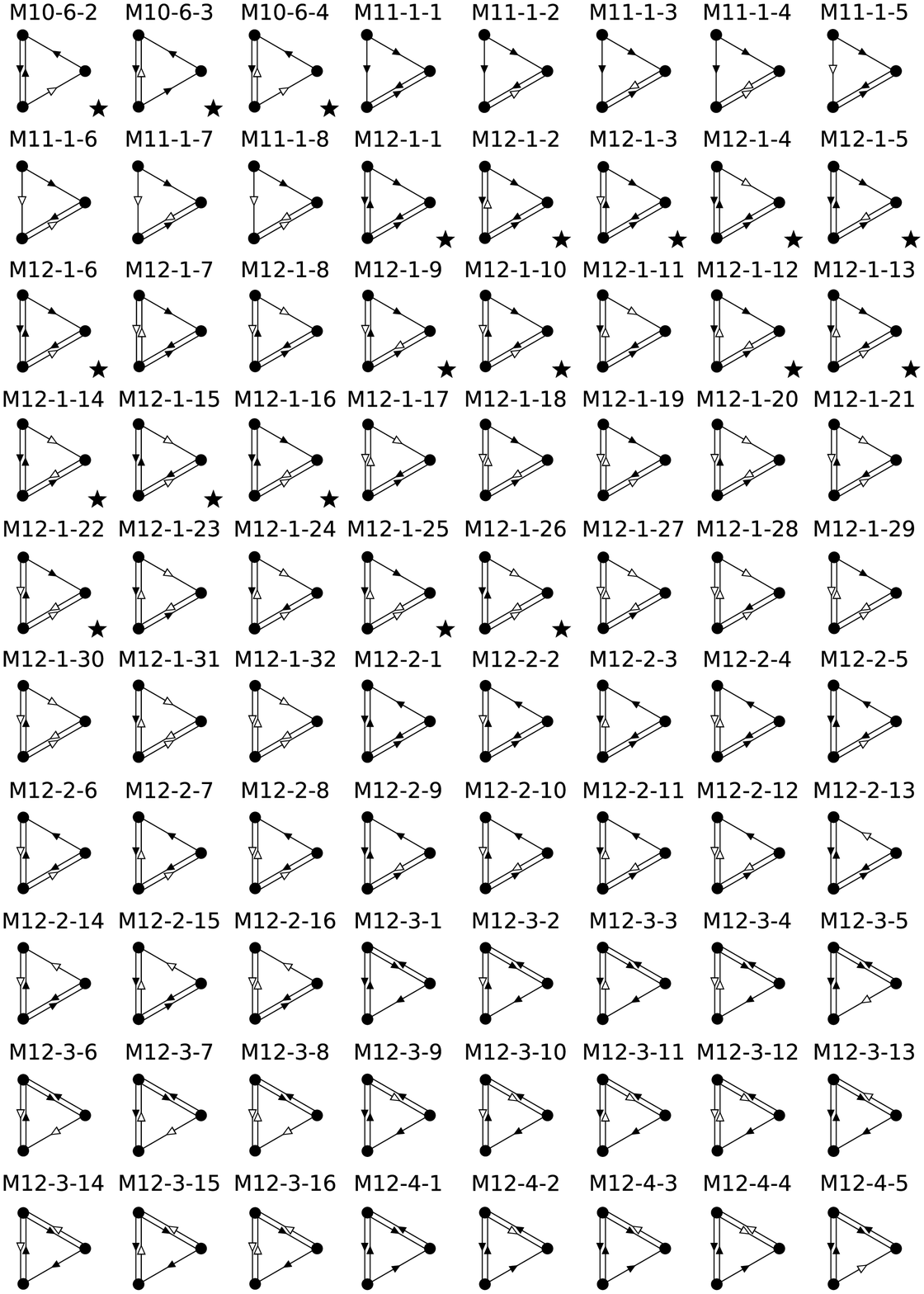}
\caption{List 2: These patterns are included for calculation in the setup of Fig.~\ref{fig: setup}, while those with ★ marks at lower right are excluded from calculation in the setup of Fig.~\ref{fig: setup2}.}
\label{fig: regu2}
\end{center}
\end{figure}

\begin{figure}[H]
\vspace{2cm}
\begin{center}
\includegraphics[width = 14cm]{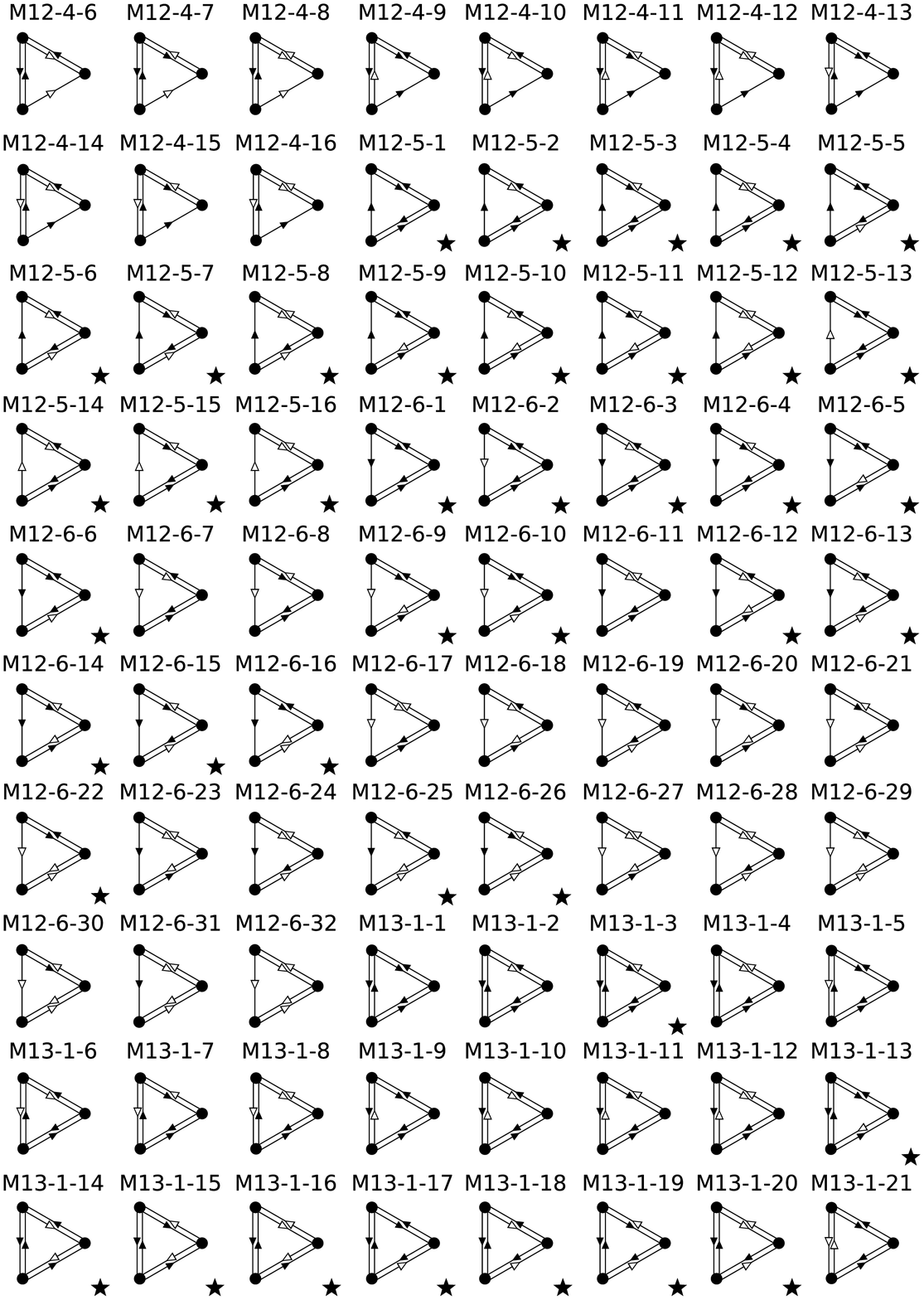}
\caption{List 3: These patterns are included for calculation in the setup of Fig.~\ref{fig: setup}, while those with ★ marks at lower right are excluded from calculation in the setup of Fig.~\ref{fig: setup2}.}
\label{fig: regu3}
\end{center}
\end{figure}

\begin{figure}[H]
\vspace{2cm}
\begin{center}
\includegraphics[width = 14cm]{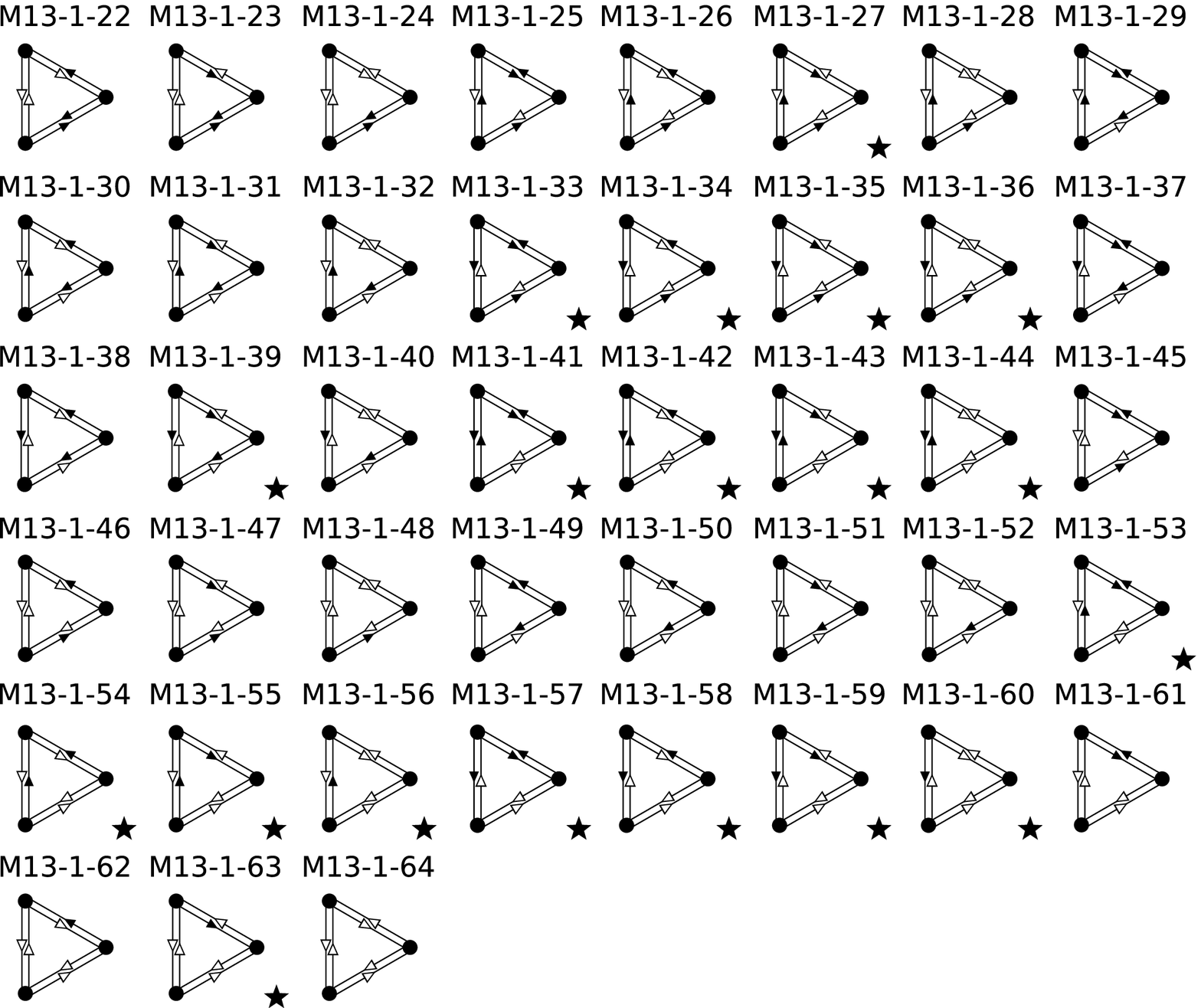}
\caption{List 4: These patterns are included for calculation in the setup of Fig.~\ref{fig: setup}, while those with ★ marks at lower right are excluded from calculation in the setup of Fig.~\ref{fig: setup2}.}
\label{fig: regu4}
\end{center}
\end{figure}

\begin{figure}[H]
\begin{center}
\includegraphics[width = 14cm]{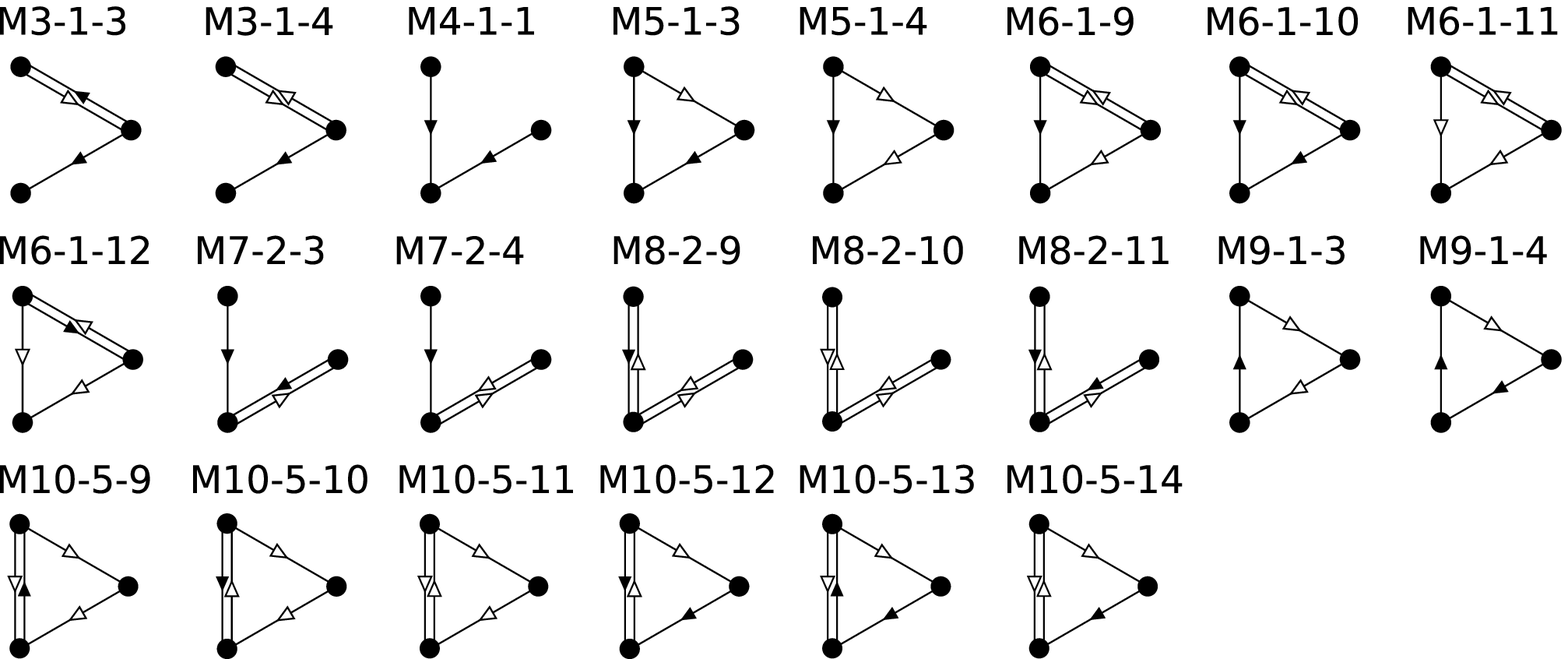}
\caption{List 5: These patterns are included for calculation only in the setup of Fig.~\ref{fig: setup2}.}
\label{fig: regu5}
\end{center}
\end{figure}

In this Supplemental Material, we list all the three-node patterns that are included for our calculation.
Patterns in Fig.~\ref{fig: regu1} to Fig.~\ref{fig: regu4} represent those included for calculation in the setup of Fig.~\ref{fig: setup} in the main text.
We exclude the patterns which have ★ marks at lower right from our calculation in the setup of Fig.~\ref{fig: setup2}.
On the other hand, patterns in Fig.~\ref{fig: regu5} represent those included for calculation only in the setup of Fig.~\ref{fig: setup2}.
An upper left node represents $X$, a middle node $Y$, and a lower left node $Z$ in each three-node pattern.
The first number of the name represents the difference of the shape (which is determined to be consistent with that in Refs. \cite{Milo2002, Alon2006}), and the second number represents the pattern's direction, and the third number represents the difference due to the different signs of edges.\\ 

\end{document}